\documentclass{article} % For LaTeX2e
\usepackage{iclr2024_conference,times}

\usepackage{hyperref}
\usepackage{url}

\usepackage{xspace}
\usepackage{booktabs}
\usepackage{caption}
\usepackage{subcaption}
\usepackage{graphicx}
\usepackage{float}
\usepackage[normalem]{ulem}
\usepackage{multirow}
\usepackage{amsmath}
\usepackage{enumitem}

\usepackage{balance}
\usepackage[mathscr]{euscript} 
\usepackage{mathtools}
\usepackage{wrapfig}
\usepackage{algorithm}
\usepackage[noend]{algorithmic}
\usepackage{xifthen}% provides \isempty test

\usepackage{mkolar_definitions}

\newcommand{\zeshel}{\textsc{ZeShEL}\xspace}
\newcommand{\beir}{\textsc{BeIR}\xspace}

\newcommand{\yugioh}{\textsf{YuGiOh}\xspace}
\newcommand{\starTrek}{\textsf{Star Trek}\xspace}

\newcommand{\scidocs}{\textsf{SciDocs}\xspace}
\newcommand{\hotpotqa}{\textsf{Hotpot-QA}\xspace}

\newcommand{\query}{\ensuremath{q}\xspace}
\newcommand{\dataItem}{\ensuremath{i}\xspace}
\newcommand{\testQuery}{\ensuremath{q_{\texttt{test}}}\xspace}

\newcommand{\querySpace}{\mathcal{Q}\xspace}
\newcommand{\itemSpace}{\mathcal{I}\xspace}

\newcommand{\nItems}{\lvert \itemSpace \rvert\xspace}

\newcommand{\queryTestData}{\mathcal{Q}_\text{test}}
\newcommand{\queryTrainData}{\mathcal{Q}_\text{train}}
\newcommand{\queryTestSize}{\lvert \queryTestData \rvert}
\newcommand{\queryTrainSize}{\lvert \queryTrainData \rvert}

\newcommand{\model}[1]{f_{#1}\xspace}

\newcommand{\eCrossenc}{\textsc{[emb]-CE}\xspace}

\newcommand{\kDistill}{k_d}

\newcommand{\tfidf}{\textsc{tf-idf}\xspace}
\newcommand{\bmtwofive}{\textsc{bm25}\xspace}

\newcommand{\baseDualEncoder}{\ensuremath{\textsc{DE}_{\textsc{src}}}\xspace}
\newcommand{\distillDualEncoder}{\ensuremath{\textsc{DE}_{\textsc{dstl}}}\xspace}

\newcommand{\adaCUR}[1]{ \ensuremath{\textsc{adaCUR}_{#1}\xspace} }

\newcommand{\ceBudget}{\mathcal{B}_\textsc{ce}}
\newcommand{\nRounds}{\mathscr{R}}
\newcommand{\roundIter}{r}
\newcommand{\nAnchorItemsPerRound}{k_s}

\newcommand{\ce}{\ensuremath{f}}

\newcommand{\trueKnnSet}{\ensuremath{\mathscr{N}}}
\newcommand{\argtopk}{\mathop{\mathrm{\arg top}k}}
\newcommand{\pairs}{\ensuremath{\mathscr{P}}}
\newcommand{\trainsplit}{\ensuremath{\textrm{train}}}

\newcommand{\approxce}{\ensuremath{\hat{f}}}
\newcommand{\approxKnnSet}{\ensuremath{\hat{\mathscr{N}}}}
\newcommand{\dimensionality}{\ensuremath{d}}
\newcommand{\qvector}{\ensuremath{\mathbf{u}}}
\newcommand{\ivector}{\ensuremath{\mathbf{v}}}
\DeclarePairedDelimiter{\norm}{\|}{\|}
\newcommand{\adaptiveSelection}[1]{\mathscr{A}_{#1}}
\newcommand{\adaptiveSelectionEmbeds}[1]{V_{\adaptiveSelection{#1}}}
\newcommand{\adaptiveSelectionScores}[1]{\mathbf{a}_{#1}}

\newcommand{\exact}{\textsc{Exact}\xspace}
\newcommand{\rnr}[1]{\textsc{RnR}\textsubscript{#1}\xspace}
\newcommand{\propInf}[2]{\textsc{Axn}\if\relax\detokenize{#1}\relax\else\textnormal{\textsubscript{#1,#2}}\fi\xspace}
\newcommand{\matrixFact}[1]{\textsc{MF}\textsubscript{#1}\xspace}
\newcommand{\inductive}{\textsc{Ind}\xspace}
\newcommand{\transductive}{\textsc{Trns}\xspace}
\newcommand{\sparseMat}{\ensuremath{G}\xspace}
\newcommand{\tour}{\textsc{TouR}\xspace}
\newcommand{\tourCE}{\textsc{TouR-ce}\xspace}
\newcommand{\tourMSE}{\textsc{TouR-mse}\xspace}
\newcount\Comments  % 0 suppresses notes to selves in text
\definecolor{purple}{rgb}{0.5,0,1}
\definecolor{teal}{rgb}{0.33,0.65,0.55}
\definecolor{green}{rgb}{0.1,0.65,0.1}
\newcommand{\kibitz}[2]{\ifnum\Comments=1\textcolor{#1}{#2}\fi}

\title{Adaptive Retrieval and Scalable Indexing for $k$-NN Search with Cross-Encoders}

\author{Nishant Yadav$^{1\thanks{Now at Google DeepMind}}$, Nicholas Monath$^2$, Manzil Zaheer$^2$, Rob Fergus$^2$, Andrew McCallum$^1$ \\
\hspace{2.5cm}$^1$ University of Massachusetts Amherst, $^2$ Google DeepMind
}

\iclrfinalcopy % Uncomment for camera-ready version, but NOT for submission.
\begin{document}

\maketitle

\begin{abstract}
Cross-encoder (CE) models which compute similarity
by jointly encoding a query-item pair perform better than
using dot-product with embedding-based models (dual-encoders) 
at estimating query-item relevance.
Existing approaches perform $k$-NN search with cross-encoders by 
approximating the CE similarity with a vector embedding space fit either with 
dual-encoders (DE) or CUR matrix factorization.
DE-based retrieve-and-rerank approaches suffer from poor 
recall as DE generalizes poorly to new domains and
the test-time retrieval with DE is decoupled
from the CE.
While CUR-based approaches can be more accurate than
the DE-based retrieve-and-rerank approach, such approaches
require a prohibitively large number of CE calls
to compute item embeddings, thus making it impractical 
for deployment at scale.
In this paper, we address these shortcomings with our proposed sparse-matrix factorization based method
that efficiently computes latent query and item representations to 
approximate CE scores and performs $k$-NN search with the approximate CE similarity.
In an offline indexing stage, we compute item embeddings by
factorizing a sparse matrix containing query-item CE scores
for a set of train queries.
Our method produces a high-quality approximation while 
requiring only a fraction of CE similarity calls 
as compared to CUR-based methods, and allows for leveraging 
DE models to initialize the embedding space while 
avoiding compute- and resource-intensive 
finetuning of DE via distillation.
At test time, we keep item embeddings fixed and perform retrieval over multiple rounds, 
alternating between a) estimating the test query embedding by 
minimizing error in approximating CE scores of items retrieved thus far,
and b) using the updated test query embedding for retrieving more items in the next round.
Our proposed $k$-NN search method can achieve up to 5\% and 54\% improvement 
in $k$-NN recall for $k=1$ and 100 respectively over the widely-used DE-based retrieve-and-rerank approach.
Furthermore, our proposed approach to index the items by aligning item embeddings with the CE 
achieves up to 100$\times$ and 5$\times$ speedup over CUR-based and 
dual-encoder distillation based approaches respectively while matching or improving
$k$-NN search recall over baselines.
\end{abstract}

\section{Introduction}
Efficient and accurate nearest neighbor search is paramount for 
retrieval~\citep{menon2022defense,rosa2022defense,qu-etal-2021-rocketqa}, 
classification in large output spaces 
(e.g., entity linking~\citep{ayoola2022refined,logeswaran-etal-2019-zero,wu-etal-2020-scalable}), 
non-parametric models~\citep{das2022knowledge,wang-etal-2022-training}, 
and many other such applications in machine learning~\citep{pmlr-v162-goyal22a,izacard2022few,bahri2020deep}. 
The accuracy and efficiency of nearest neighbor search depends on a combination of factors 
(1) the computational cost of pairwise distance comparisons between datapoints, 
(2) preprocessing time for constructing a nearest neighbor index 
(e.g., dimensionality reduction \citep{indyk2000dimensionality}, 
quantization~\citep{ge2013optimized,guo2020accelerating}, 
data structure construction~\citep{beygelzimer2006cover,malkov2018efficient,zaheer2019sg}), 
and (3) the time taken to query the index to retrieve the nearest neighbor(s).

Similarity functions such as cross-encoders which take a pair of 
data points as inputs and directly output a scalar score, have achieved 
state-of-the-art results on numerous tasks (e.g., QA \citep{qu-etal-2021-rocketqa, thakur2021beir}, 
entity linking \citep{logeswaran-etal-2019-zero}). 
However, these models are exceptionally computationally expensive since these are typically parameterized by several layers of neural models such as transformers~\citep{vaswani2017attention},
and scoring each item for a given query requires a forward pass of the large parametric model, 
making them impractical similarity functions to use directly in nearest neighbor indices~\citep{yadav2022efficient}. 
Initial work has approximated search with cross-encoders (CE) for a given test query 
using a heuristic retrieve-and-rerank approach that uses a separate model 
to retrieve a subset of items followed by re-ranking using the CE. 
Prior work performs the initial retrieval using dot-product of sparse query/item embedding 
from models such as BM25, or 
dense query/item embeddings from models such as dual-encoders (DE) which are typically
trained on the same task and data as the CE.
To support search with CE, recent 
work~\citep{yadav2022efficient,yadav2023efficient} improves upon heuristic retrieve-and-rerank approaches, by directly learning an embedding space that approximates the CE score function. These approaches use CUR decomposition~\citep{mahoney2009cur} to compute (relatively) low-dimensional embeddings for queries and items. The item embeddings are computed by
scoring each item against a set of \emph{anchor/train} queries. At test-time, the test query embedding is computed by
using CE scores of the test query against a set of (adaptively-chosen) \emph{anchor} items.

Both DE-based retrieve-and-rerank and CUR-based 
methods are not well suited for a typical application setting 
in $k$-NN search -- building an index on a new set of 
targets with a given (trained) similarity function. 
The DE-based approach has several disadvantages in this setting. 
DE models show poor generalization to new domains and thus require
additional fine-tuning on the target domain to improve performance~\cite{yadav2022efficient,thakur-etal-2021-augmented}.  
This can be both resource-intensive as well time-consuming. 
Furthermore, it requires access to the parameters (not just embedding outputs) 
of the DE, which might not be possible if the DE is provided by an API service. 
On the other hand, while CUR-based approaches outperform retrieve-and-rerank 
approaches without additional fine-tuning of DE, they require computing a dense 
score matrix by scoring each item against a set of anchor/train queries. 
This does not scale well with the number of items. 
For instance, for a domain with 500 anchor/train queries and 10K items, 
it takes around 10 hours\footnote{On an Nvidia 2080ti GPU with 12 GB memory using batch size=50} 
to compute the dense query-item score matrix with a CE parameterized using 
\texttt{bert-base}~\citep{yadav2022efficient}. By simple extrapolation, 
indexing 5 million items using 500 queries would take around 5000 GPU hours.

In this paper, we propose a sparse-matrix factorization-based 
approach to improve the efficiency of fitting an embedding space to approximate the cross-encoder for $k$-NN search.
Our proposed approach significantly reduces the offline 
indexing cost as compared to existing approaches by constructing 
a sparse matrix containing cross-encoder scores between 
a set of training queries~$(\querySpace_\trainsplit)$ 
and all the items~$(\itemSpace)$, 
and using efficient matrix factorization methods to 
produce a set of item embeddings that are aligned with
the cross-encoder.
At test-time, our proposed approach, \propInf{}{}, 
computes a test query embedding to approximate
cross-encoder scores between the test query and items,
and performs retrieval using approximate cross-encoder scores.
\propInf{}{} performs retrieval over multiple rounds while 
keeping the item embedding fixed and incrementally refining the test query embedding using
cross-encoder scores of the items retrieved in previous rounds.
In the first round, the cross-encoder is used to score the test query
against a small number of items chosen uniformly at random 
or baseline retrieval methods such as dual-encoder or BM25.
In each subsequent round, \propInf{}{} alternates between 
(a) updating the test query embedding to improve 
the approximation of the cross-encoder score of items retrieved so far, and
(b) retrieving additional items using the improved approximation of the cross-encoder,
and computing the exact cross-encoder scores for the retrieved items.
Finally, the retrieved items are ranked based on exact cross-encoder scores
and the top-$k$ items returned as the $k$-nearest neighbors for the given test query.

We perform an empirical evaluation of our method using cross-encoder models 
trained for the task of entity linking and
information retrieval on \zeshel~\citep{logeswaran-etal-2019-zero} and \beir~\citep{thakur2021beir} benchmark respectively.
Our proposed $k$-NN search method can be used together with dense item
embeddings produced by any method such as baseline dual-encoder models 
and still yield up to 5\% and 54\% improvement in $k$-NN recall for $k=$1 and 100 respectively 
over retrieve-and-rerank style inference with the same dual-encoder.
Furthermore, our proposed approach to align item embeddings with the cross-encoder 
achieves up to 100$\times$ and 5$\times$ speedup over CUR-based approaches and 
training dual-encoders via distillation-based respectively while matching or improving
test-time $k$-NN search recall over baseline approaches.

\section{Proposed Approach}

\paragraph{Task Description}
 A cross-encoder model $\ce: \querySpace \times \itemSpace \rightarrow \RR$ 
 maps a query-item pair  $(\query, \dataItem) \in \querySpace \times \itemSpace$ to a scalar similarity.
 We consider the task of similarity search with the cross-encoder, in 
 particular finding the $k$-nearest neighbors items for a given query $\query$ from a fixed set of items $\itemSpace$: 
 \begin{equation}
     \trueKnnSet(\query) \triangleq \argtopk_{\dataItem \in \itemSpace} \ce(\query,\dataItem)
     \vspace{-0.2cm}
 \end{equation}
where $\argtopk$ returns the indices of the top $k$ scoring items of the function. 
Exact $k$-NN search with a cross-encoder would require $\mathcal{O}(\nItems)$
cross-encoder calls as an item needs to be jointly encoded with the test query
in order to compute its score. 
Since cross-encoders are typically parameterized using deep neural
models such as transformers~\citep{vaswani2017attention}, 
$\mathcal{O}(\nItems)$ calls to the cross-encoder
model can be prohibitively expensive at test time. 
Therefore, we tackle the task of 
approximate $k$-NN search with cross-encoder models. 
Let $\approxce(\cdot,\cdot)$ denote the approximation to the cross-encoder 
that is learned using exact cross-encoder scores
for a sample of query-item pairs.
We refer to the approximate $k$-nearest neighbors as 
$\approxKnnSet(\query)\triangleq \argtopk_{\dataItem \in \itemSpace} \approxce(\query,\dataItem)$ 
and measure the quality of the approximation using nearest neighbor recall: 
$\frac{\lvert \approxKnnSet(\query) \cap \trueKnnSet(\query)\rvert}{\lvert \trueKnnSet(\query)\rvert}$ 

In this work, we assume black-box access to the cross-encoder\footnote{Approximating a neural scoring function by compressing, 
approximating, quantizing the scoring function is widely studied but outside 
the scope of this paper.}, 
access to the set of items and train queries from the target domain,
and a base dual-encoder~(\baseDualEncoder) trained on the same
task and source data as the cross-encoder.
In~\S\ref{subsec:offline_sparse_mf_indexing}, we first present our proposed sparse-matrix factorization
based method to compute item embeddings in an offline step.
In~\S\ref{subsec:test_time_sparse_mf_inference}, we present
an online approach to compute a test query embedding
to approximate the cross-encoder scores and perform 
$k$-NN search using the approximate cross-encoder scores.

\subsection{Proposed Offline Indexing of Items}
\label{subsec:offline_sparse_mf_indexing}

In this section, we describe our proposed
approach to efficiently align the item embeddings with the cross-encoder 
where efficiency is measured in terms of the number of 
training samples (query-item pairs) required to be gathered and scored 
using the cross-encoder and wall-clock time to fit an approximation of 
the cross-encoder model.
We consider an approximation of the cross-encoder 
with an inner-product space where a query~($\query$) and an 
item~($\dataItem$) are represented with  $\dimensionality$-dimensional 
vectors $\qvector_\query \in \RR^\dimensionality$ and 
$\ivector_\dataItem\in \RR^\dimensionality$ respectively. 
$k$-NN search using this approximation corresponds to solving
the following vector-based nearest neighbor search:
\begin{equation}
    \approxKnnSet(\query)\triangleq \argtopk_{\dataItem \in \itemSpace} \qvector_\query \ivector_\dataItem^\intercal.
\end{equation}
This vector-based $k$-nearest neighbor search can potentially be made more 
efficient using data structures such as cover trees \citep{beygelzimer2006cover}, HNSW~\citep{malkov2018efficient}, 
or any of the many other highly effective vector nearest neighbor search 
indexes \citep{guo2020accelerating,johnson2019billion}. 
The focus of our work is not on a new way to make the vector nearest 
neighbor search more efficient, but rather to develop  
efficient and accurate methods of fitting the embedded representations of 
$\qvector_{\query}$ and $\ivector_{\dataItem}^\intercal$ to approximate the cross-encoder scores.

Let $\sparseMat \in \RR^{|\querySpace_\trainsplit| \times |\itemSpace|}$  
denote the pairwise similarity matrix containing the exact cross-encoder
over the pairs of training queries ($\querySpace_\trainsplit$) and items ($\itemSpace$).
We assume that \sparseMat is \emph{partially observed} or incomplete, that is only a very small subset of the query-item pairs ($\pairs_\trainsplit$) are observed in $\sparseMat$. 
Let $U \in \RR^{|\querySpace_\trainsplit| \times \dimensionality}$
and $V \in \RR^{|\itemSpace| \times \dimensionality}$ be 
matrices such that each row corresponds to the embedding of a 
query $\query \in \querySpace_\trainsplit$ and an item $\dataItem \in \itemSpace$ respectively. 
We optimize the following widely-used objective for matrix completion to estimate $U$ and $V$ via stochastic gradient descent:
\begin{equation}
    \min_{\substack{U\in \RR^{|\querySpace_\trainsplit|\times \dimensionality}, V \in \RR^{\nItems \times \dimensionality}}} \norm{ (\sparseMat - UV^\intercal)_{\pairs_\trainsplit}}_2 \label{eq:sparse_mf}
\end{equation}
where $(\cdot)_{\pairs_\trainsplit}$ denotes projection on the set of 
observed entries in $\sparseMat$.
There are two important considerations: 
(1) how to select with values of \sparseMat to observe 
(and incur the cost of running the cross-encoder model), and 
(2) how to compute/parameterize the matrices $U$ and $V$.

\paragraph{Constructing Sparse Matrix \sparseMat}
Given a set of items~($\itemSpace$) and train queries~($\queryTrainData$), 
we construct the sparse matrix~\sparseMat
by selecting $\kDistill$ items $\itemSpace_\query \subset \itemSpace$ 
for each query $\query \in \querySpace_\trainsplit$ either 
uniformly at random or using top-$\kDistill$ items from a baseline 
retrieval method such as the base dual-encoder (\baseDualEncoder). 
This approach requires $\kDistill \lvert \querySpace_\trainsplit \rvert$ calls to the cross-encoder.
We also experiment with an approach that selects
$\kDistill$ queries $\querySpace_\dataItem \subset \querySpace_\trainsplit$ 
for each item $\dataItem \in \itemSpace$, 
and thus requires $\kDistill \nItems$ calls to the cross-encoder.

\paragraph{Parameterizing and Training $U$ and $V$} 
\begin{itemize}[topsep=0pt,itemsep=0ex,partopsep=1ex,parsep=1ex,leftmargin=*]
    \item \textbf{Transductive}~(\matrixFact{\transductive}): In this setting, $U$ and
    $V$ are trainable parameters and are learned by optimizing the 
    objective in Eq.~\ref{eq:lin_reg_query_emb}.  
    $U$ and $V$ can be optionally initialized using query and item
    embeddings from the base dual-encoder (\baseDualEncoder).
    Note that this parameterization requires scoring each item against at least
    a small number of queries to update the embedding of an item from its initialized value, 
    thus requiring scoring of $\mathcal{O}(\nItems)$ query-item pairs to construct the sparse matrix \sparseMat.
    Such an approach may not scale well with the number of items as
    the number of cross-encoder calls to construct $\sparseMat$ and the number of trainable parameters are both linear in the number of items.
    For instance, when $\nItems$ = 5 million, $\dimensionality=1000$,   $V$ would contain 5 billion trainable parameters.
    
    \item \textbf{Inductive}~(\matrixFact{\inductive}): In this setting, we train parametric models
    to produce query and item embeddings $U$ and $V$ from (raw)
    query and item features such as textual descriptions of queries and items. 
    Unlike transductive approaches, inductive matrix factorization approaches can 
    produce embeddings for unseen queries and items, and thus can be used to 
    produce embeddings for items not scored against any train query in matrix \sparseMat 
    as well as embeddings for test queries $\testQuery \notin \querySpace_\trainsplit$.
    Prior work typically uses \baseDualEncoder (a DE trained on the same task and source domains
    as the CE) and finetunes \baseDualEncoder
    on the target domain via distillation using the CE.
    However, training all parameters of such parametric encoding models via 
    distillation can be compute- and resource-intensive as these models
    are built using several layers of neural models such as transformers.
    Recall that our goal is to efficiently build an accurate approximation of the 
    CE on a given target domain. 
    Thus, to improve the efficiency of fitting the approximation of the CE, 
    we propose to train a shallow MLP model (using data from the target domain) that takes query/item embeddings from \baseDualEncoder as input and outputs 
    updated embeddings while keeping \baseDualEncoder parameters frozen.
    
\end{itemize}

\subsection{Proposed Test-Time $k$-NN Search Method: \propInf{}{} }
\label{subsec:test_time_sparse_mf_inference}

At test-time, we need to perform $k$-NN search for a test query 
$\testQuery \notin \querySpace_\trainsplit$, and thus need to
compute an embedding for the test query in order to approximate
cross-encoder scores and perform retrieval with the approximate scores.
Note that computing the test query embedding by factorizing the matrix \sparseMat at \emph{test-time} 
while including the test query $\testQuery$ is computationally infeasible. 
Thus, an ideal solution would be to compute item representations 
in an offline indexing step, and
compute the test query embedding \emph{on-the-fly} while keeping
item embeddings fixed.
A potential solution is to use a parametric model such as \baseDualEncoder or \matrixFact{\inductive}
to compute test query embedding, perform retrieval using inner-product scores
between test query and item embeddings, and finally, re-rank the retrieved items 
using the cross-encoder.
While such a retrieve-and-rerank approach can work, the retrieval
step on such an approach is decoupled from the re-ranking model, 
and thus may result in poor recall.

In this work, we propose an adaptive approach \propInf{}{}, which 
stands for "\textbf{A}daptive \textbf{Cross}-Encoder \textbf{N}earest Neighbor Search". 
As described in Algorithm~\ref{alg:adaptive_test_time_inference},
\propInf{}{} performs retrieval over $\nRounds$ rounds while incrementally 
refining the cross-encoder approximation for $\testQuery$ by 
updating $\qvector_{\testQuery}$, the embedding for $\testQuery$. 
The test-time inference latency (and throughput) depends largely on 
the number of cross-encoder calls made at test time as each cross-encoder call 
requires a forward pass through a large neural model. 
Thus, we operate under a fixed computational budget which allows for up to 
$\ceBudget$ cross-encoder calls at test-time.

\begin{algorithm}[!t]
\caption{\propInf{}{} - Test-time $k$-NN Search Inference }
\begin{algorithmic}[1]
\footnotesize
\STATE \textbf{Input:} $\query$: Test query, $V \in \RR^{\nItems \times d} $ Item Embeddings, $\nRounds$: Number of iterative search rounds, $\nAnchorItemsPerRound$: Number of items to retrieve in each round,  $\model{\theta}$: Cross-Encoder (CE) model
\STATE \textbf{Output:} $\hat{S}$: Approximate scores of $\query$ with all items,
$\adaptiveSelection{\nRounds}$: Retrieved items with CE scores in $\adaptiveSelectionScores{\nRounds}$.
% \item[]
\STATE $\adaptiveSelection{1} \gets \textsc{Init}(\itemSpace, \nAnchorItemsPerRound)$ \hfill $\rhd$ \textcolor{gray}{Initial set of items} \label{alg_line:init_items_first_round}

\STATE $\adaptiveSelectionScores{1} \gets [\model{\theta}(\query, \dataItem)]_{\dataItem \in \adaptiveSelection{1} }$ \hfill $\rhd$ \textcolor{gray}{CE scores of $\query$ with items in $\adaptiveSelection{1}$}
\STATE $\qvector_{\query} \gets \text{Solve-Linear-Regression}(V, \adaptiveSelection{1},  \adaptiveSelectionScores{1}  )$ \hfill $\rhd$ \textcolor{gray}{Compute query embedding by solving Eq.\ref{eq:lin_reg_query_emb}} \label{alg_line:lin_reg_query_emb_first_round}
\FOR {$\roundIter \gets 2 \text{ to } \nRounds$}
    \STATE $\hat{S}^{(\roundIter)} \gets \qvector_{\query} \times V^\intercal $ \hfill $\rhd$ \textcolor{gray}{Update approx. scores} \label{alg_line:update_approx_ce_score}

    \STATE $\adaptiveSelection{\roundIter} \gets \adaptiveSelection{\roundIter-1} \cup \argtopk_{\dataItem \in \itemSpace\setminus \adaptiveSelection{\roundIter-1}, k=\nAnchorItemsPerRound }  \hat{S}^{(\roundIter)}_{\dataItem} $ 
    \hfill $\rhd$ \textcolor{gray}{Retrieve $\nAnchorItemsPerRound$ new items}
    \label{alg_line:lin_reg_item_select}
    
    \STATE $\adaptiveSelectionScores{\roundIter} \gets \adaptiveSelectionScores{\roundIter-1} \oplus [\model{\theta}(\query, \dataItem)]_{\dataItem \in \adaptiveSelection{\roundIter} \setminus \adaptiveSelection{\roundIter-1} }$ \hfill $\rhd$ \textcolor{gray}{Compute CE scores of new items} \label{alg_line:update_exact_ce_scores}
    \STATE $\qvector_{\query} \gets \text{Solve-Linear-Regression}(V, \adaptiveSelection{\roundIter},  \adaptiveSelectionScores{\roundIter}  )$ \hfill $\rhd$ \textcolor{gray}{Compute query embedding by solving Eq.\ref{eq:lin_reg_query_emb}} \label{alg_line:lin_reg_query_emb}
\ENDFOR

\STATE $\hat{S} \gets \qvector_\query \times V^\intercal $  \hfill $\rhd$ \textcolor{gray}{Compute approx. scores}
    
\STATE \textbf{return} $\hat{S}, \adaptiveSelection{\nRounds}, \adaptiveSelectionScores{\nRounds}$ 
\end{algorithmic}
\label{alg:adaptive_test_time_inference}
\end{algorithm}

Let $\adaptiveSelection{\roundIter}$ be the cumulative set of items chosen up to round $\roundIter$.
In the first round ($\roundIter=1$), we select $\ceBudget/\nRounds$ items either uniformly at random
or using separate retrieval models such as dual-encoders or BM25
and compute the exact cross-encoder scores of these items for the given test query.
We compute the test query embedding $\qvector_{\testQuery}$ by solving the following
system of linear equations
\begin{equation}
     \adaptiveSelectionEmbeds{\roundIter} \qvector_{\testQuery} =   \adaptiveSelectionScores{\roundIter} \label{eq:lin_reg_query_emb}
\end{equation}
where $\adaptiveSelectionEmbeds{\roundIter} \in \RR^{ \lvert \adaptiveSelection{\roundIter} \rvert \times d }$ 
contains embeddings for items in $\adaptiveSelection{\roundIter}$, and $\adaptiveSelectionScores{\roundIter}$
contains cross-encoder scores for $\testQuery$ paired with items in $\adaptiveSelection{\roundIter}$.
In round $\roundIter > 1$, we select additional $\ceBudget/\nRounds$ items from 
$\itemSpace \setminus \adaptiveSelection{\roundIter-1}$
using inner-product of test query embedding $\qvector_{\testQuery}$ and item 
embeddings $\ivector_\dataItem$~(line~\ref{alg_line:lin_reg_item_select} in Alg.~\ref{alg:adaptive_test_time_inference}).
% Formally,
\begin{equation}
    \adaptiveSelection{\roundIter} = \adaptiveSelection{\roundIter-1} \cup \argtopk_{\dataItem \in \itemSpace\setminus \adaptiveSelection{\roundIter-1}, k=\ceBudget/\nRounds } \qvector_{\testQuery} \ivector_\dataItem^\intercal \label{eq:lin_reg_item_select}
    % \vspace{-0.1cm}
\end{equation}
After computing $\adaptiveSelection{\roundIter}$, we compute CE scores for new items 
chosen in round $\roundIter$, and we update the test query embedding 
$\qvector_{\testQuery}$ by solving Eq.~\ref{eq:lin_reg_query_emb}
with the latest set of items $\adaptiveSelection{\roundIter}$ which includes additional 
items selected in round $\roundIter$.
Note that solving for $\qvector_{\testQuery}$ in Eq~\ref{eq:lin_reg_query_emb}
is akin to solving a linear regression problem with embeddings of items
in $\adaptiveSelection{\roundIter}$ as features and cross-encoder scores of the items
as regression targets.
We solve Eq.~\ref{eq:lin_reg_query_emb} analytically to get 
$\qvector_{\testQuery}~=~(\adaptiveSelectionEmbeds{\roundIter}^\intercal \adaptiveSelectionEmbeds{\roundIter})^\dagger \adaptiveSelectionEmbeds{\roundIter}^\intercal \adaptiveSelectionScores{\roundIter} $ where $M^\dagger$ denotes pseudo-inverse of a matrix $M$.

At the end of $\nRounds$ rounds, we obtain $\adaptiveSelection{\nRounds}$ containing $\ceBudget$ items, 
all of which have been scored using the cross-encoder model. We return
top-$k$ items from this set sorted based on exact cross-encoder scores as the set of approximate $k$-NN for given test query $\testQuery$ 
\begin{equation}
    \approxKnnSet(\testQuery) = \argtopk_{\dataItem \in \adaptiveSelection{\nRounds}} \ce(\testQuery, \dataItem)
\end{equation}
\paragraph{Regularizing Test Query Embedding}
The system of equation in Eq~\ref{eq:lin_reg_query_emb} in round $\roundIter$ contains $\lvert \adaptiveSelection{\roundIter} \rvert$ equations with $d$ variables and is an under-determined system when $\lvert \adaptiveSelection{\roundIter} \rvert < d$. 
In such a case, there exist infinitely many solutions to  
Eq~\ref{eq:lin_reg_query_emb} and the test query 
embedding $\qvector_{\testQuery}$ can achieve zero approximation 
error on items in $\adaptiveSelection{\roundIter}$, and may show poor generalization
when estimating cross-encoder scores for items in $ \itemSpace \setminus \adaptiveSelection{\roundIter}$.
Since the approximate scores are used to select the additional set
of items in round $\roundIter+1$ (line~\ref{alg_line:lin_reg_item_select} in Alg.~\ref{alg:adaptive_test_time_inference}), 
such poor approximation affects the additional set of items chosen, and 
subsequently, it may affect the overall retrieval quality in certain settings.
To avoid such overfitting, we compute the final test query embedding as:
\begin{equation}
   \qvector_{\testQuery} = (1 - \lambda) \qvector_{\testQuery}^{\text{(LinReg)}} + \lambda  \qvector^{\text{(param)}}_{\testQuery}  \label{eq:weighted_comb_lin_reg}
\end{equation}
where $\qvector_{\testQuery}^{\text{(LinReg)}}$ is the analytical solution 
to the linear system in Eq.~\ref{eq:lin_reg_query_emb} 
and $\qvector^{\text{(param)}}_{\testQuery}$ is the test query 
embedding obtained from a parametric model such as a dual-encoder 
or an inductive matrix factorization model. We tune the 
weight parameter $\lambda \in [0,1]$ on the dev set.

\section{Experiments}
In our experiments, we evaluate proposed approaches
and baselines on the task of finding $k$-nearest neighbors
for cross-encoder (CE) models as well as on downstream tasks. 
We use cross-encoders trained for the downstream task of zero-shot entity 
linking and zero-shot information retrieval and 
present extensive analysis of the effect of various design choices 
on the offline indexing latency and the test-time retrieval recall.

\paragraph{Experimental Setup}
We run experiments on two datasets/benchmarks -- \zeshel~\citep{logeswaran-etal-2019-zero},  
a zero-shot entity linking benchmark, and \beir benchmark~\citep{thakur2021beir},
a collection of information retrieval datasets for evaluating 
zero-shot performance of IR models. 
We use separate CE models for \zeshel and \beir datasets, 
trained using ground-truth labeled data from the corresponding dataset.
For evaluation, we use two test domains from \zeshel dataset --\yugioh and \starTrek 
with 10K and 34K items (entities) respectively, and
we use \scidocs and \hotpotqa datasets from \beir with 25K and 5M items (documents) respectively.
These domains were \emph{not} part of the data used to
train the corresponding cross-encoder models.
Following the precedent set by previous 
work~\citep{yadav2022efficient,yadav2023efficient}, we create a 
train/test split uniformly at random for each \zeshel domain. 
For datasets from \beir, we use pseudo-queries
released as part of the benchmark as train queries
and test on queries in the official test split in \beir benchmark.
We use queries in the train split to 
train proposed matrix factorization models or 
baseline DE models via distillation,
and we evaluate on the corresponding domain's test split. 
We refer interested readers to Appendix~\ref{apndx_sec:training_details}
for more details about datasets, cross-encoder training, and model architecture.

\vspace{-0.2cm} 

\paragraph{Baselines}
We compare with the following retrieve-and-rerank baselines,
denoted by \rnr{X}, where top-scoring items wrt baseline scoring method $X$ are retrieved 
and then re-ranked using the CE.
\begin{itemize}[topsep=0pt,itemsep=0ex,partopsep=0.5ex,parsep=0.5ex,leftmargin=*]
    \item \textbf{\tfidf}:  It computes the similarity score for a query-item pair using the dot-product of sparse query/item vectors containing TF-IDF weights. 
    \item \textbf{Dual-Encoders (DE)}:
    It computes query-item scores using the dot-product of dense embeddings
    produced by encoding queries and items separately. 
    We experiment with two DE models.
    \begin{itemize}[topsep=0pt,itemsep=0ex,partopsep=0.5ex,parsep=0.5ex,leftmargin=*]
        \item \baseDualEncoder: DE trained on the same \emph{source} data and downstream task as the cross-encoder model. This model is \emph{not} trained or finetuned on the target domains
        used for evaluation in this work.
        \item \distillDualEncoder: This corresponds to \baseDualEncoder
        further finetuned via distillation using the
        cross-encoder model on the \emph{target} domain i.e. the domain used for evaluation. 
    \end{itemize}
\end{itemize}
We also compare with \adaCUR{}~\citep{yadav2023efficient}, 
a CUR-based approach that computes a dense matrix with CE 
scores between training queries and all items to index the items, and performs adaptive retrieval at test time. 
We use \adaCUR{X} to denote inference with \adaCUR{} method when items in the 
first round are chosen using method $X \in \{\baseDualEncoder, \tfidf\}$.
We refer readers to Appendix~\ref{apndx_sec:training_details} for implementation details for all baselines and proposed approaches. 
\vspace{-0.3cm}

\paragraph{Proposed Approach}
We construct the sparse matrix $G$ on the target domain 
by selecting top-scoring items wrt \baseDualEncoder for each 
query in $\queryTrainData$ followed by  
computing the CE scores for observed query-item pairs in $G$.
We use \baseDualEncoder to initialize embeddings for train queries 
and all items, followed by inductive (\matrixFact{\inductive})
or transductive (\matrixFact{\transductive}) matrix factorization 
while minimizing the objective function in~\ref{eq:sparse_mf}.
We use the same sparse matrix $G$ when training  DE
via distillation (\distillDualEncoder) on the target domain.
We use \propInf{X}{Y} to denote the proposed $k$-NN search 
method~(\S\ref{subsec:test_time_sparse_mf_inference}) when using 
method $X$ to compute item embeddings and method $Y$ 
to retrieve items in the first round.
\vspace{-0.3cm}

\paragraph{Evaluation Metrics}
Following the precedent set by previous work~\citep{yadav2022efficient, yadav2023efficient}, 
we use Top-$k$-Recall@$m$ for test queries as the evaluation metric which measures
the fraction of $k$-nearest neighbors as per the CE which 
are present in the set of $m$ retrieved items. 
For each method, we retrieve $m$ items and re-rank them using 
exact CE scores.
We also evaluate the quality of the retrieved $k$-NN items wrt the CE
on the downstream task. We use entity linking accuracy for \zeshel, 
and we use downstream task specific nDCG@10 and recall for \beir domains.

For each approach, we calculate the time taken for indexing a given set of 
items from the target domain which involves some or all of the 
following steps: a) computing query/item embeddings using \baseDualEncoder,
b) computing (dense or sparse) query-item score matrix $\sparseMat$ for $\queryTrainData$,
c) gradient-based training using $\sparseMat$ to estimate item embeddings for \matrixFact{\transductive} or parameters of models such as \distillDualEncoder and \matrixFact{\inductive}, and
d) for \distillDualEncoder and \matrixFact{\inductive}, computing updated item embeddings after training.

\subsection{Results}

\begin{figure}
    \centering
    \begin{subfigure}[b]{\textwidth}
    \centering
    \includegraphics[width=\textwidth, trim={0 10.1cm 0 0}, clip]{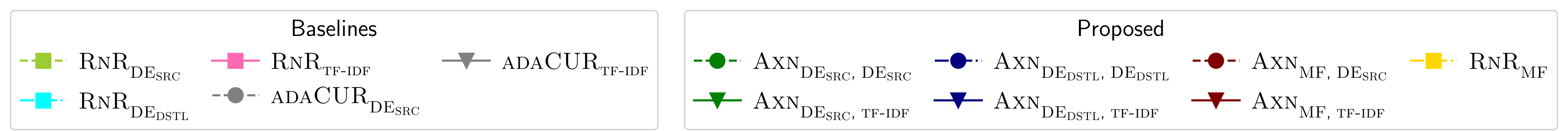}
    \end{subfigure}
    \begin{subfigure}[b]{0.48\textwidth}
    \includegraphics[width=\textwidth, trim={0 0.25cm 0 0}, clip]{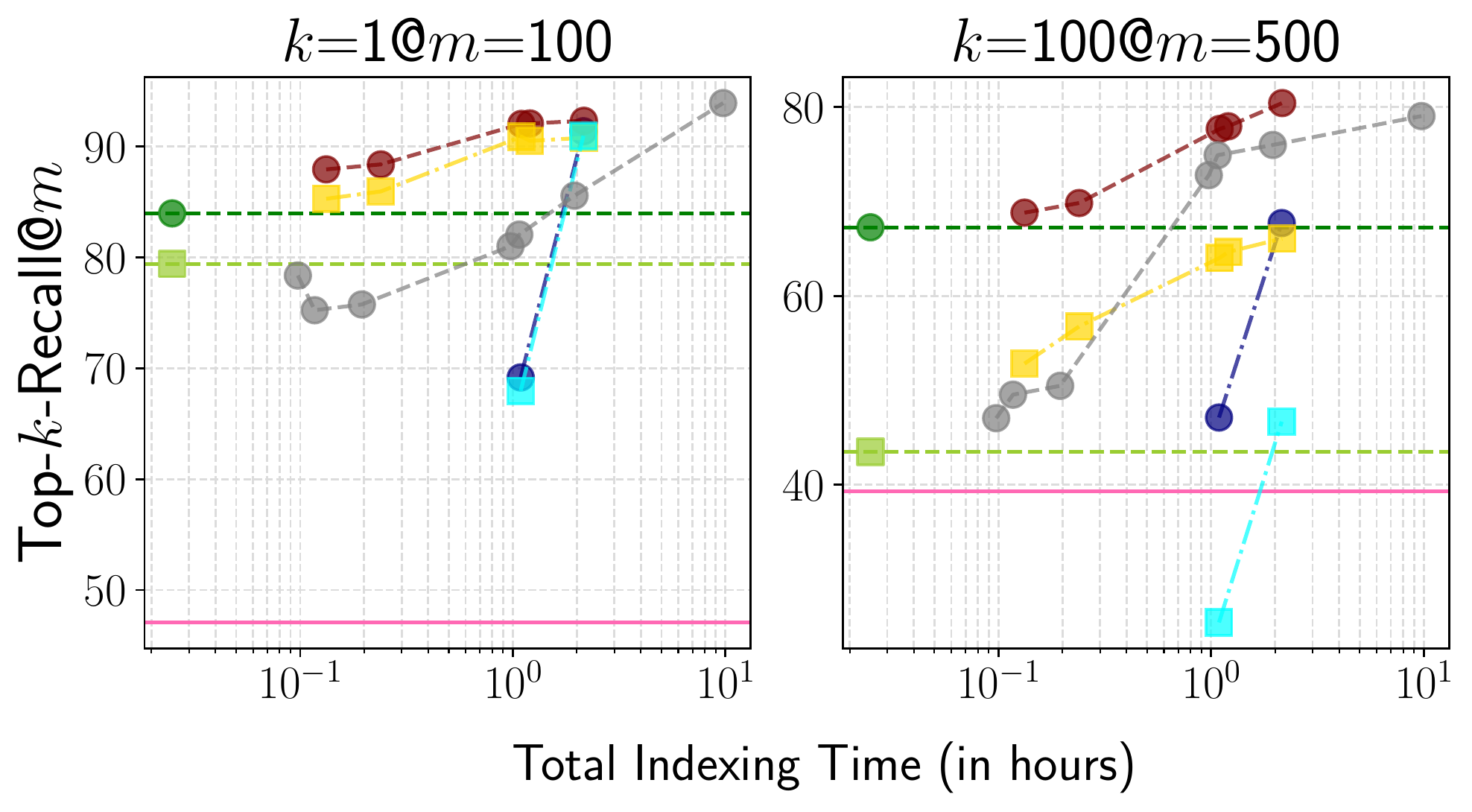}
    \caption{Domain = \yugioh, 10K items}
    \label{fig:rq_2a_recall_vs_indexing_cost_as_wall_clock_time_yugioh}    
    \end{subfigure}
    \hfill
    \begin{subfigure}[b]{0.48\textwidth}
    \includegraphics[width=\textwidth, trim={0 0.25cm 0 0}, clip]{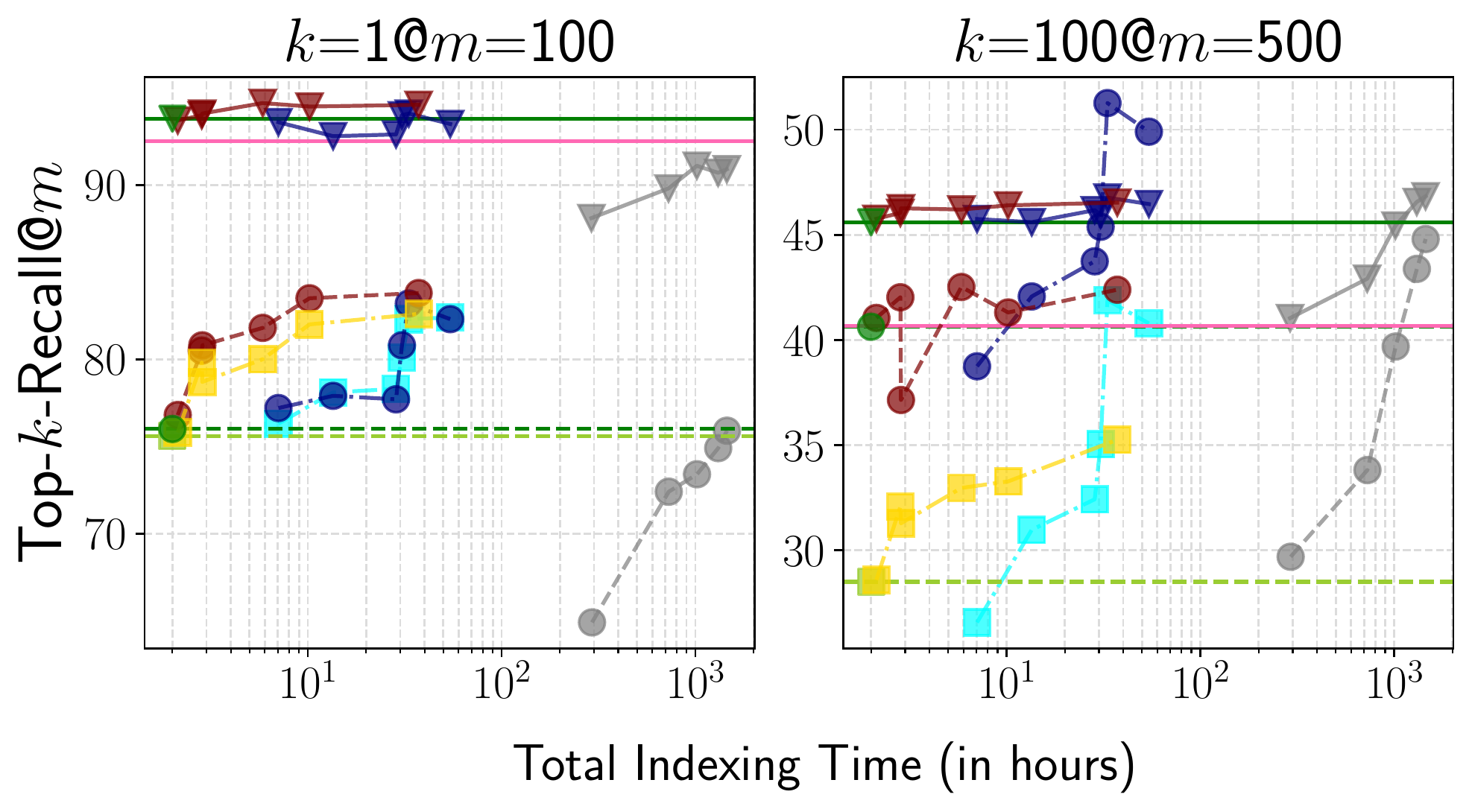}
    \caption{Domain = \hotpotqa, 5 million items}
    \label{fig:rq_2a_recall_vs_indexing_cost_wall_clock_time_hotpotqa}    
    \end{subfigure}
    \caption{
    Top-1-Recall and Top-100-Recall at inference cost budget ($m$) of 100 and 500 CE calls respectively 
    versus indexing time for various approaches. Matrix factorization approaches (\matrixFact{}) can
    be significantly faster than \adaCUR{} and training DE via distillation (\distillDualEncoder). 
    The proposed adaptive $k$-NN search method (\propInf{}{})
    provides consistent improvement over corresponding retrieve-and-rerank style inference (\rnr{}).}
    \label{fig:rq_2a_recall_vs_indexing_cost_wall_clock_time}
\end{figure}

Figure~\ref{fig:rq_2a_recall_vs_indexing_cost_wall_clock_time} shows Top-1-Recall@Inference-Cost=100 and Top-100-Recall@Inference-Cost=500 
versus the total wall-clock time taken to index the items for various approaches on \yugioh
and \hotpotqa.
\adaCUR{} can control the indexing time by 
varying $\queryTrainSize$, the number of train queries,
while \matrixFact{} and distillation-based methods can control
the indexing time by varying $\queryTrainSize$ and 
the number of items scored per train query~($\kDistill$).
For \yugioh, we use $\queryTrainSize \leq$ 500 for all methods, and for \hotpotqa, we use $\queryTrainSize \leq$ 1K for \adaCUR{} and $\queryTrainSize \leq$ 50K with other methods.

\paragraph{Proposed Inference (\propInf{}{}) vs Retrieve-and-Rerank (\rnr{})}
\propInf{}{} consistently provides
improvement over the corresponding retrieve-and-rerank (\rnr{}) baseline. 
For instance, \propInf{\baseDualEncoder}{\baseDualEncoder} provides
an improvement of 5.2\% for $k$=1 and 54\% for $k$=100 over \rnr{\baseDualEncoder} for domain=\yugioh.
Note that this performance improvement comes at \emph{no additional}
offline indexing cost and with negligible test-time overhead\footnote{We refer readers to~\S\ref{apndx_subsec:adaptive_search_overhead} for analysis of overhead incurred by \propInf{}{}}.
$\rnr{\tfidf}$ performs poorly on \yugioh while it serves
as a strong baseline for \hotpotqa, potentially due to 
differences in task, data, and CE model. 
On \hotpotqa, Top-$k$-Recall for  
$\propInf{}{}$ can be further 
improved by sampling items in the first round using
\tfidf (\propInf{Z}{\tfidf}) instead of 
\baseDualEncoder (\propInf{Z}{\baseDualEncoder}) for all indexing
methods $\text{Z} \in \{\baseDualEncoder, \distillDualEncoder, \matrixFact{} \}$. 

\begin{figure}[!h]
    \vspace{-0.3cm}
    \begin{subfigure}[b]{0.8\textwidth}
    \includegraphics[height=3cm, trim={0.25cm 0.25cm 0 0.25cm }, clip]{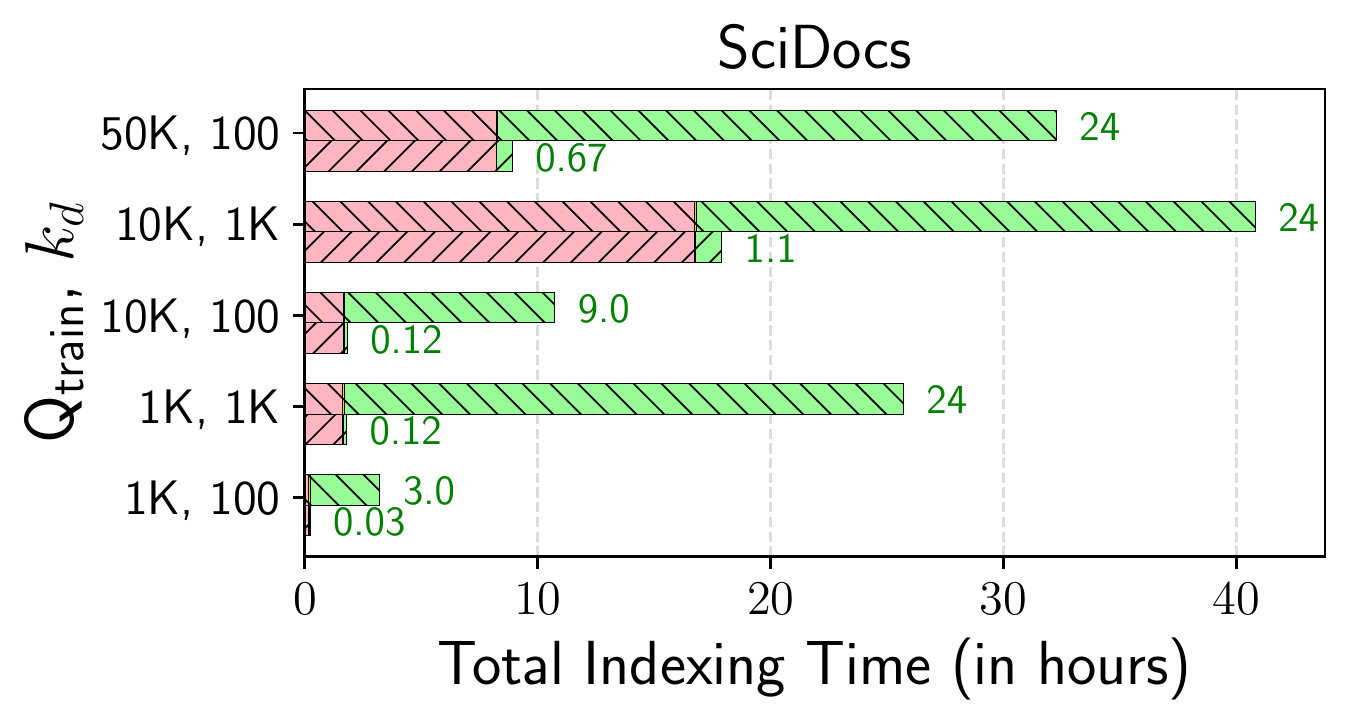}
    \includegraphics[height=3cm, trim={1cm 0.25cm 0 0.25cm }, clip]{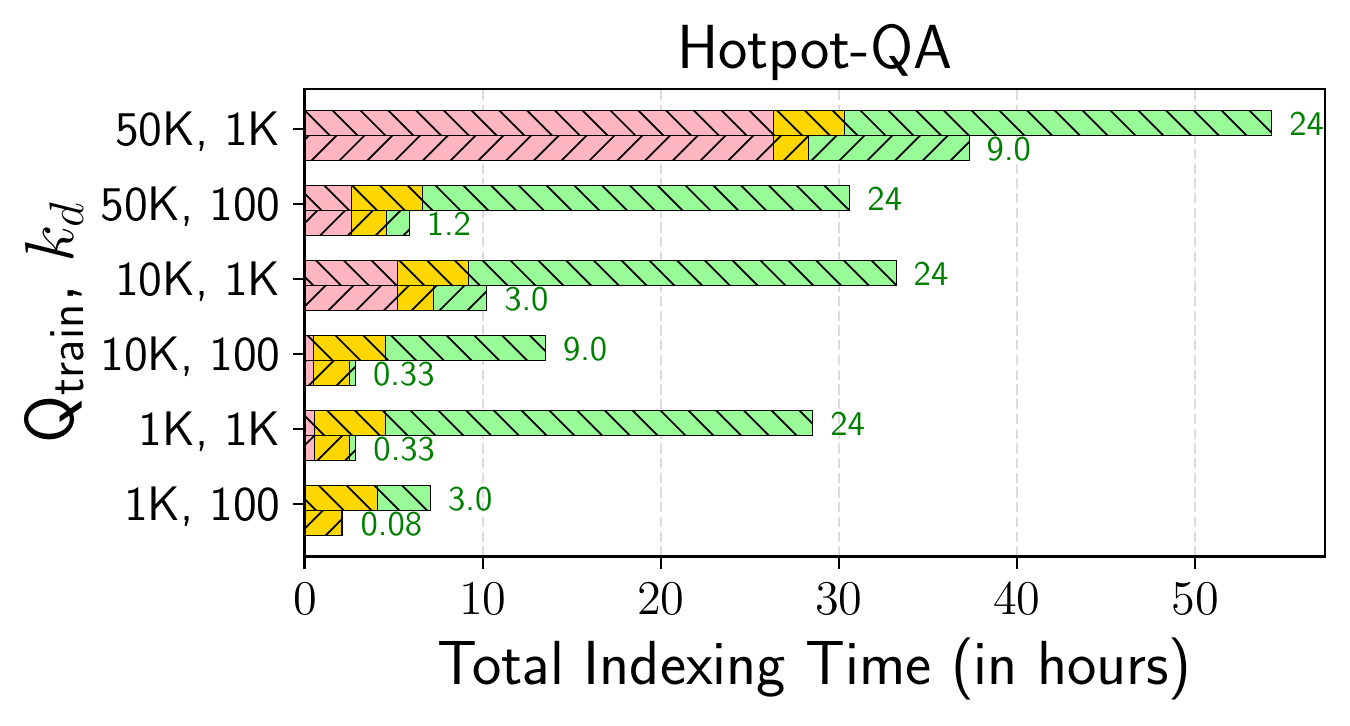}
    \end{subfigure}
    \hfill
    \begin{subfigure}[b]{0.15\textwidth}
    \includegraphics[height=3cm, trim={8.7cm 1cm 0 2.5cm}, clip]{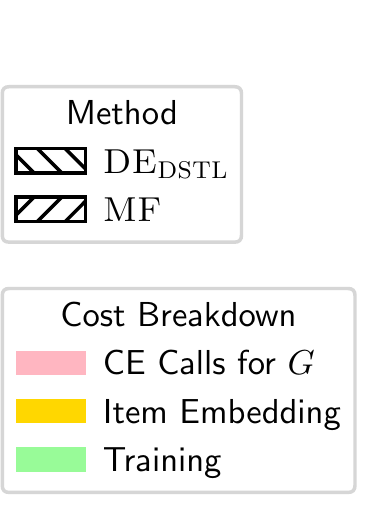}
    \phantomcaption{}
    \end{subfigure}
    \caption{Breakdown of indexing latency of \matrixFact{} and \distillDualEncoder 
    into various steps with training time shown on the right of each bar for different values of
    $\queryTrainSize$ and no. of items scored per query ($\kDistill$).
    }
    \vspace{-0.5cm}
    \label{fig:rq_2b_indexing_latency_breakdown}
\end{figure}

\paragraph{Matrix Factorization vs \distillDualEncoder}
Unsurprisingly, performance on the target domain can be further improved
by using data from the target domain to fit an embedding space to approximate the CE. 
As shown in Figure~\ref{fig:rq_2a_recall_vs_indexing_cost_wall_clock_time}, our proposed matrix factorization based approaches (\matrixFact{}) 
can be significantly more efficient than the distillation-based~(\distillDualEncoder) approaches 
while matching or outperforming \distillDualEncoder in terms of 
$k$-NN search recall in the majority of the cases.
Figure~\ref{fig:rq_2b_indexing_latency_breakdown} shows the breakdown of total indexing time of \distillDualEncoder and \matrixFact{}
for different numbers of training queries ($\queryTrainSize$) and 
number of items scored per query ($\kDistill$) using the CE
in the sparse matrix $G$.
As expected, both the time taken to compute $G$ and 
the training time increases with the number of queries 
and the number of items scored per query.
The training time does not increase proportionally after 10K queries
as we allocated a maximum training time of 24 hours for all methods.
For \matrixFact{}, the majority of the time is spent either 
in computing sparse matrix $G$ or the initial item embeddings.
While we report total GPU hours taken for CE calls to compute $G$
and initial item embeddings, these steps can be easily 
parallelized across multiple GPUs without any communication overhead.
Since \distillDualEncoder trains all parameters of a 
large parametric neural model, it requires large amounts 
of GPU memory and takes up to several hours~\footnote{We trained dual-encoders on an Nvidia RTX8000 GPU with 48 GB memory for a maximum of 24 hours.}.
In contrast, \matrixFact{}-approaches 
require significantly less memory\footnote{We used an Nvidia 2080ti with 12 GB memory for \matrixFact{}-based methods.}
and training time as these approaches train the item embeddings 
as free parameters~(\matrixFact{\transductive}) or train a shallow neural network on top 
of fixed embeddings~(\matrixFact{\inductive}) from an existing DE.
We report results for \matrixFact{\transductive} on small-scale domains (e.g. \yugioh with 10K items)
and for \matrixFact{\inductive} on large-scale domain \hotpotqa (5 million items).
We refer interested readers to Appendix~\ref{apndx_subsec:transductive_vs_inductive_mf} for comparison of \matrixFact{\transductive}
and \matrixFact{\inductive} on small- and large-scale datasets.

\vspace{-0.3cm}

\paragraph{Proposed Approaches vs \adaCUR{}}
Our proposed inference method~(\propInf{}{}) in combination with \matrixFact{}
or DE can outperform or closely match the performance
of \adaCUR{} while requiring orders of magnitude less compute for the offline
indexing stage, on both small- and large-scale datasets. 
For instance, \adaCUR{\baseDualEncoder} requires 1000+ GPU hours
for embedding 5 million items in \hotpotqa, and 
achieves Top-1-Recall@100 = 75.9 and Top-100-Recall@500 = 44.8. 
In contrast, \matrixFact{\inductive} 
with $\queryTrainSize$=10K and 100 items per query 
takes less than three hours to fit
item embeddings, and \propInf{\matrixFact{\inductive}}{\baseDualEncoder} 
achieves Top-1-Recall@100 = 80.5 and Top-100-Recall@500 = 42.6.

\begin{figure}[!h]
    \centering
    \begin{subfigure}[b]{\textwidth}
    \centering
    \includegraphics[width=\textwidth, trim={0 10.1cm 0 0.2cm}, clip]{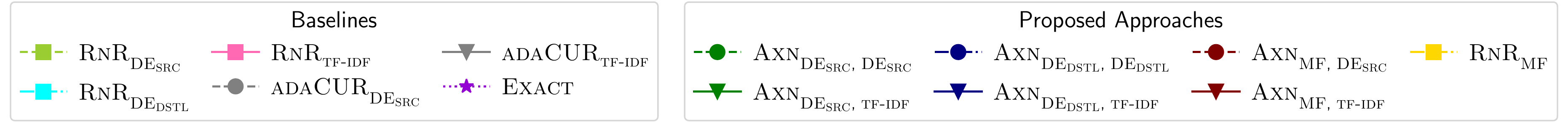}
    \end{subfigure}
    \begin{subfigure}[b]{0.24\textwidth}
    \includegraphics[width=\textwidth, trim={0 0.2cm 0 0.25cm}, clip]{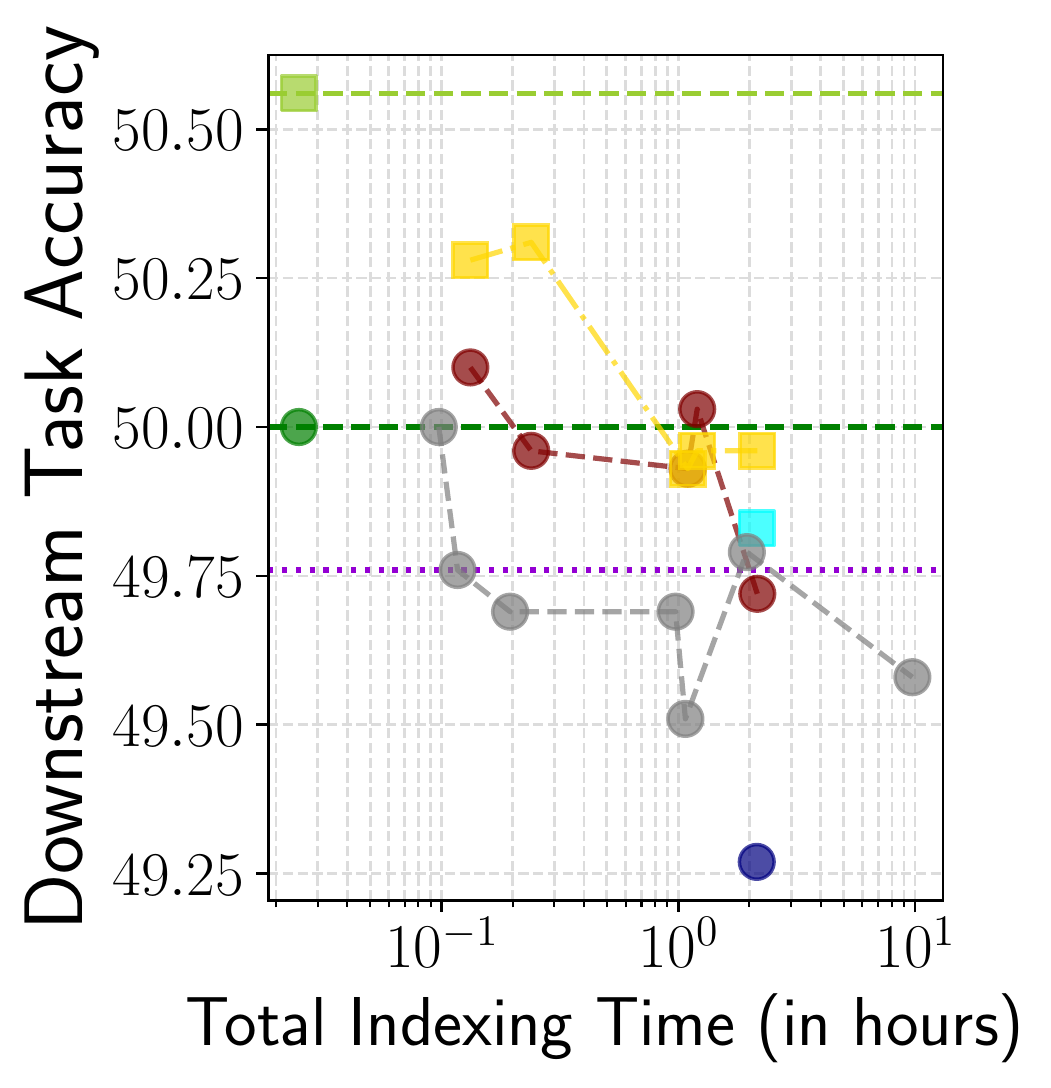}
    \caption{\yugioh }
    \label{fig:rq_3_gt_eval_accuracy_vs_indexing_cost_as_wall_clock_time_yugioh}    
    \end{subfigure}
    \hfill
    \begin{subfigure}[b]{0.24\textwidth}
    \includegraphics[width=\textwidth, trim={0 0.2cm 0 0.25cm}, clip]{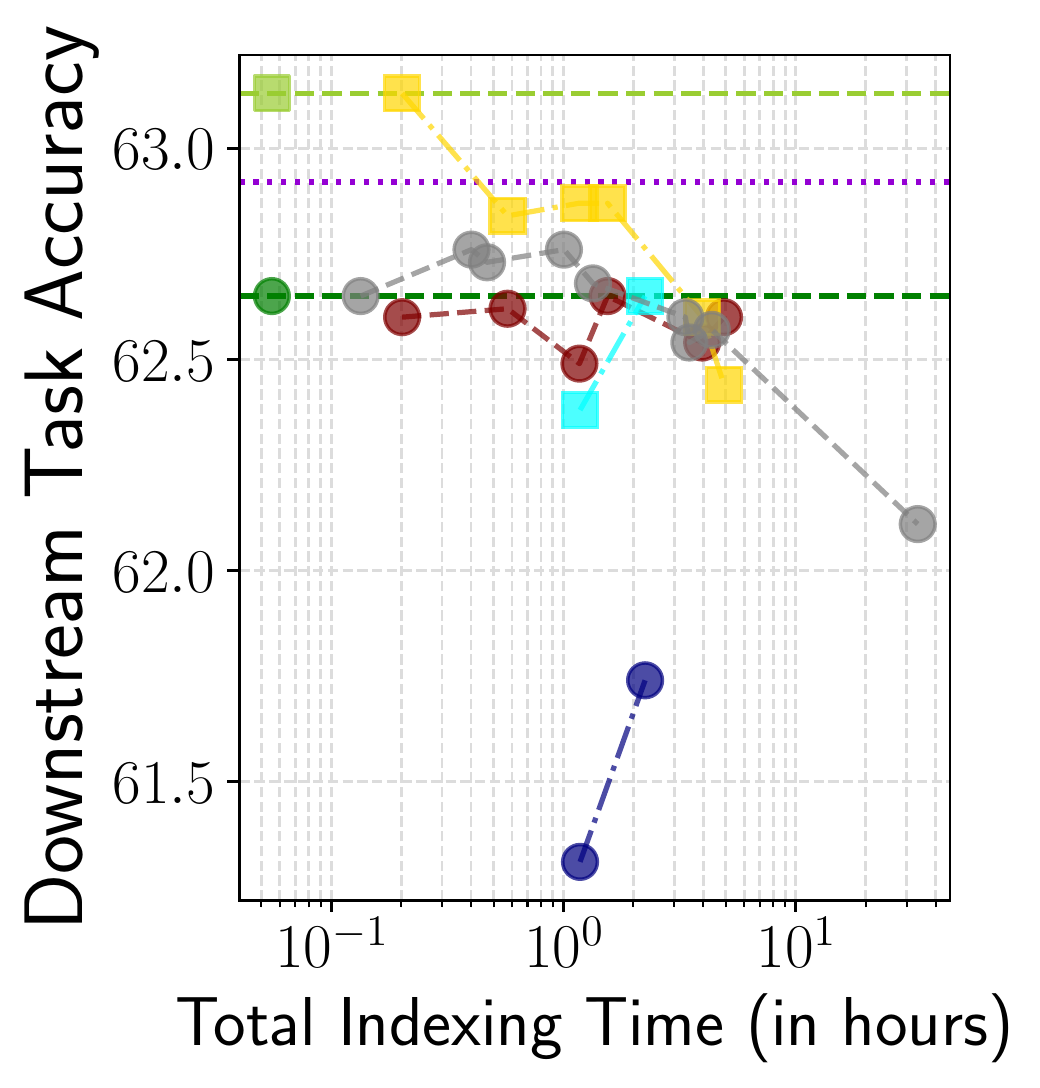}
    \caption{\starTrek }
    \label{fig:rq_3_gt_eval_accuracy_vs_indexing_cost_as_wall_clock_time_star_trek}    
    \end{subfigure}
    \hfill
    \begin{subfigure}[b]{0.24\textwidth}
    \includegraphics[width=\textwidth, trim={0 0.2cm 0 0.25cm}, clip]{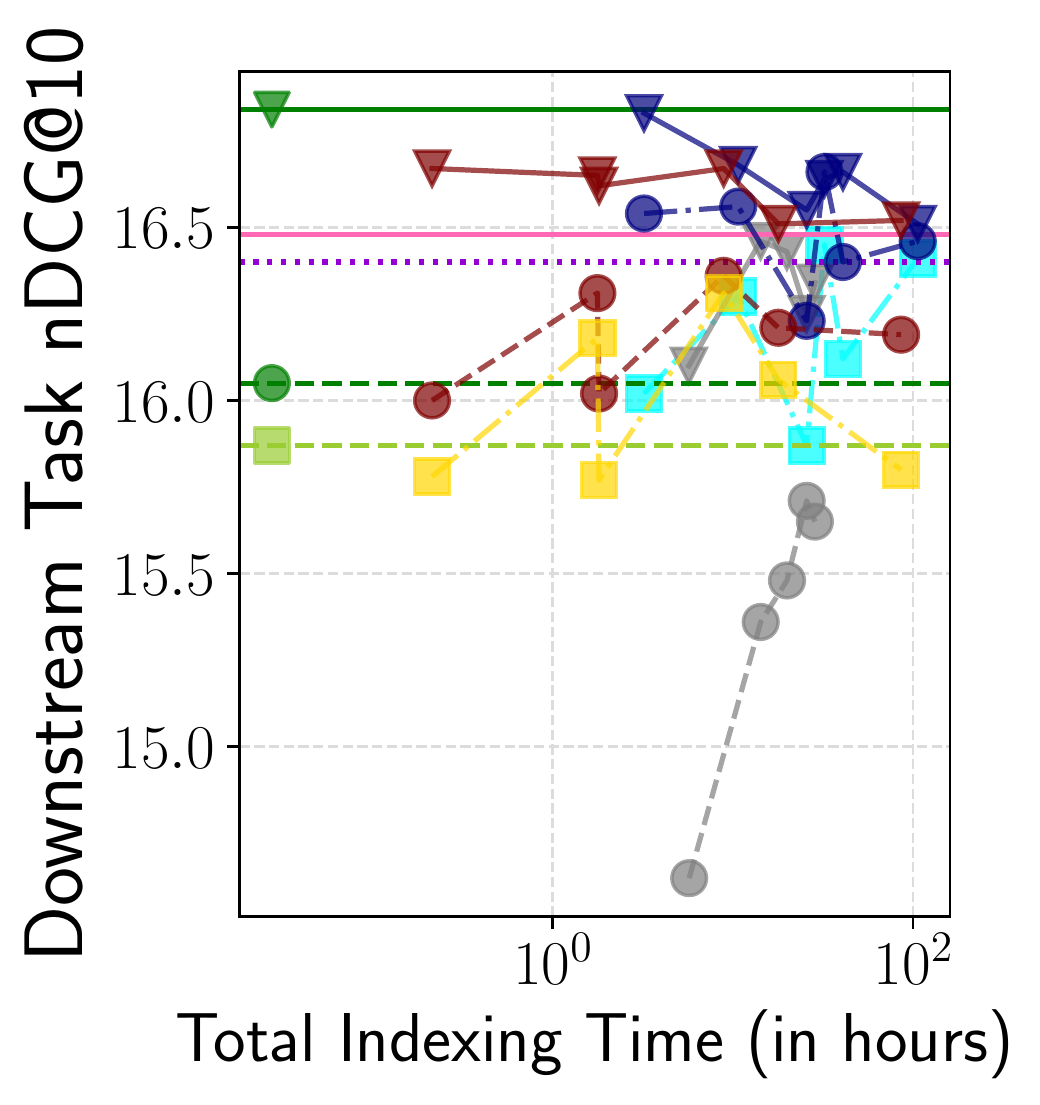}
    \caption{\scidocs}
    \label{fig:rq_3_gt_eval_ndcg_vs_indexing_cost_wall_clock_time_scidocs} 
    \end{subfigure}
    \hfill
    \begin{subfigure}[b]{0.24\textwidth}
    \includegraphics[width=\textwidth, trim={0 0.2cm 0 0.25cm}, clip]{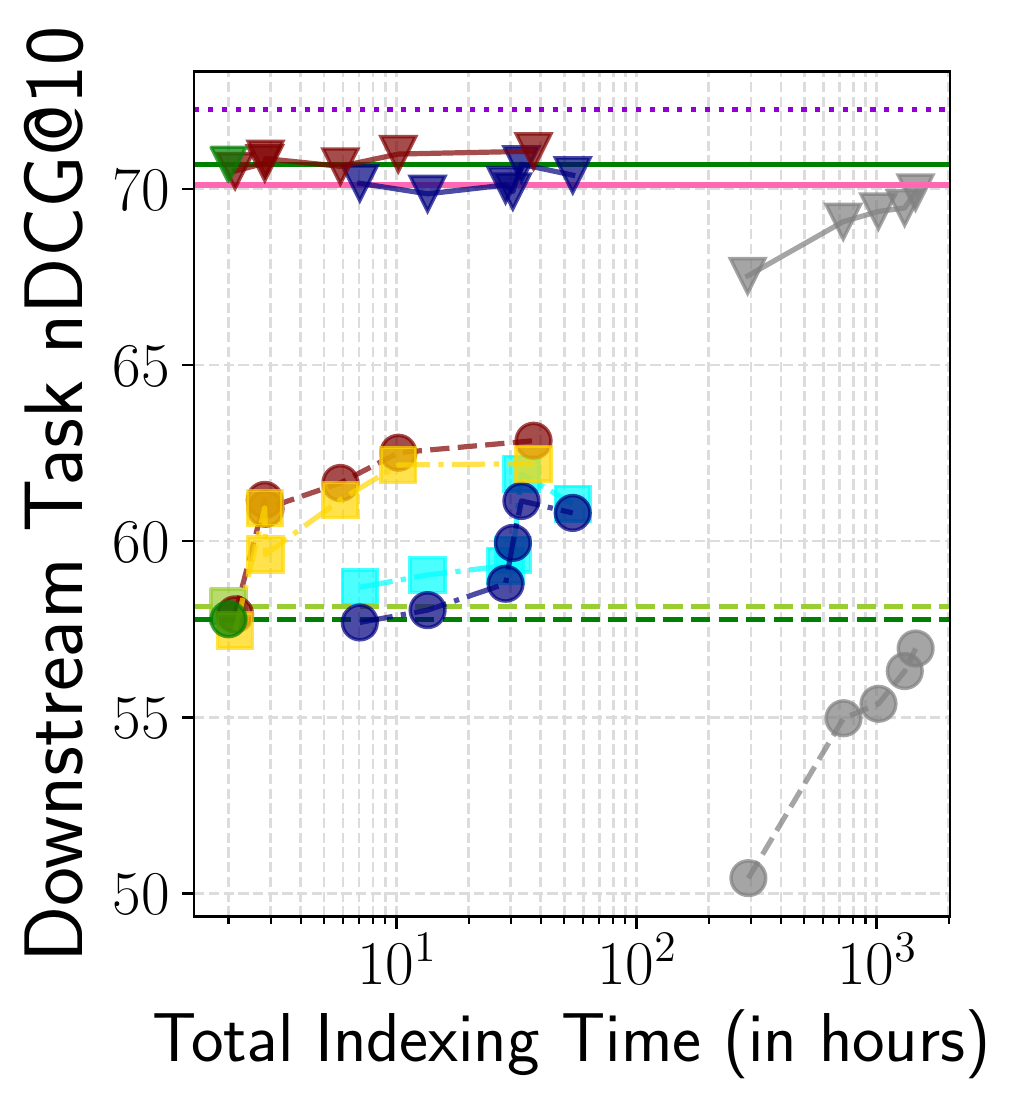}
    \caption{\hotpotqa}
    \label{fig:rq_3_gt_eval_ndcg_vs_indexing_cost_wall_clock_time_hotpotqa} 
    \end{subfigure}
    
    \caption{Downstream task performance versus indexing time for proposed and baseline approaches on different domains. All methods use a fixed inference cost budget of 100 cross-encoder calls.  
    }
    \label{fig:rq_3_gt_eval_vs_indexing_cost_wall_clock_time}
    \vspace{-0.6cm}
\end{figure}

% \vspace{-0.2cm}
\paragraph{Downstream Task Performance}
Figure~\ref{fig:rq_3_gt_eval_vs_indexing_cost_wall_clock_time} shows 
downstream task performance on proposed and baseline approaches 
including \exact which performs exact brute-force search using CE at test-time.
For \hotpotqa, we observe that improvement in $k$-NN search accuracy 
results in improvement in downstream task performance with 
\exact brute-force performing the best.
We observe a different trend on \scidocs, \yugioh, and \starTrek where
\exact search results in suboptimal performance as compared to \rnr{\baseDualEncoder}.
For instance, \rnr{\baseDualEncoder} achieves accuracy of 50.6
while the accuracy of \exact is 49.8 on the downstream task of entity linking
on \yugioh. 
We believe that this difference in trends in $k$-NN search 
performance and downstream task performance could be due to differences
in the training setup of the corresponding CE (i.e. 
the loss function and negatives used during training, 
see Appendix~\ref{apndx_subsec:cross_encoder_training} for details) 
as well as the nature of the task and data. 
While beyond the scope of this paper, it would be interesting
to explore different loss functions and  training strategies such as using negative items mined using $k$-NN search strategies
proposed in this work
to improve the robustness and generalization capabilities of 
cross-encoders and minimize such discrepancies in
$k$-NN search and downstream task performance.

We refer readers to Appendix~\ref{apndx_sec:sparse_mf_ablations_and_analysis} for 
an analysis of the overhead incurred by \propInf{}{}~(\S\ref{apndx_subsec:adaptive_search_overhead}), 
a comparison of \propInf{}{} with pseudo-relevance feedback based approaches~(\S\ref{apndx_subsec:query_embed_ablation}),  
an analysis of design choices for our proposed approach~(\S\ref{apndx_subsec:transductive_vs_inductive_mf},\ref{apndx_subsec:sparse_mat_construction_ablation}), and
results on other downstream evaluation metrics for \beir.

\section{Related Work}

\paragraph{Approximating Similarity Function}
Matrix factorization methods have been widely used
for computing low-rank approximation of 
dense distance and kernel matrices~\citep{musco2017sublinear, bakshi2018sublinear, indyk2019sample},
non-PSD matrices~\citep{archan2022sublinear} 
as well as for estimating missing entries in sparse matrices~\citep{koren2009matrix, luo2014efficient, yu2014parallel, mehta2017review, xue2017deep}.
In this work, we focus on methods for factorizing sparse matrices
instead of dense matrices
as computing each entry in the matrix (i.e. CE score for a query-item pair) 
requires a forward-pass through an expensive neural model.
An essential assumption for matrix completion methods 
is that the underlying matrix $M$ is low-rank, thus enabling
recovery of the missing entries while only observing a small
fraction of entries in $M$~\citep{candes2012exact, nguyen2019low}.
Theoretically, such matrix completion methods require $\Omega(nr)$ 
samples to recover an $m \times n$ matrix of rank $r$ with 
$m \leq n$~\citep{krishnamurthy2013low, xu2015cur}. The sample complexity can be 
improved in the presence of features describing rows and columns of the
matrix, often referred to as side information~\citep{jain2013provable, xu2013speedup, zhong2019provable}. 
Inductive matrix completion~(\matrixFact{\inductive}) approaches leverage such query and item
features to improve the sample complexity and also enable generalization to unseen queries (rows) and items (columns).
Training dual-encoder (DE) models via distillation using a cross-encoder (CE),
where the DE consumes raw query and item features 
(such as query/item description) and produces query/item
embeddings, can be seen as solving an inductive matrix factorization problem.
A typical training objective for training DE involves
minimizing the discrepancy between CE (teacher model)
and DE (student model) scores on observed entries in the sparse
matrix~\citep{Hofsttter2020ImprovingEN, reddi2021rankdistil, thakur-etal-2021-augmented}. 
Recent work has explored different strategies for
distillation-based training of DE such as curriculum learning
based methods~\citep{zeng2022curriculum}, joint training of CE and DE
to mutually improve the performance of both models~\citep{liu2021trans, ren2021rocketqav2}.
Inductive MF methods~(\matrixFact{\inductive}) used in this work also share similar motivations to adapters~\citep{houlsby2019parameter} which
introduce a small number of trainable parameters  
between layers of the model, and may reduce 
training time and memory requirements
in certain settings~\citep{ruckle-etal-2021-adapterdrop}.
\matrixFact{\inductive} used in this work only 
trains a shallow MLP on top of query/item embeddings from DE while keeping DE parameters frozen, 
and does not introduce 
any parameters in the DE.

\vspace{-0.2cm}
\paragraph{Nearest Neighbor Search}
$k$-NN search has been widely studied in applications where
the inputs are described as vectors in 
$\RR^\dimensionality$~\citep{clarkson2006nearest,li2019approximate},
and the similarity is computed using simple (dis-)similarity
functions such as inner-product~\citep{johnson2019billion, guo2020accelerating} and
$\ell_2$-distance~\citep{kleinberg1997two, chavez2001searching, hjaltason2003index}.
These approaches typically work by speeding up
each distance/similarity 
computation~\citep{jegou2010product, hwang2012fast, zhang2014composite, yu2017greedy, bagaria2021bandit} 
as well as constructing tree-based~\citep{beygelzimer2006cover, dong2019learning} 
or graph-based data structures~\citep{malkov2018efficient, wang2021comprehensive, groh2022ggnn}
over the given item set to efficiently
navigate and prune the search space to find (approximate) 
$k$-NN items for a given query. 
Recent work also explores such graph-based~\citep{boytsov2019accurate, tan2020fast, tan2021fast, macavaney2022adaptive},  or 
tree-based~\citep{boytsov2019pruning} data structures for non-metric
and parametric similarity functions. 
Another line of work explores model quantization~\citep{nayak2019bit, liu2021improving} 
and early-exit strategies~\citep{xin-etal-2020-early, xin-etal-2020-deebert}
to approximate the neural model while speeding up each forward pass through
the model and reducing its memory footprint.
It would be interesting to study if such data structures 
and approaches to speed up cross-encoder score computation can be 
combined with matrix factorization based approaches proposed in this work
to further improve recall-vs-cost trade-offs for $k$-NN search with cross-encoders.

\vspace{-0.2cm}
\paragraph{Pseudo-Relevance Feedback (PRF)}
Similar to PRF-based methods in information retrieval~\citep{rocchio1971relevance, lavrenko2017relevance}, our proposed
$k$-NN search method $\propInf{}{}$ refines the test query representation
using model-based feedback. In our case, we use the cross-encoder scores
of items retrieved in the previous round as feedback to update the test query
representation. PRF-based approaches have been widely used in information
retrieval for retrieval with sparse\citep{li-etal-2018-nprf, mao2020generation, mao-etal-2021-generation} and dense embeddings~\citep{yu2021improving, wang2021pseudo}. 
We refer readers to Appendix~\S\ref{apndx_subsec:query_embed_ablation}
for comparison with a recent PRF-based method~\citep{sung-etal-2023-optimizing}.

\section{Conclusion}

In this paper, we present an approach to perform $k$-NN search 
with cross-encoders by efficiently approximating the cross-encoder
scores using dot-product of learned test query and item embeddings.
In the offline indexing step, we compute item embeddings to index a given set of items
from a target domain by factorizing a sparse query-item score matrix, leveraging 
existing dual-encoder models to initialize the item embeddings while avoiding 
computationally-expensive distillation-based training
of dual-encoder models.
At test time, we compute the test query embedding to approximate 
cross-encoder scores of the given test query for a small set of adaptively-chosen items, and perform retrieval with the approximate cross-encoder
scores.
We perform extensive empirical analysis on two zero-shot retrieval benchmarks
and show that our proposed approach provides significant improvement
in test-time $k$-NN search recall-vs-cost tradeoffs 
while still requiring significantly less compute resources 
for indexing items from a target domain as compared to previous approaches.

\section*{Acknowledgments}

We thank members of UMass IESL for helpful discussions and feedback. This work was supported
in part by the Center for Data Science and the Center for Intelligent Information Retrieval, in part
by the National Science Foundation under Grant
No. NSF1763618, in part by the Chan Zuckerberg
Initiative under the project “Scientific Knowledge
Base Construction”, in part by International Business Machines Corporation Cognitive Horizons
Network agreement number W1668553, in part by
Amazon Digital Services, and in part using highperformance computing equipment obtained under
a grant from the Collaborative R\&D Fund managed
by the Massachusetts Technology Collaborative.
Any opinions, findings, conclusions, and recommendations expressed in this material are those of
the authors and do not necessarily reflect those of
the sponsor(s).

% Entries for the entire Anthology, followed by custom entries

\bibliography{references}
\bibliographystyle{iclr2024_conference}

\appendix

\section{Training and Implementation Details}
\label{apndx_sec:training_details}

\begin{table*}[!ht]
    \small
    \centering
    \begin{tabular}{l l|c c c c }
        \toprule
        Dataset & Domain                  &  $\nItems$ & $(\lvert \queryTrainData \rvert/ \lvert \queryTestData \rvert)$ Splits & Train Query~($\queryTrainData$) Type \\
        \midrule
        \zeshel  & \yugioh                &   10,031     & (100/3274), (500/2874), (2000/1374) & Real Queries\\
        \zeshel  & \starTrek              &   34,430     & (100/4127), (500/3727), (2000/2227) & Real Queries\\
        \midrule
        \beir  & \scidocs               &   25,657       & \{1K, 10K, 50K\}/1000 & Pseudo-Queries \\
        \beir  & \hotpotqa              &    5,233,329  & \{1K, 10K, 50K\}/1000  & Pseudo-Queries \\
        \bottomrule
    \end{tabular}
    \caption{
    Statistics on number of items ($\itemSpace$), number of queries in train~$(\queryTrainData)$ and test~$(\queryTestData)$ splits for each domain. 
    Following the precedent set by~\citet{yadav2022efficient}, 
    we create train/test split by
    splitting the queries in each \zeshel domain uniformly at random,
    and experiment with three values of $\queryTrainSize \in \{100, 500, 2000 \}$. 
    For \beir domains, we use pseudo-queries released as part of the benchmark as train queries~($\queryTrainData$) and run $k$-NN evaluation on test-queries from the official test-split~(as per \beir benchmark) of these domains.
    For HotpotQA, we use the first 1K queries out of a total of 7K test queries and we use all 1K test queries for SciDocs.
    }
    
    \label{tab:dataset_stats}
\end{table*}
\subsection{Training Cross-Encoder Models}
\label{apndx_subsec:cross_encoder_training}

In our experiments, we use \eCrossenc, a cross-encoder model
variant proposed by~\citet{yadav2022efficient} that
jointly encodes a query-item pair and computes the final score
using dot-product of contextualized query and item embeddings
extracted after joint encoding.

\paragraph{\zeshel Dataset}
For \zeshel, we use the cross-encoder model checkpoint\footnote{ \href{https://huggingface.co/nishantyadav/emb_crossenc_zeshel}{https://huggingface.co/nishantyadav/emb\_crossenc\_zeshel}} 
released by~\citet{yadav2022efficient}. 
We refer readers to \citet{yadav2022efficient} for further details on parameterization and training of the cross-encoder.

\paragraph{\beir Benchmark}
For \beir, we use the cross-encoder model checkpoint\footnote{ \href{https://huggingface.co/nishantyadav/emb_crossenc_msmarco_miniLM}{https://huggingface.co/nishantyadav/emb\_crossenc\_msmarco\_miniLM}}
trained on MS-MARCO dataset and released by~\citet{yadav2023efficient}. 
The cross-encoder model is parameterized
using a 6-layer \textsc{Mini-LM}\footnote{\href{https://huggingface.co/sentence-transformers/all-MiniLM-L6-v2}{https://huggingface.co/sentence-transformers/all-MiniLM-L6-v2}} 
model~\citep{wang2020minilm} and uses the dot-product based scoring mechanism
for cross-encoders proposed by~\citet{yadav2022efficient}.

\subsection{Training Dual-Encoder and Matrix Factorization Models}

For \beir datasets, we train matrix factorization models and \distillDualEncoder using sparse matrix $\sparseMat$
containing number of train queries $\queryTrainSize \in \{\text{1K, 10K, 50K} \}$ with 
number of items per query $\kDistill \in \{100, 1000\}$.
For \zeshel datasets, we use $\queryTrainSize \in \{\text{100, 500, 2000}\}$ with 
the number of items per query $\kDistill \in \{100, 1000\}$ for matrix factorization models
and $\kDistill \in \{25, 100\}$ for training \distillDualEncoder model.
Table~\ref{tab:dataset_stats} shows train/test splits used for each domain.

\subsubsection{Training Dual-Encoder Models}
\label{apndx_subsec:dual_encoder_training}

We train dual-encoder models on Nvidia RTX8000 GPUs with 48 GB GPU memory.

\paragraph{\zeshel dataset}
We report results for DE baselines as reported in~\citet{yadav2022efficient}.
The DE models were initialized using \texttt{bert-base-uncased} and
contain separate query and item encoders, thus resulting in a total of $2\times110M$ parameters.
The DE models are trained using cross-entropy loss
to match the DE score distribution with the CE score distribution. 
We refer readers to~\citet{yadav2022efficient} for details related to 
training of DE models on \zeshel dataset.

\paragraph{\beir benchmark}
For \beir domains, we use a dual-encoder model checkpoint\footnote{\href{https://www.sbert.net/docs/pretrained-models/msmarco-v2.html}{\texttt{msmarco-distilroberta-base-v2}: www.sbert.net/docs/pretrained-models/msmarco-v2.html}}
released as part of \texttt{sentence-transformer} repository as \baseDualEncoder, unless specified otherwise.
This DE model was initialized using 
\texttt{distillbert-base}~\citep{sanh2019distilbert} model and trained on MS-MARCO dataset
which contains 40 million (query, positive document (item), negative document (item)) triplets
using triplet ranking loss.
This \baseDualEncoder is not trained on target domains \scidocs and \hotpotqa
used for running $k$-NN experiments in this paper.
We finetune \baseDualEncoder via distillation on the target domain to get the \distillDualEncoder model.
Given a set of training queries $\queryTrainData$ from the target domain,
we retrieve top-100 or top-1000 items for each query, score the items with the cross-encoder
model and train the dual-encoder by minimizing cross-entropy loss between
predicted query-item scores (using DE) and target query-item scores (obtained 
using CE).
We train \distillDualEncoder using AdamW~\citep{loshchilov2017decoupled} 
optimizer with learning rate 1e-5 and accumulating gradient over 4 steps. We trained for 10 epochs when using top-100 items per query and
for 4 epochs when using top-1000 items per query. We allocate a maximum time of 24 hours
for training.

\subsubsection{Matrix-Factorization Models}
We train both transductive~($\matrixFact{\transductive}$) and inductive~($\matrixFact{\inductive}$) 
matrix factorization models on Nvidia 2080ti GPUs with 12 GB GPU memory for all datasets
with the exception that we trained $\matrixFact{\transductive}$ for \hotpotqa on Nvidia A100 GPUs with 80 GB GPU memory.
We use AdamW optimizer~\citep{loshchilov2017decoupled} with learning rate and number of epochs as shown in Table~\ref{apndx_tab:sparse_mat_hyper_param_details}.
Training $\matrixFact{\transductive}$ on \hotpotqa required 80 GB GPU memory as
it involved training 768-dimensional embeddings for 5 million items which roughly translates
to around 4 billion trainable parameters, and we used AdamW optimizer with stores additional
memory for each trainable parameter. 
For smaller datasets with the number of items of the order of 50K,
smaller GPUs with 12 GB memory sufficed.

For inductive matrix factorization~(\matrixFact{\inductive}), we train a 2-layer MLP with skip-connection on top of query and item embeddings from \baseDualEncoder.
For a given input embedding $x_\text{in} \in \RR^{\dimensionality}$, we compute the output embedding $x_\text{out} \in \RR^{\dimensionality}$ as
\begin{align*}
    x'_\text{out} &= b_2 + W_2^{\intercal} \text{gelu}(b_1 + W_1^{\intercal} x_\text{in}) \\
    x_\text{out}  &= \sigma(w_\text{skip}) x'_\text{out} + (1 - \sigma(w_\text{skip})) x  
\end{align*}
where $W_1 \in \RR^{\dimensionality \times 2\dimensionality}, b_1 \in \RR^{2\dimensionality}, W_2 \in \RR^{2\dimensionality \times \dimensionality}, b_2 \in \RR^{\dimensionality}, w_\text{skip} \in \RR$ are learnable parameters and $\sigma(.)$ is the sigmoid function.
We initialize  $w_\text{skip}$ with -5 and use default PyTorch initialization
for other parameters.  
We trained separate MLP models
for queries and items. 
We would like to highlight that a simple 2-layer MLP \emph{without} 
the skip connection i.e. using $x'_\text{out}$ as the final output embedding 
performed poorly in our experiments and it did not generalize well to unseen queries and items.

\begin{table}[!ht]
    \centering
    \small
    \begin{tabular}{c c c c}
    \toprule
     Domain         &  \matrixFact{} Type & Learning Rate & Number of Epochs \\
     \midrule
     \scidocs       &  \matrixFact{\transductive}        & 0.005         & 4 if $(\queryTrainSize,\kDistill) \in \{\text{(10K,1K), (50K, 1K}\}$ else 10 \\
     \scidocs       &  \matrixFact{\inductive}           & 0.005         & 10 if $(\queryTrainSize,\kDistill) \in \{\text{(10K,1K), (50K, 1K}\}$ else 20 \\
     \midrule
     
     \hotpotqa      & \matrixFact{\transductive}        & 0.001         & 4 if $(\queryTrainSize,\kDistill) \in \{\text{(10K,1K), (50K, 1K}\}$ else 10 \\
     \hotpotqa      & \matrixFact{\inductive}           & 0.001         & 10 if $(\queryTrainSize,\kDistill) \in \{\text{(10K,1K), (50K, 1K}\}$ else 20 \\
     \midrule
     
     \yugioh        & \matrixFact{\transductive}        & 0.001         & 20\\
     \midrule
     
     \starTrek      & \matrixFact{\transductive}        & 0.001         & 20 \\
    \bottomrule
    \end{tabular}
    \caption{Hyperparameters for transductive~(\matrixFact{\transductive}) and inductive~(\matrixFact{\inductive}) matrix factorization models for different number of training queries~($\queryTrainSize$) and number of items per train query~($\kDistill$) in sparse matrix $\sparseMat$.}
    \label{apndx_tab:sparse_mat_hyper_param_details}
\end{table}

\subsection{\tfidf }
For \beir datasets, we use \bmtwofive with parameters as reported in~\citet{thakur2021beir} and for \zeshel, we use \tfidf with default parameters from Scikit-learn~\citep{scikit-learn}, as reported in \citet{yadav2022efficient}.

\subsection{Test-Time Inference with \propInf{}{}, \adaCUR{}, and \rnr{}}
For \rnr{X}, we retrieve top-scoring items using dot-product of query and item embeddings
computed using baseline retrieval method $X$ and re-rank the retrieved items using the cross-encoder model.
For $\rnr{\matrixFact{\transductive}}$, we use dense query embedding from base dual-encoder
model \baseDualEncoder for test-queries $\testQuery \notin \queryTrainData$ along with item embeddings learnt using transductive matrix factorization to retrieve-and-rerank items for
the given test query.

For both \adaCUR{} and \propInf{}{}, we use $\nRounds=10$ for domains in \beir and $\nRounds=5$ for domains 
in \zeshel unless stated otherwise.
For \beir datasets, we tune \propInf{}{} weight parameter $\lambda$ (in eq~\ref{eq:weighted_comb_lin_reg}) on the dev set. We refer interested
readers to \S\ref{apndx_subsec:query_embed_ablation} for the effect of $\lambda$ on final performance.
For \zeshel, we report results for $\lambda=0$.
For \hotpotqa, we restrict our $k$-NN search with \propInf{X}{Y} and \adaCUR{Y} to top-10K items wrt method $Y$, $Y \in \{\baseDualEncoder, \tfidf\}$.
For other domains, we do not use any such heuristic and search over all items.

\begin{figure}[!t]
    \centering    
    \begin{subfigure}[b]{0.3\textwidth}
        \centering
        \includegraphics[width=\textwidth]{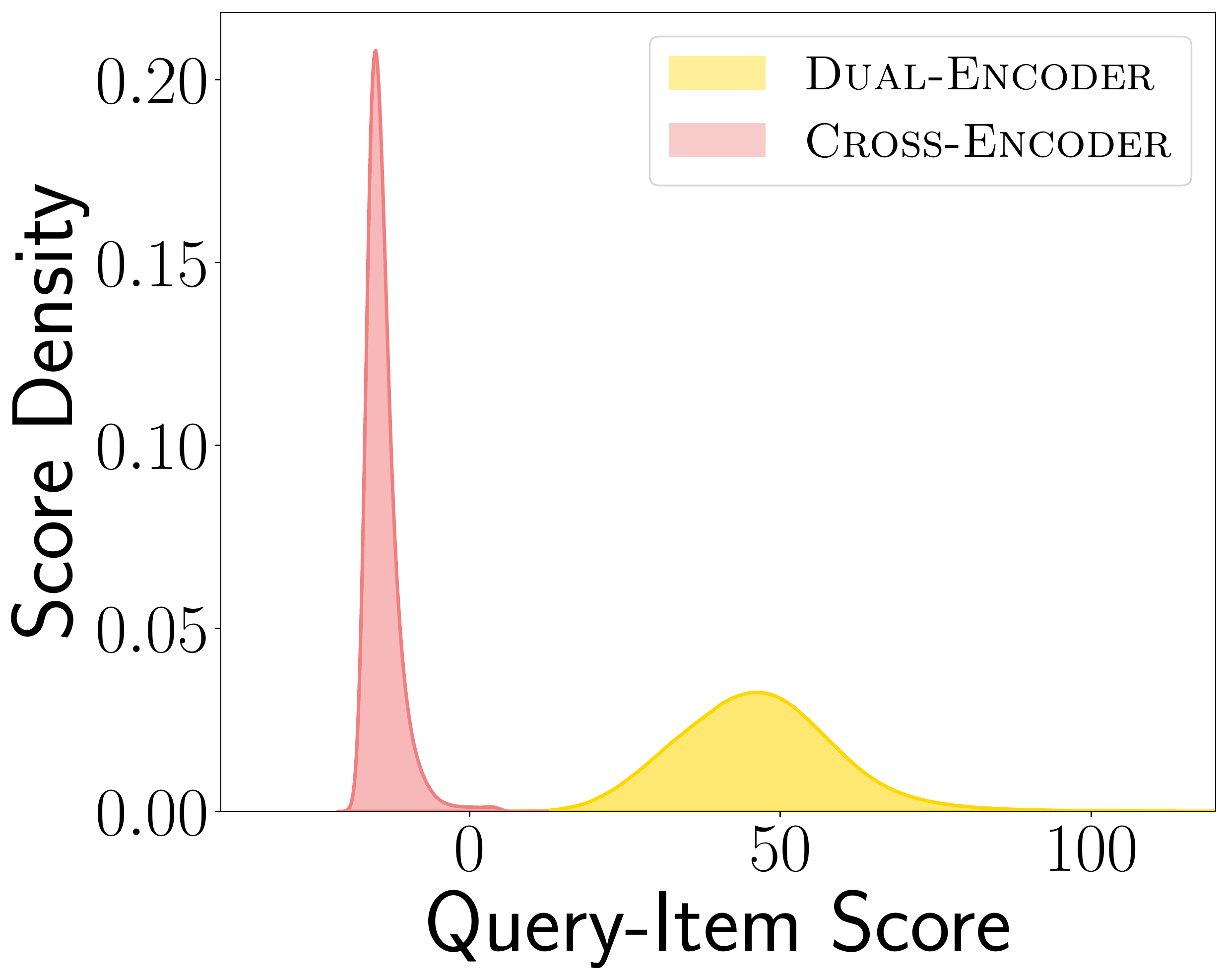}    
        \caption{\scidocs}
        \label{apndx_fig:rq_0_score_distribution_scidocs}
    \end{subfigure}
    \hspace{2cm}
    \begin{subfigure}[b]{0.3\textwidth}
        \centering
        \includegraphics[width=\textwidth]{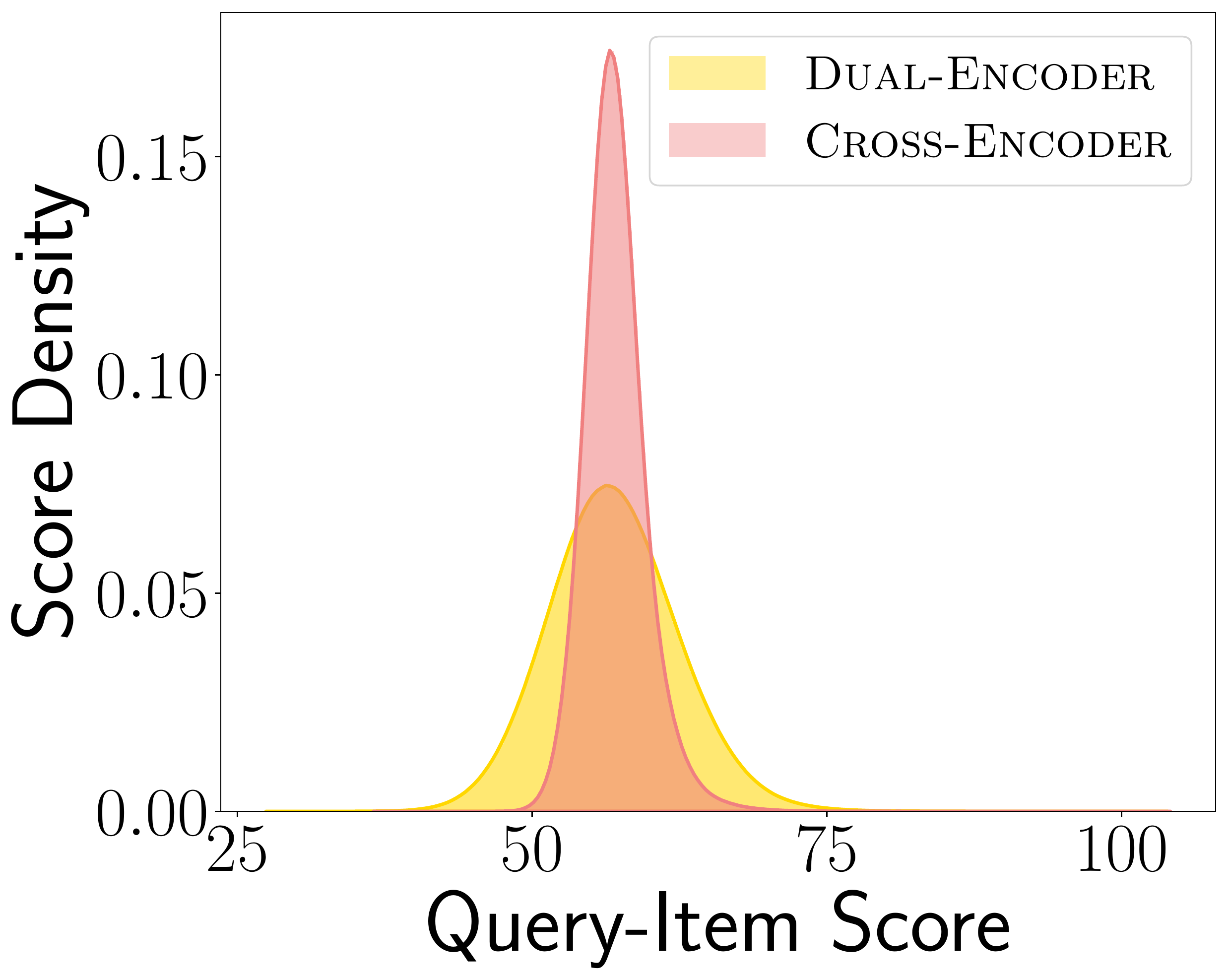}
        \caption{\yugioh}
        \label{apndx_fig:rq_0_score_distribution_yugioh}
    
    \end{subfigure}
    \caption{
    Score distribution for cross-encoder (CE) and dual-encoder (DE) models on \scidocs for \beir and \yugioh from \zeshel. For each domain, we use cross-encoder and dual-encoder models trained on the corresponding task. See \S\ref{apndx_subsec:cross_encoder_training} for details on cross-encoder training and  \S\ref{apndx_subsec:dual_encoder_training} for dual-encoder training.}
    \label{apndx_fig:rq_0_score_distribution}
\end{figure}

\paragraph{Cross-Encoder Score Normalization for \propInf{}{}}
Figure~\ref{apndx_fig:rq_0_score_distribution_scidocs} shows query-item score distribution for the cross-encoder model and \baseDualEncoder
on \scidocs datasets from \beir benchmark.
For cross-encoder models trained on \beir dataset, we observe that the cross-encoder
and \baseDualEncoder model produce query-item scores in significantly different ranges.
Since \baseDualEncoder is used to initialize the embedding space for
matrix factorization approaches, this resulted in a mismatch in the range of the target score distribution from the cross-encoder in sparse matrix $\sparseMat$ and the initial predicted score distribution from \baseDualEncoder.
Consequently, using raw cross-encoder scores while training \matrixFact{} models and while computing test query embedding 
by solving the linear regression problem in Eq~\ref{eq:lin_reg_query_emb} leads to a poor approximation of the cross-encoder.
To alleviate this issue, we normalize the cross-encoder scores to match the 
score distribution from \baseDualEncoder model using two parameters $\alpha, \beta \in \RR$.
\[s_\text{final}(q,i) =  \beta(s_\text{init}(q,i) - \alpha)\] 
where $s_\text{init}(q,i)$ and $s_\text{final}(q,i)$ are initial and normalized cross-encoder scores, and $\alpha$ and $\beta$ are
estimated by re-normalizing cross-encoder distribution to match dual-encoder score distribution using 100 training queries.
Note that such score normalization does not affect the final ranking 
of items.

We do \emph{not} perform any such normalization for \zeshel datasets the cross-encoder and
\baseDualEncoder model output scores in similar ranges as shown in Figure~\ref{apndx_fig:rq_0_score_distribution_yugioh}.

\section{Additional Results and Analysis}
\label{apndx_sec:sparse_mf_ablations_and_analysis}

\subsection{Overhead of Adaptive Retrieval with \propInf{}{}}
\label{apndx_subsec:adaptive_search_overhead}

\begin{figure}[!ht]
    \centering
    \begin{subfigure}[b]{0.49\textwidth}
    \includegraphics[width=\textwidth]{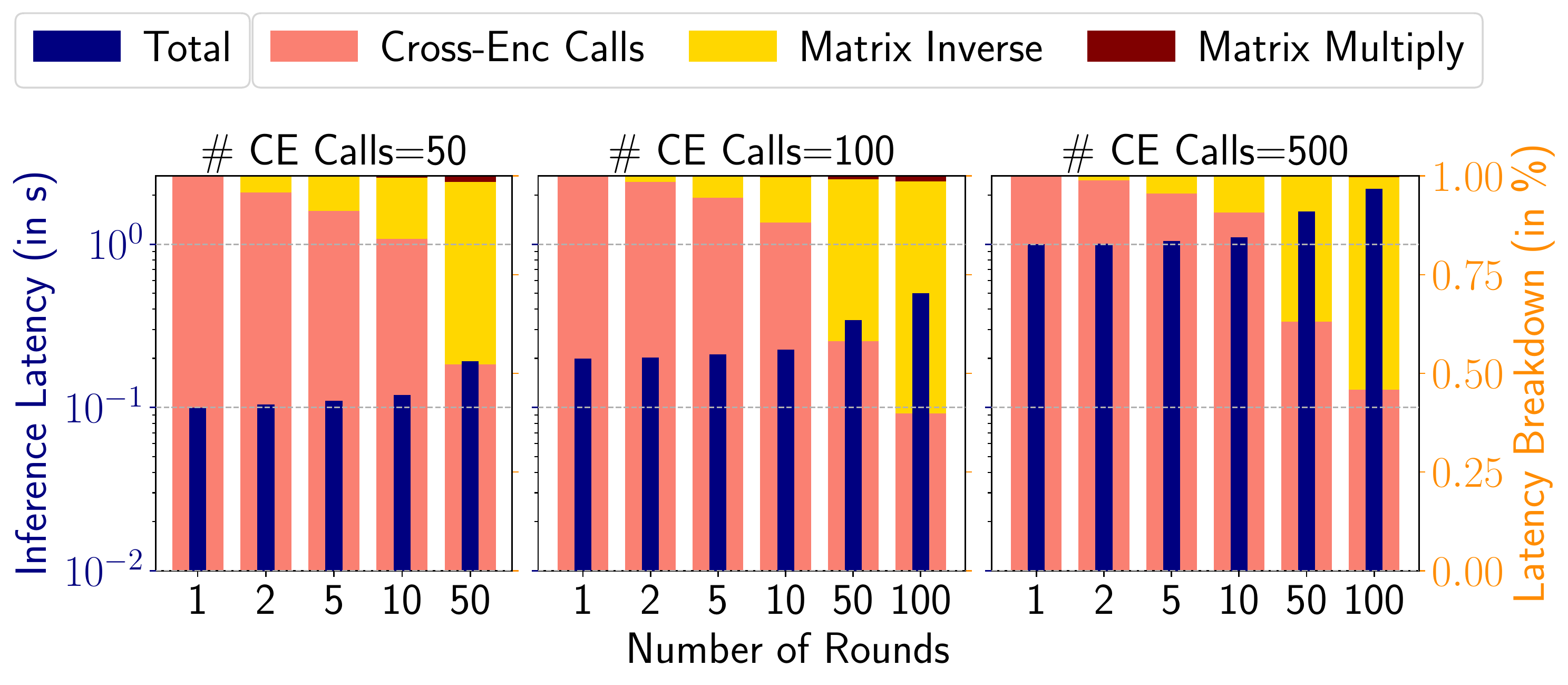}
    \caption{ \adaCUR{\baseDualEncoder} }
    \label{apndx_fig:rq_4_inference_latency_breakdown_hotpotqa_adacur}
    \end{subfigure}
    \hfill
    \begin{subfigure}[b]{0.49\textwidth}
    \includegraphics[width=\textwidth]{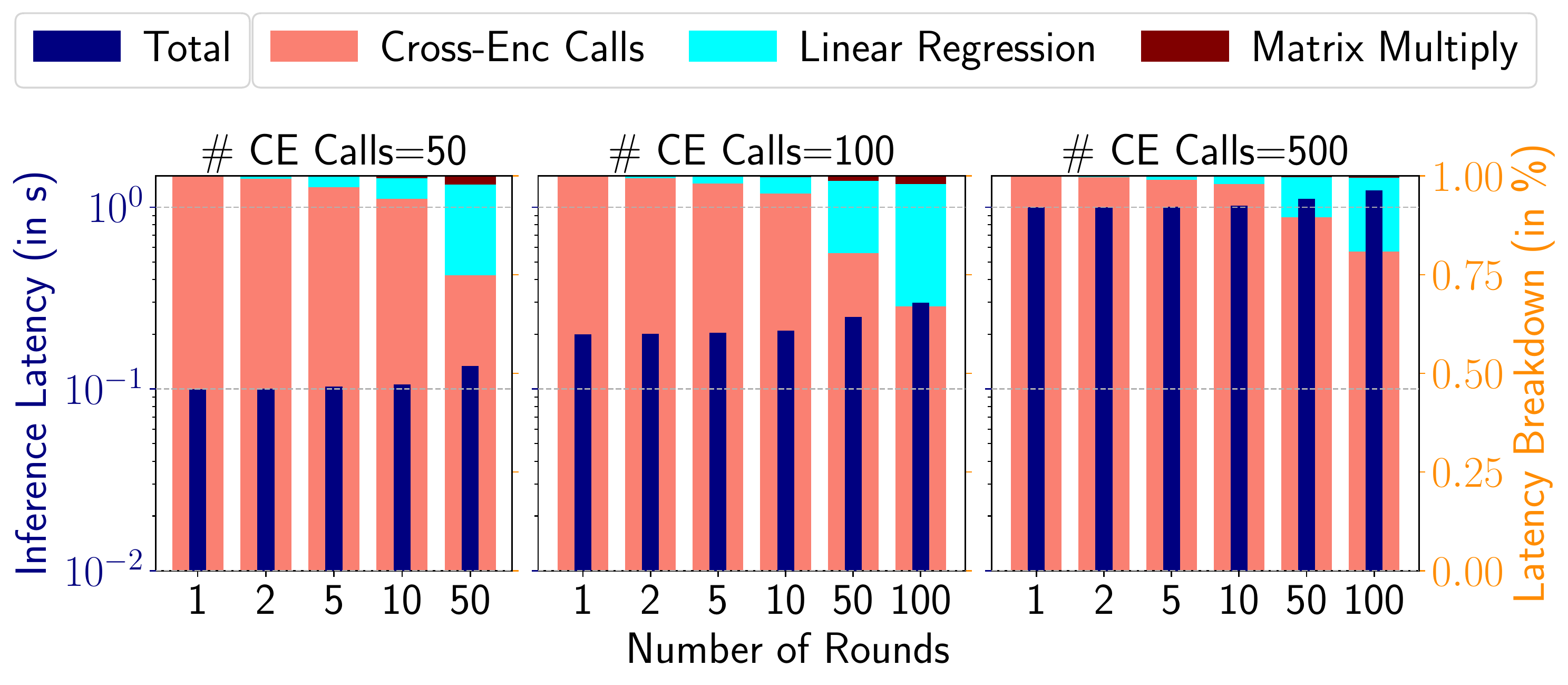}
    \caption{\propInf{\baseDualEncoder}{\baseDualEncoder}}
    \label{apndx_fig:rq_4_inference_latency_breakdown_hotpotqa_linreg}
    \end{subfigure}
    
    \caption{Breakdown of inference latency for \adaCUR{\baseDualEncoder} and \propInf{\baseDualEncoder}{\baseDualEncoder} under different test-time CE call budgets for domain=\hotpotqa. See~\S\ref{apndx_subsec:adaptive_search_overhead} for detailed discussion.
    }
    \label{apndx_fig:rq_4_inference_latency_breakdown_hotpotqa}
\end{figure}

\begin{figure}[!ht]
    \centering
    \begin{subfigure}[b]{0.49\textwidth}
    \includegraphics[width=\textwidth]{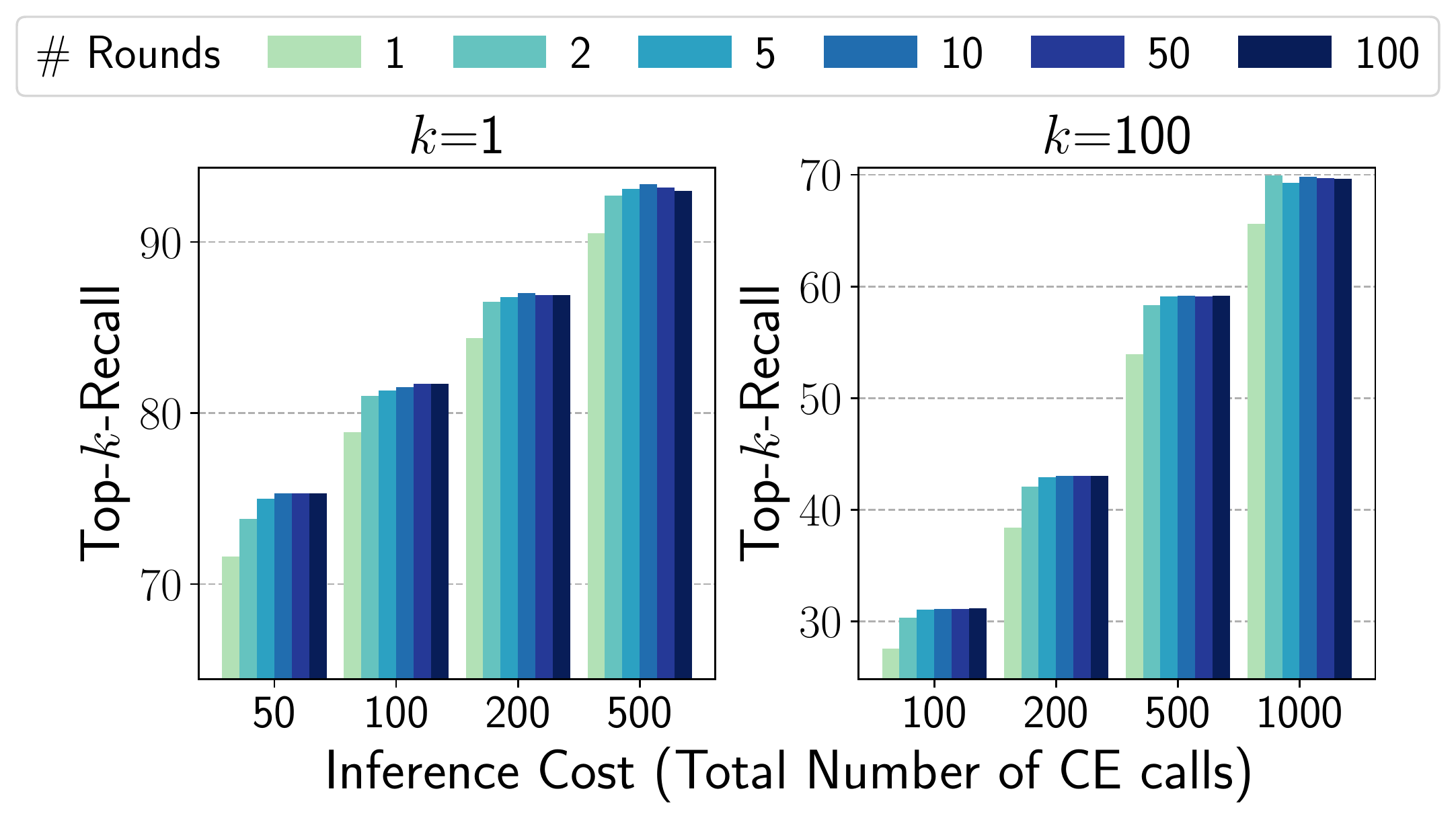}
    \caption{ \scidocs }
    \label{apndx_fig:rq_4_recall_vs_n_steps_scidocs}
    \end{subfigure}
    \hfill
    \begin{subfigure}[b]{0.49\textwidth}
    \includegraphics[width=\textwidth]{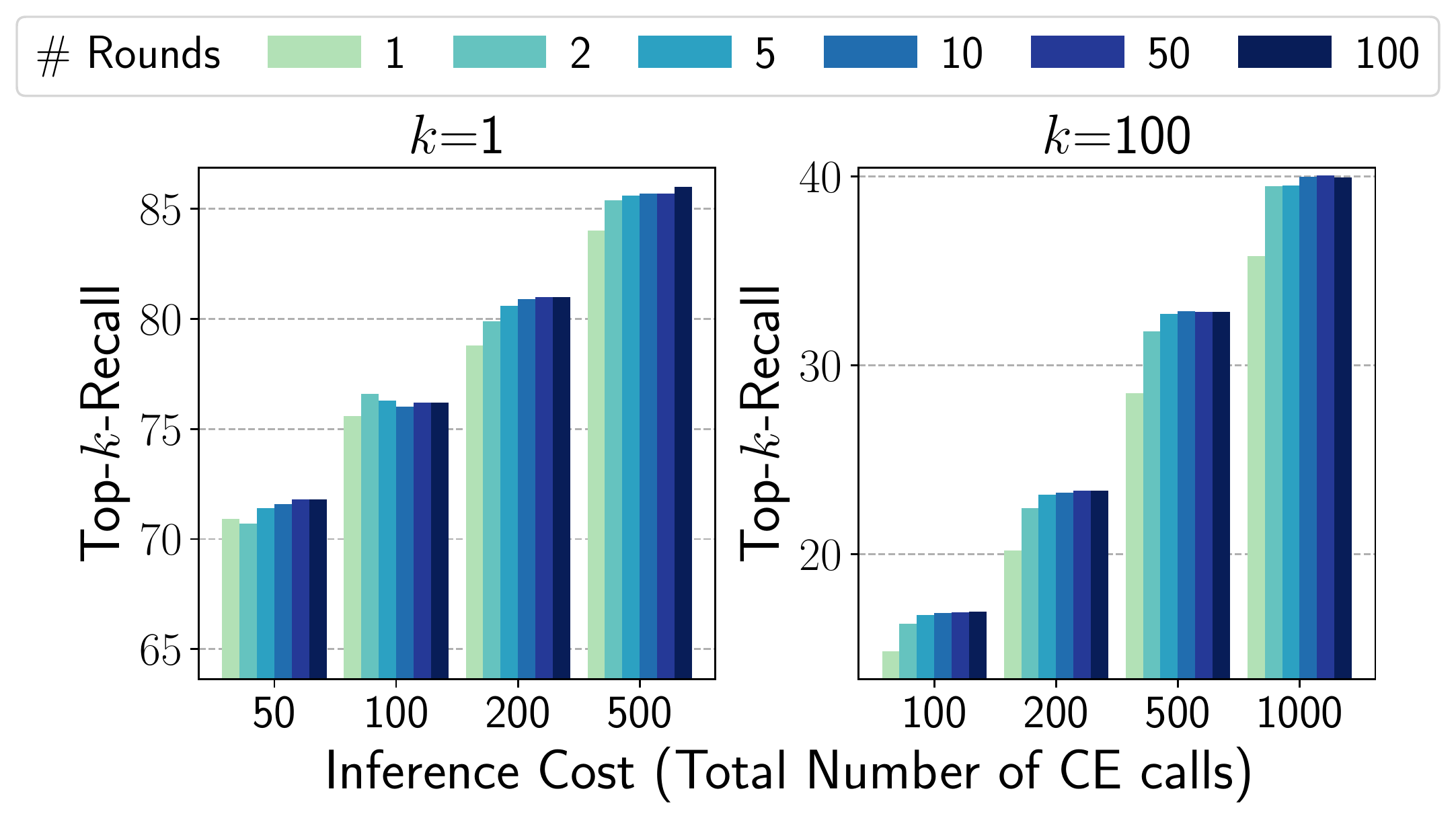}
    \caption{\hotpotqa}
    \label{apndx_fig:rq_4_recall_vs_n_steps_hotpotqa}
    \end{subfigure}
    \caption{Top-$k$-Recall versus number of rounds  for \propInf{\baseDualEncoder}{\baseDualEncoder}
    under different test-time cross-encoder call budgets for domains \hotpotqa and \scidocs. Number of rounds~($\nRounds$) = 1 corresponds to retrieve-and-rerank style inference with \baseDualEncoder i.e. \rnr{\baseDualEncoder}. Top-$k$-Recall generally improves with the number of rounds and saturates around 5 to 10 rounds. 
    }
    \label{apndx_fig:rq_4_recall_vs_n_steps}
\end{figure}

Figures~\ref{apndx_fig:rq_4_inference_latency_breakdown_hotpotqa_adacur} and~\ref{apndx_fig:rq_4_inference_latency_breakdown_hotpotqa_linreg} 
show total inference latency for \adaCUR{} and \propInf{}{}
for varying number of rounds~($\nRounds$) at different cross-encoder (CE) calls budgets.
The secondary y-axis in Figure~\ref{apndx_fig:rq_4_inference_latency_breakdown_hotpotqa} shows the breakdown of the
inference latency into three main steps in Algorithm~\ref{alg:adaptive_test_time_inference} -
(a) CE Calls: computing CE scores for retrieved items (line~\ref{alg_line:update_exact_ce_scores}),
(b) solving linear regression problem to update test query embedding for \propInf{}{}(line~\ref{alg_line:lin_reg_query_emb}) 
(c) Matrix Multiply: updating approximate scores for all items~(line~\ref{alg_line:update_approx_ce_score}) 
followed by retrieving items using approximate scores.
In case of \adaCUR{}, computing query embedding in step (b)
involves computing the pseudo-inverse of a matrix instead of solving
a linear regression problem.

As shown in Figure~\ref{apndx_fig:rq_4_inference_latency_breakdown_hotpotqa}, the overhead of adaptive retrieval is negligible
for $\nRounds = 5$ to 10, and the overhead increases linearly with the number of rounds. \propInf{\baseDualEncoder}{\baseDualEncoder} for $\nRounds=1$ corresponds to \rnr{\baseDualEncoder}, retrieve-and-rerank style inference using \baseDualEncoder.
We observe that \propInf{}{} incurs less overhead than 
\adaCUR{} under the same test-time CE call budget.
Each CE call takes an amortized time of $\sim$2 ms\footnote{On an Nvidia 2080ti GPU with 12 GB memory for a 6-layer Mini-LM~\citep{wang2020minilm} based model.} when computing CE scores with a batch-size of up to 50 for domain=\hotpotqa.
While the time complexity of updating the approximate scores is linear in the number of items,
we observe that this step can be significantly sped up using GPUs/TPUs, and use of efficient 
vector-based $k$-NN search methods. 
In this work, to get an efficient implementation for large domains such as \hotpotqa, 
we first shortlist 10K items for the test query using 
the baseline retrieval method (e.g. \baseDualEncoder), 
and only update the approximate scores for those 10K 
during inference using brute-force computation of scores for all 10K items.
Further, note that the approximate scores are only used for retrieving 
items~(line~\ref{alg_line:lin_reg_item_select} in Alg.~\ref{alg:adaptive_test_time_inference}), and this operation 
can also be implemented on CPUs using efficient vector-based $k$-NN search methods~\citep{malkov2018efficient, guo2020accelerating} 
without the need for brute-force computation of approximate scores
for all items.

\subsection{Comparing different query embedding methods}
\label{apndx_subsec:query_embed_ablation}

\begin{figure}[!ht]
    \centering
    \begin{subfigure}[b]{\textwidth}
    \centering
    \includegraphics[width=0.8\textwidth, trim={0 10.1cm 0 0}, clip]{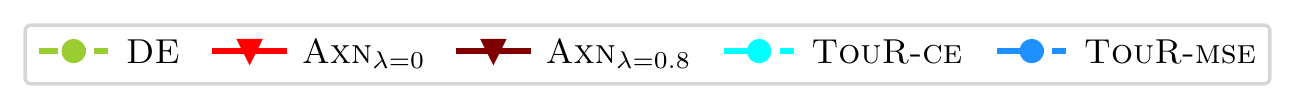}
    \end{subfigure}
    \begin{subfigure}[b]{\textwidth}
    \includegraphics[width=0.48\textwidth]{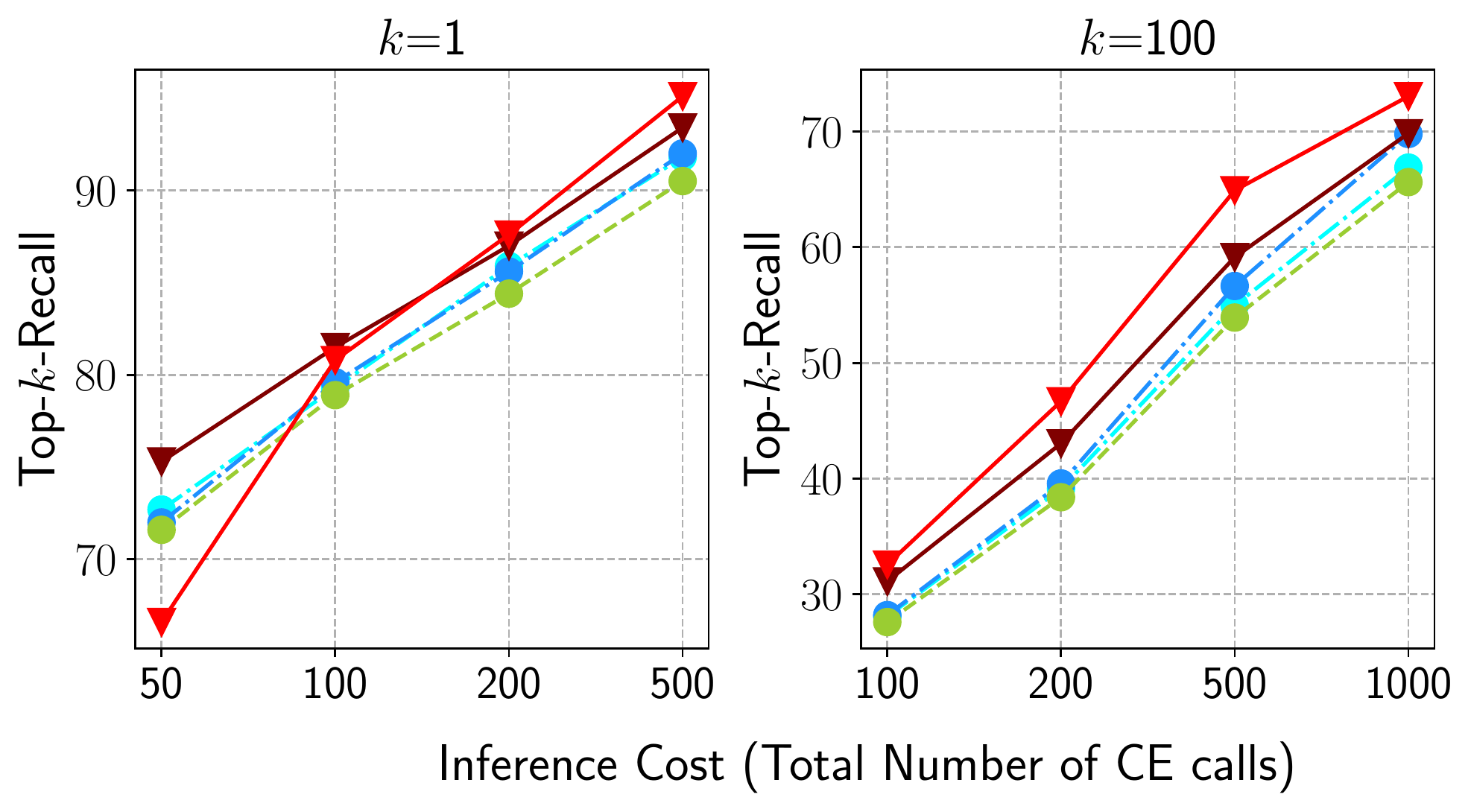}
    \includegraphics[width=0.24\textwidth]{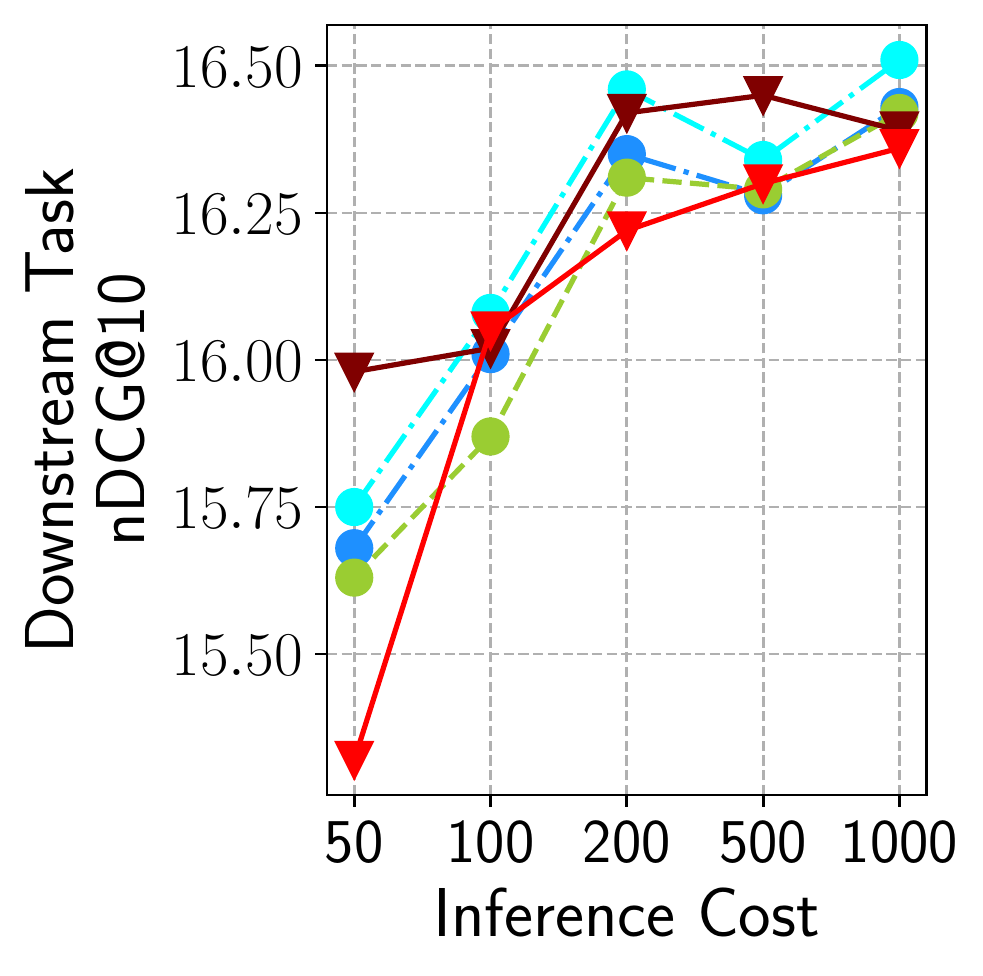}    
    \includegraphics[width=0.24\textwidth]{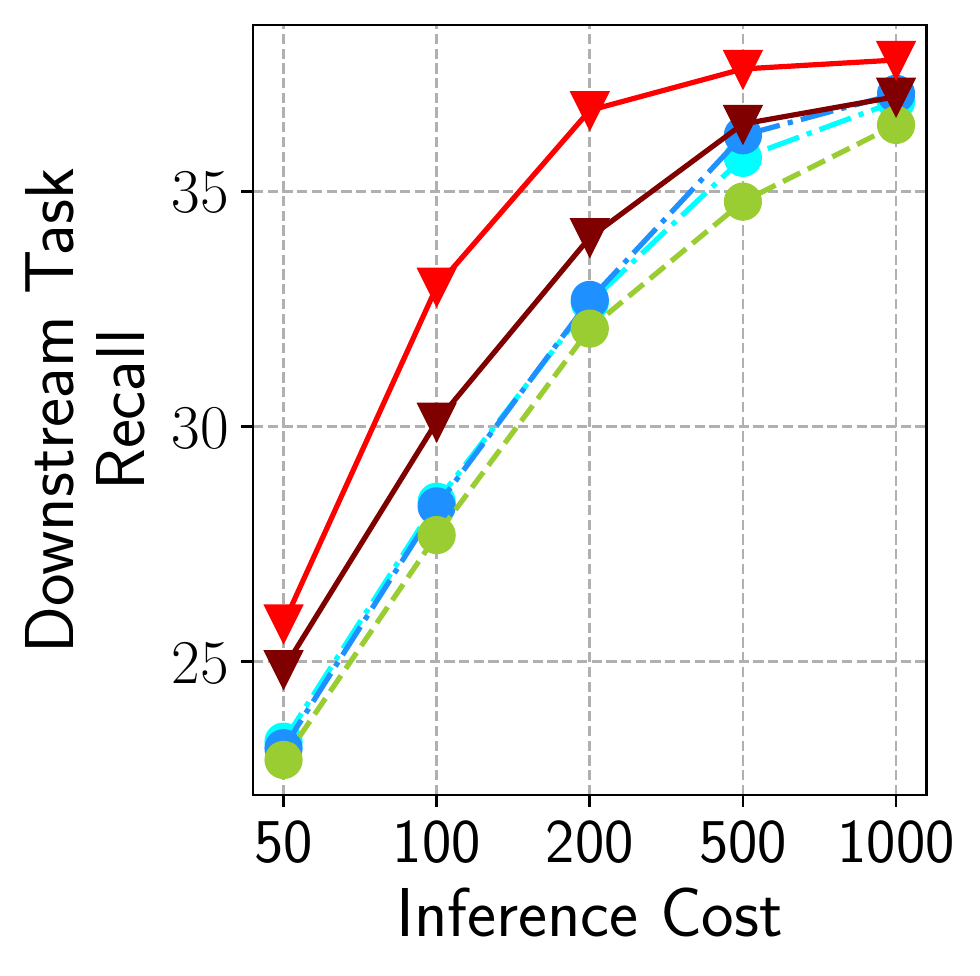}    
    \vspace{-0.2cm}
    \caption{\scidocs}
    \end{subfigure}
    
    \begin{subfigure}[b]{\textwidth}
    \includegraphics[width=0.48\textwidth]{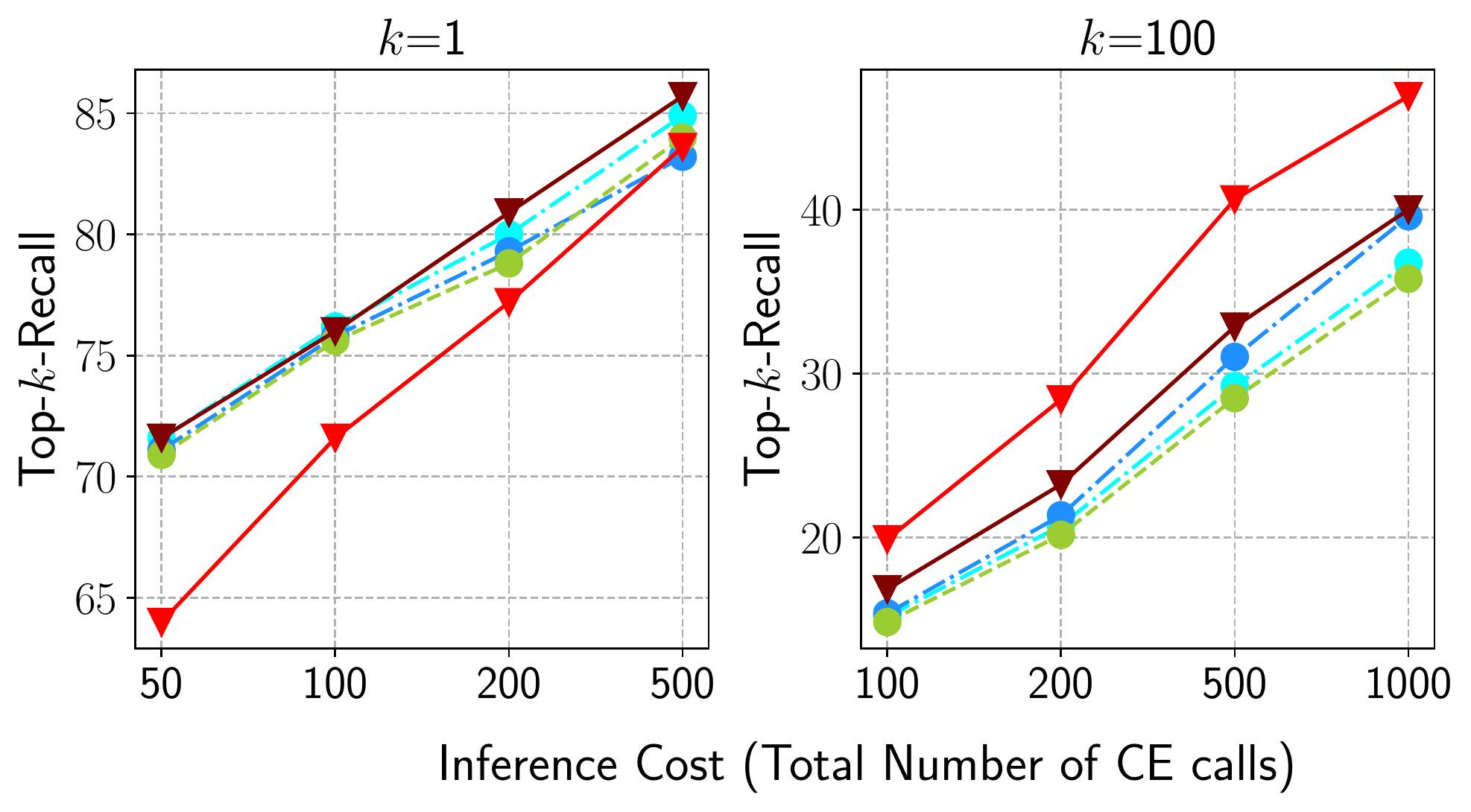}
    \includegraphics[width=0.24\textwidth]{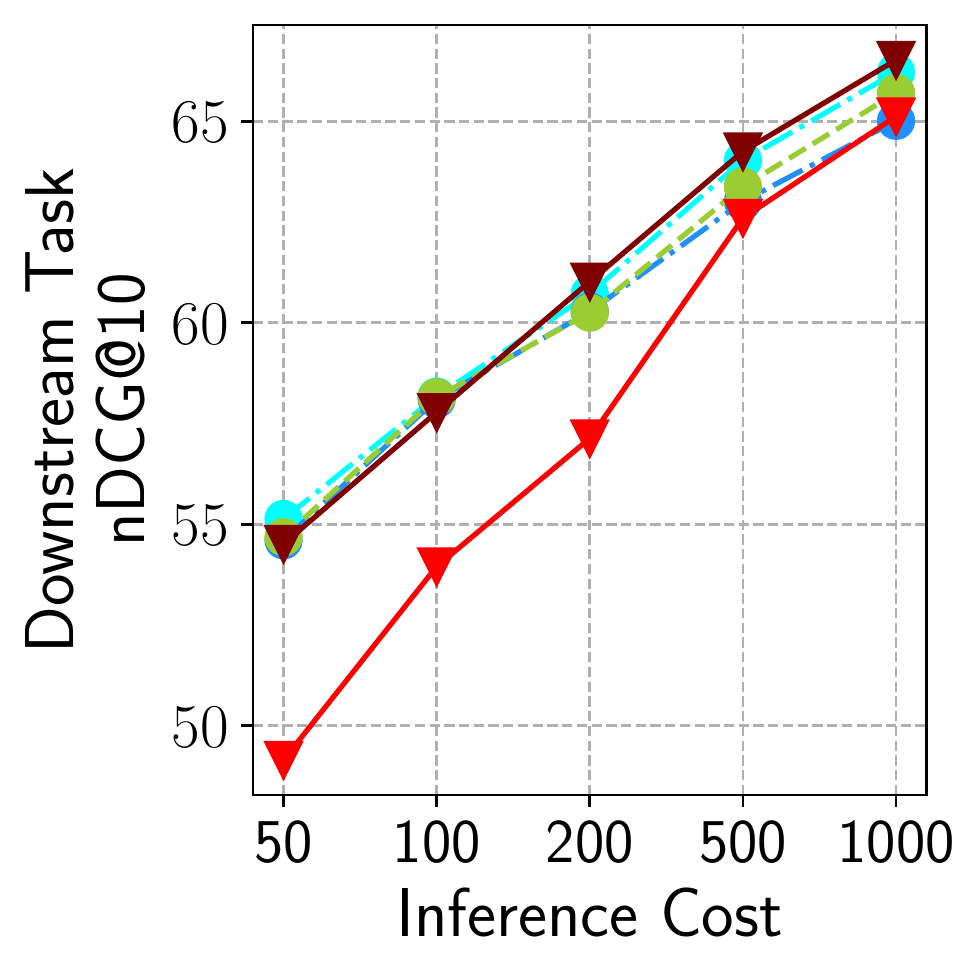}    
    \includegraphics[width=0.24\textwidth]{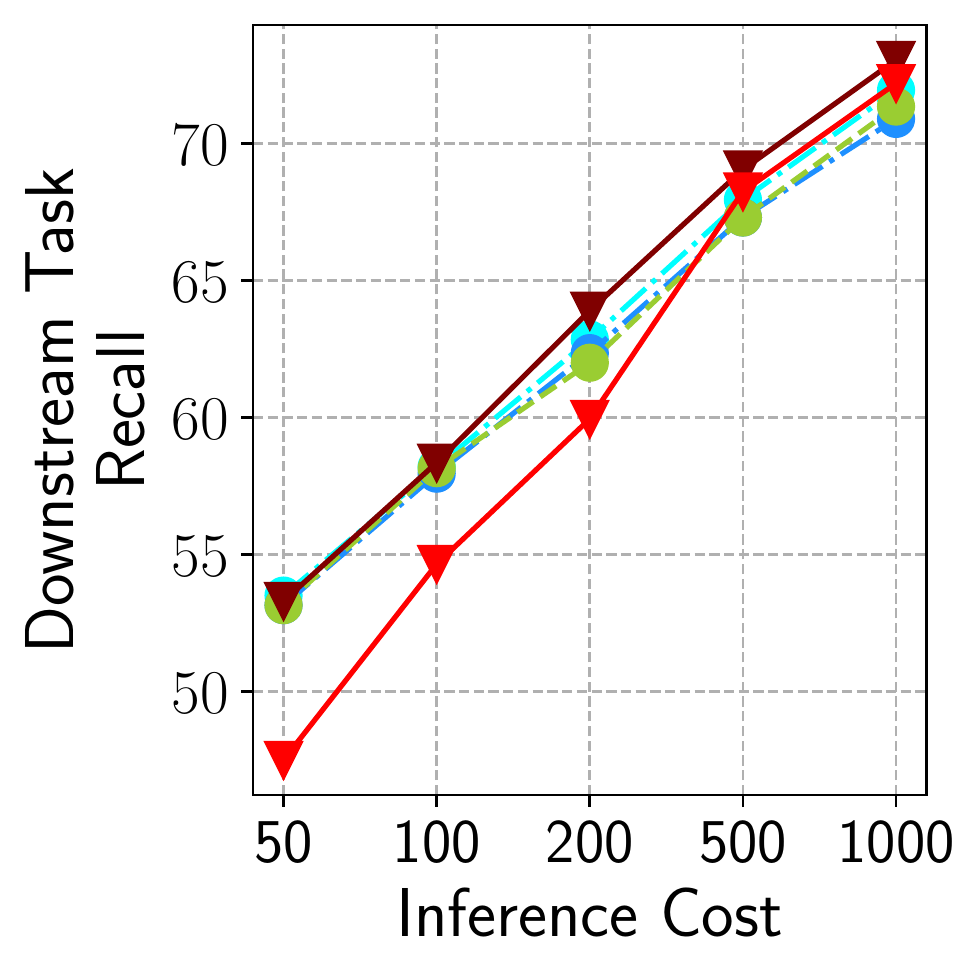}    
    \vspace{-0.2cm}
    \caption{\hotpotqa}
    \end{subfigure}
    \caption{Top-$k$-Recall versus inference cost for different test query embedding methods on domains \scidocs and \hotpotqa. See~\S\ref{apndx_subsec:query_embed_ablation} for detailed discussion. }
    \label{apndx_fig:rq_5_test_query_ablation_hotpotqa}
\end{figure}

\begin{figure}[!ht]
    \centering
    \includegraphics[width=0.8\textwidth]{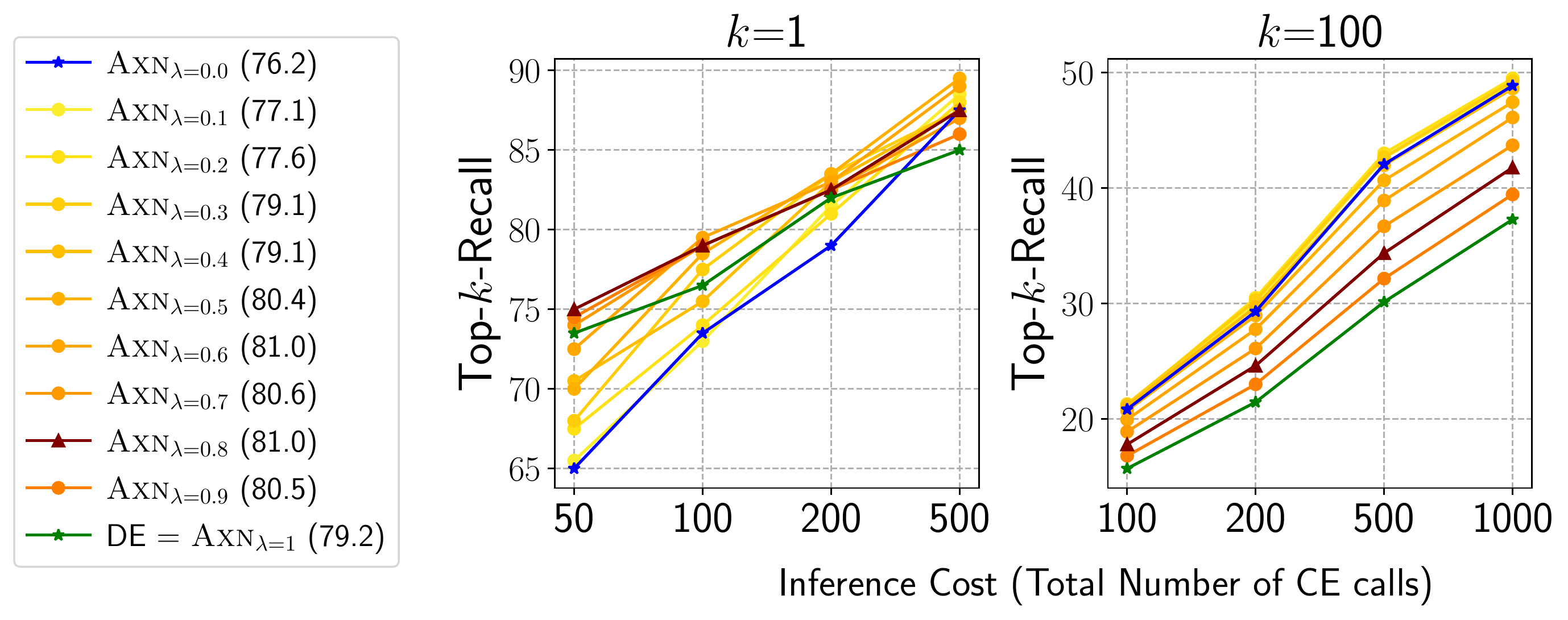}
    \caption{Top-$k$-Recall for \propInf{\baseDualEncoder}{\baseDualEncoder} for different 
    values of $\lambda$ parameter in eq~\ref{eq:weighted_comb_lin_reg}. 
    We use  200 queries from the validation set in \hotpotqa and 
    the value in parentheses in the legend denotes average Top-1-Recall, 
    averaged over different test-time inference cost budgets. 
    For $k=1$, using $\lambda=0.8$ yields the best performance 
    and for $k=100$, we use $\lambda=0$ unless specified otherwise.}
    \label{apndx_fig:rq_5_test_query_lambda_dev_set_hotpotqa}
\end{figure}

Our proposed $k$-NN search method shares a similar motivation 
to pseudo-relevance feedback (PRF) methods that aim to improve the 
quality of retrieval by updating the initial query representation
using heuristic or model-based feedback on retrieved items.
We show results for \tour~\citep{sung-etal-2023-optimizing}, a recent PRF-based method that, 
similar to our method, also optimizes the test query representations 
using retrieval results while utilizing the CE call budget of $\ceBudget$ CE calls over $\nRounds$ rounds.  
However, unlike \propInf{}{}, \tour
uses a single gradient-based update to query embedding to minimize
KL-Divergence (\tourCE) or mean-squared error (\tourMSE) between approximate and exact scores for top-$\ceBudget/\nRounds$ items in each round.
In contrast, \propInf{}{} computes the analytical solution to 
the least-square problem in Eq.~\ref{eq:lin_reg_query_emb} in each round,
and optionally computes a weighted sum with the test query 
embedding from a dense parametric model such as a dual-encoder using weight $\lambda \in [0,1]$ in Eq.~\ref{eq:weighted_comb_lin_reg}.
For \tourCE, we use learning rate =0.1 (chosen from \{0.1, 0.5, 1.0\}) and for \tourMSE, we use learning rate = 1e-3 (chosen from \{1e-2, 1e-3, 1e-4\}).

Figure~\ref{apndx_fig:rq_5_test_query_ablation_hotpotqa} shows 
Top-$k$-Recall and downstream task metrics versus test-time inference CE cost budget~($\ceBudget$)  for \propInf{\baseDualEncoder}{\baseDualEncoder} under two settings of the weight parameter, $\lambda=0$ and 0.8, and for $\baseDualEncoder$ and \tour baselines.
For both \scidocs and \hotpotqa, $\propInf{}{}_{\lambda=0.8}$ performs better than $\propInf{}{}_{\lambda=0}$ for $k$-NN search when $k=1$ while $\lambda=0$ works better for searching for $k$=100 nearest neighbors.
\tour and \propInf{}{} achieve similar Top-1-Recall at smaller inference costs with \propInf{}{} performing marginally better
than \tour at larger cost budgets. However, for $k=100$,
$\propInf{}{}_{\lambda=0}$ achieves significantly better
recall than \tour.
We observe mixed trends for downstream task metrics. For instance,
$\propInf{}{}_{\lambda=0.8}$ and \tour baselines yield similar
performance for nDCG@10 on both \scidocs and \hotpotqa and
for downstream task recall on \hotpotqa while
$\propInf{}{}_{\lambda=0}$ performs better than all baselines 
on downstream task recall for \scidocs.

\subsection{Transductive versus Inductive Matrix Factorization}
\label{apndx_subsec:transductive_vs_inductive_mf}
\begin{figure}[!ht]
    \centering    
    \begin{subfigure}[b]{0.7\textwidth}
        \centering
        \includegraphics[width=\textwidth, trim={0 11cm 0 0.2cm}, clip]{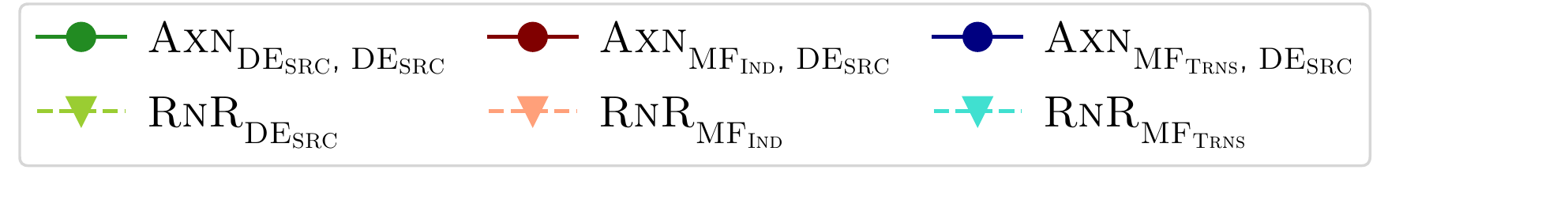}    
    \end{subfigure}
    \begin{subfigure}[b]{0.48\textwidth}
        \centering
        \includegraphics[width=\textwidth, trim={0 0 0 3cm}, clip]{apndx/rq_6_test_q=scidocs_all_0-1000_testq_method=all_nq_train=-1_kNN_recall_vs_index_cost=time_taken_total_topk=1_at_100_100_at_500.pdf}    
    \caption{\scidocs}
    \end{subfigure}
    \hfill
    \begin{subfigure}[b]{0.48\textwidth}
        \centering
        \includegraphics[width=\textwidth, trim={0 0 0 3cm}, clip]{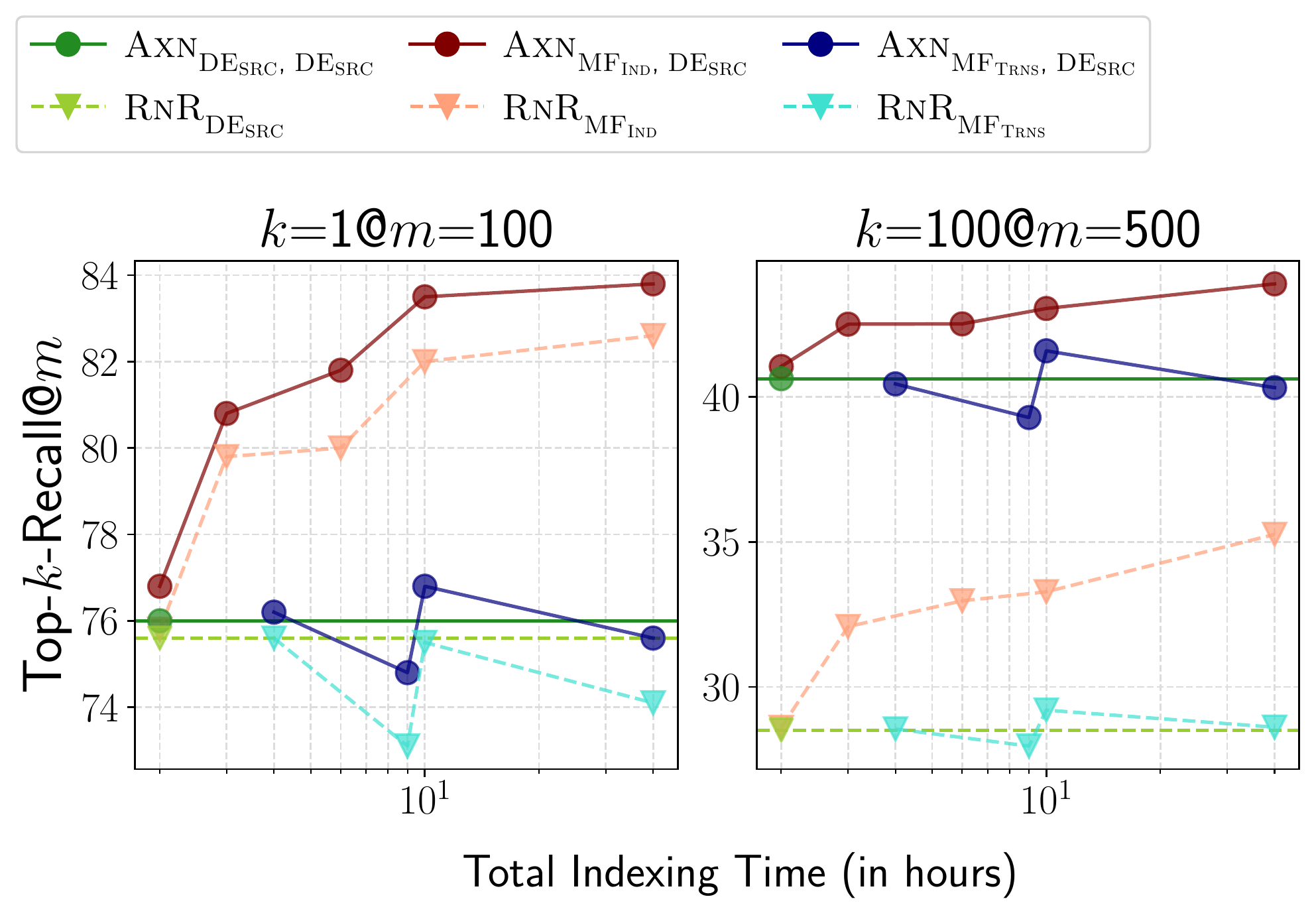}    
    \caption{\hotpotqa}
    \end{subfigure}
    \caption{
    Top-$k$-Recall versus indexing time for transductive~($\matrixFact{\transductive}$) and
    inductive~($\matrixFact{\inductive}$) matrix factorization for \scidocs and \hotpotqa. 
    We report Top-1-Recall and Top-100-Recall at fixed inference cost budget ($m$) of 100 and 500 CE calls respectively. See~\S\ref{apndx_subsec:transductive_vs_inductive_mf} for detailed discussion.}
    \label{apndx_fig:rq_6_inductive_vs_transductive_mf_ablation}
\end{figure}

Figure~\ref{apndx_fig:rq_6_inductive_vs_transductive_mf_ablation} shows
Top-$k$-Recall versus indexing time for \baseDualEncoder, and transductive~(\matrixFact{\transductive}) and inductive~(\matrixFact{\inductive})
matrix factorization in combination with two test-time inference methods: proposed inference method~(\propInf{}{}) and retrieve-and-rerank style~(\rnr{}) inference. 
We construct the sparse matrix \sparseMat by selecting
top-$\kDistill$ items for each train query using \baseDualEncoder,
and report results for $\queryTrainSize \in \{1K, 10K, 50K\}$ and 
$\kDistill \in \{100, 1000 \}$. We use \baseDualEncoder to initialize
the query and item embeddings for \matrixFact{} methods.

Recall that \matrixFact{\transductive} trains item embeddings as free-parameters,
and thus requires scoring an item against a small number of 
train queries in order to update the item embedding.
For this reason, \matrixFact{\transductive} performs marginally better 
than or at par with \matrixFact{\inductive} on small-scale data \scidocs with 25K items, 
as selecting even for $\queryTrainSize = 1000, \kDistill = 100$, results
in each item being scored with four queries on average. 
However, \matrixFact{\transductive} performs poorly for large-scale
data \hotpotqa (with 5 million items) due to the increased sparsity of matrix \sparseMat, providing marginal to no improvement over \baseDualEncoder.
In contrast, \matrixFact{\inductive} provides consistent improvement over
\baseDualEncoder on \hotpotqa.

\subsection{Effect of Sparse Matrix Construction Strategy}
\label{apndx_subsec:sparse_mat_construction_ablation}

\begin{table}[!ht]
    \centering
    \scriptsize{
    \begin{tabular}{c l|r  r r | r r r }
    \toprule
    & & \multicolumn{3}{c|}{\hotpotqa} & \multicolumn{3}{c}{\scidocs} \\
    Sparse Matrix &  $\queryTrainSize$, $k_d$ & \multicolumn{1}{c}{Time to}   &  \multicolumn{2}{c|}{Train-Time} & \multicolumn{1}{c}{Time to}   &  \multicolumn{2}{c}{Train-Time} \\
    Construction Strategy & &  \multicolumn{1}{c}{compute \sparseMat} & \matrixFact{\inductive} & \matrixFact{\transductive} &  \multicolumn{1}{c}{compute \sparseMat} & \matrixFact{\inductive} & \matrixFact{\transductive} \\
     \midrule
    \multirow{6}{*}{$\kDistill$ items per query} &  1K, 100    &  3 mins & 5 mins (20)   & -    &  10 mins  & 5 mins (10)    & 1.5 mins  (10)\\
     & 1K, 1000   &  31 mins    & 20 mins (20)  & -                 &  1.6 hrs    & 20 mins (20)    & 7 mins  (10)\\
     & 10K, 100   &  30 mins    & 20 mins (20)  & 1.2 hrs (10)      &  1.6 hrs    & 20 mins  (20)   & 7.5 mins  (10)\\
     & 10K, 1000  &  5.2 hrs    & 3 hrs (20)    & 3.2 hrs (~4 )    &  16.7 hrs   & 3.2 hrs (20)    & 1.1 hrs  (~~4)\\
     & 50K, 100   &  2.6 hrs    & 1.2 hrs (20)  & 4.1 hrs (10)      &  8.3 hrs    & 1.3 hrs (20)    & 0.6 hrs  (10) \\
     & 50K, 1000  &  26.3 hrs   & 9 hrs (10)    & 16 hrs (~~4)      &  82 hrs     & 14 hrs (10)     & 3.7 hrs  (~~4)\\
     \midrule
     \multirow{3}{*}{$\kDistill$ queries per item} & 50K, 2     & 5.8 hrs & 3hrs (20) & 7.5 hrs (10)  & 5 mins     & 3 mins (20) & 6.5 mins  (20)\\
     & 50K, 5     & 12.7 hrs    & 8hrs (20) & 8.5 hrs (~~4)     & 14 mins         & 5 mins (20)    & 9 mins  (20)\\
     & 50K, 10     & 23 hrs     & 9hrs (10)  & 16 hrs (~~4)     & 26 mins        & 6 mins (20)   & 10 mins  (20)\\
    \bottomrule
    \end{tabular}
    }
    \caption{Breakdown of indexing latency for transductive~\matrixFact{\transductive} and inductive~\matrixFact{\inductive} matrix factorization methods on \scidocs and \hotpotqa. For each setting, we show the number of epochs for training the model in parentheses.
    Total indexing time also includes the time taken to compute initial query and item embeddings using \baseDualEncoder.  Computing item embeddings takes 90 seconds for \scidocs (with 25K items) and $\sim$2 hours for \hotpotqa (with 5 million items) on an Nvidia 2080ti GPU with 12 GB GPU memory.}
    \label{apndx_tab:rq_7_indexing_time_breakdown}
\end{table}

\begin{figure}[!ht]
    \centering    
    \begin{subfigure}[b]{0.8\textwidth}
        \centering
        \includegraphics[width=\textwidth, trim={0 18.5cm 0 7cm}, clip]{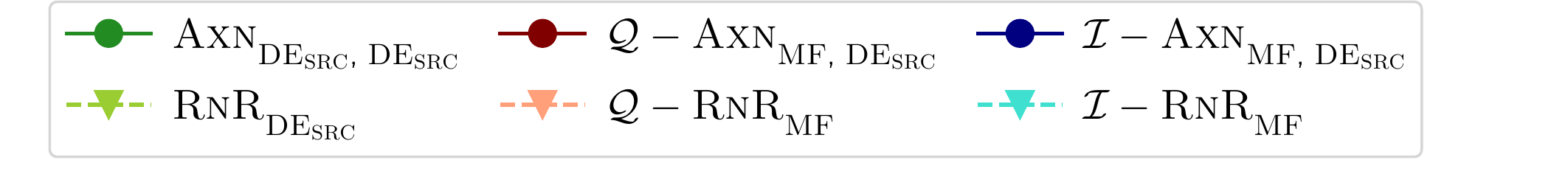}  
    \end{subfigure}
    \begin{subfigure}[b]{\textwidth}
    \begin{subfigure}[b]{0.48\textwidth}
        \centering
        \renewcommand\thesubfigure{\alph{subfigure}-1}
        \includegraphics[width=\textwidth, trim={0 7cm 0 10cm}, clip]{apndx/rq_7_test_q=scidocs_all_0-1000_testq_method=all_nq_train=-1_transductive_kNN_recall_vs_index_cost=time_taken_total_topk=1_at_100_100_at_500.pdf}
    \caption{\scidocs}
    \end{subfigure}
    \hfill
    \begin{subfigure}[b]{0.48\textwidth}
        \centering
        \addtocounter{subfigure}{-1}
        \renewcommand\thesubfigure{\alph{subfigure}-2}
        \includegraphics[width=\textwidth, trim={0 7cm 0 10cm}, clip]{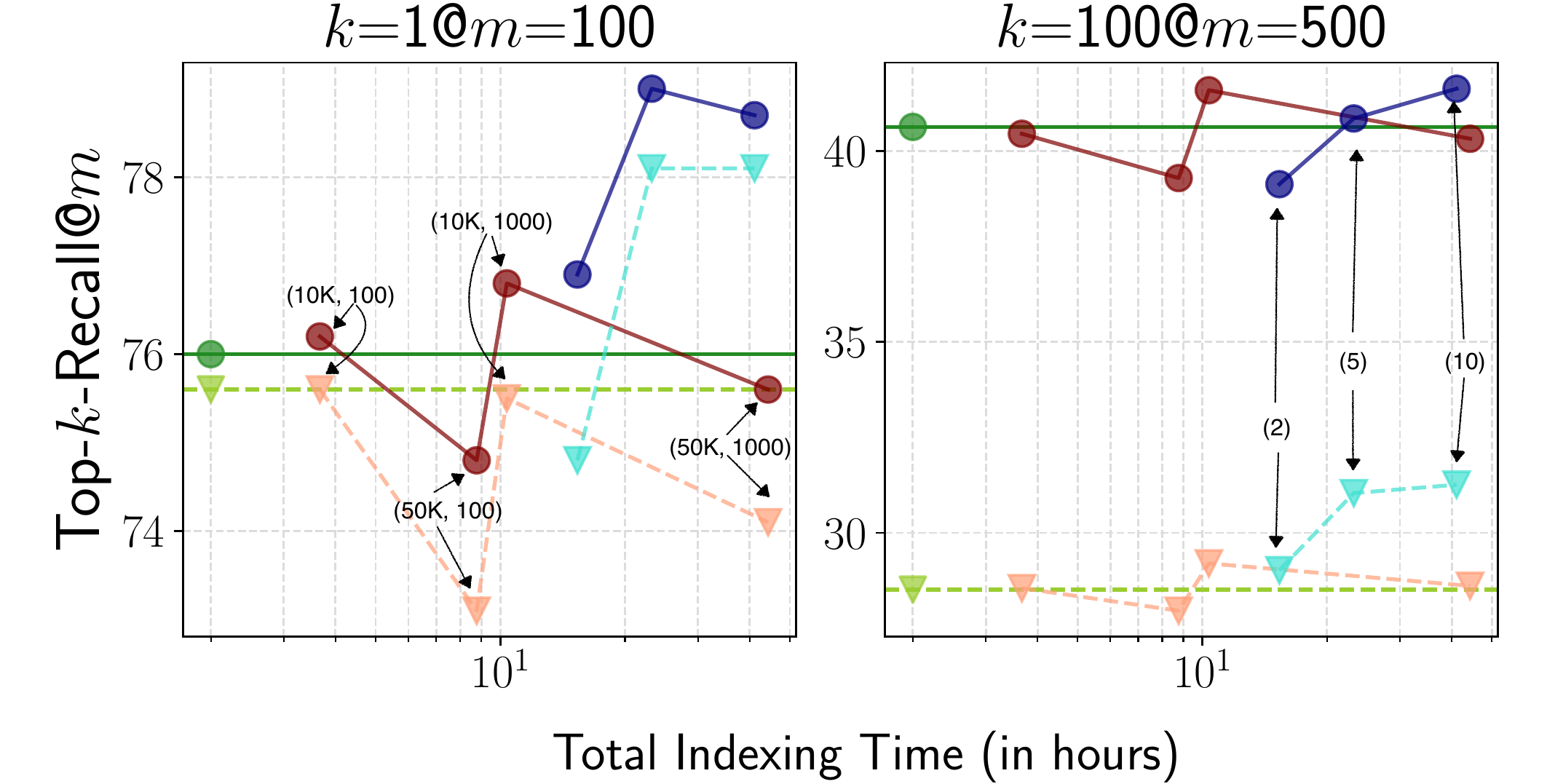}    
    \caption{\hotpotqa}
    \end{subfigure}
    \addtocounter{subfigure}{-1}
    \caption{Transductive Matrix Factorization~(\matrixFact{\transductive})}
    \end{subfigure}

    \begin{subfigure}[b]{\textwidth}
    \begin{subfigure}[b]{0.48\textwidth}
        \centering
        \renewcommand\thesubfigure{\alph{subfigure}-1}
        \includegraphics[width=\textwidth,trim={0 7cm 0 10cm}, clip]{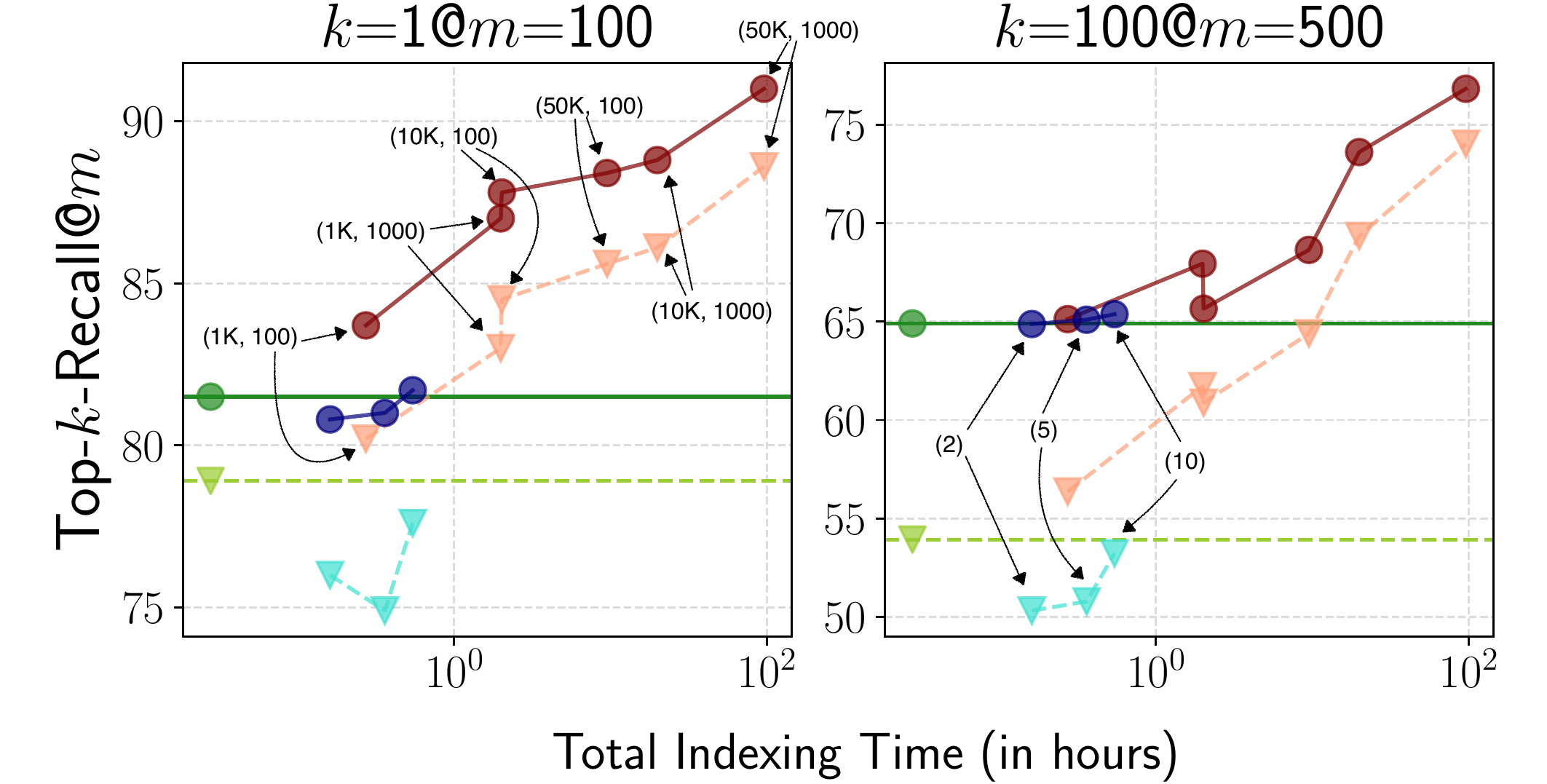}    
    \caption{\scidocs}
    \end{subfigure}
    \hfill
    \begin{subfigure}[b]{0.48\textwidth}
        \centering
        \addtocounter{subfigure}{-1}
        \renewcommand\thesubfigure{\alph{subfigure}-2}
        \includegraphics[width=\textwidth, trim={0 7cm 0 10cm}, clip]{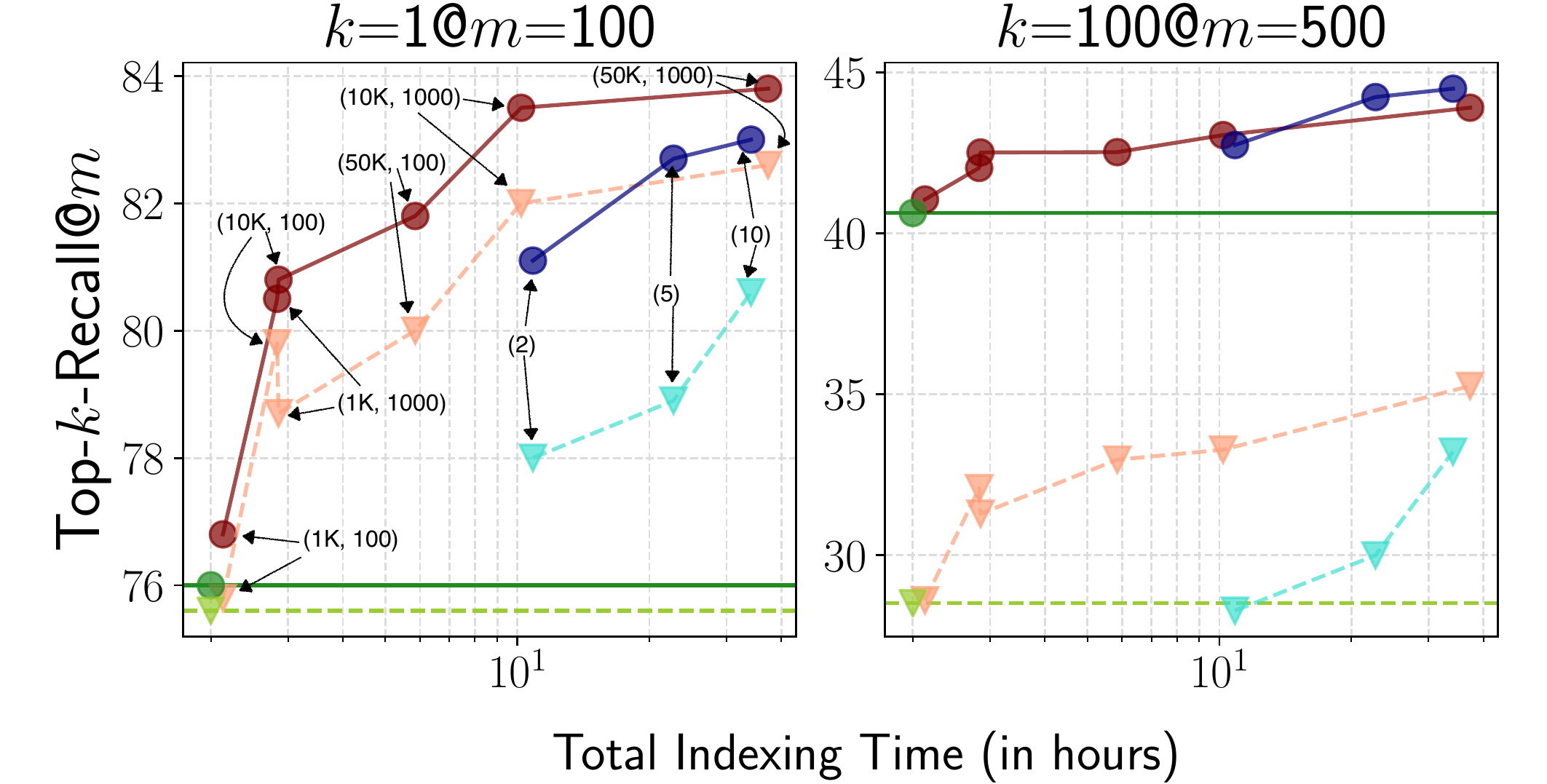}    
    \caption{\hotpotqa}
    \end{subfigure}
    \addtocounter{subfigure}{-1}
    \caption{Inductive Matrix Factorization~(\matrixFact{\inductive})}
    \end{subfigure}
    
    \caption{
    Top-1-Recall and Top-100-Recall at fixed inference cost budget ($m$) of 100 and 500 cross-encoder calls respectively versus indexing time (in hours) for different strategies of constructing sparse matrix $\sparseMat$. $\mathcal{Q}-*$ indicates that $\sparseMat$ is constructed by selecting a fixed number of items per \emph{query} in $\queryTrainData$, and $\mathcal{I}-*$ indicates that \sparseMat is constructed by selecting fixed number of queries per \emph{item} in $\itemSpace$. For $\mathcal{Q}-*$ approaches, the text annotations indicate 
    ($\queryTrainSize, \kDistill$) pairs where $\queryTrainSize$ is the number of anchor/train queries and $\kDistill$ is the number of items per query in the sparse matrix \sparseMat. For $\mathcal{I}-*$ approaches, the text annotations indicate the number of queries per item in the sparse matrix \sparseMat. 
    See~\S\ref{apndx_subsec:sparse_mat_construction_ablation} for detailed discussion.}
    \label{apndx_fig:rq_7_sparse_mat_construction_ablation}
\end{figure}

Figure~\ref{apndx_fig:rq_7_sparse_mat_construction_ablation} shows
Top-$k$-Recall versus indexing time for
and \matrixFact{} with two different strategies to construct
sparse matrix $G$ and Table~\ref{apndx_tab:rq_7_indexing_time_breakdown} shows the time taken 
to construct the sparse matrix $G$ and the time taken to train the matrix factorization model.
$\mathcal{Q}-*$ indicates that $\sparseMat$ is constructed by selecting a fixed number of $\kDistill$ items per \emph{query} in $\queryTrainData$, and $\mathcal{I}-*$ indicates that \sparseMat is constructed by selecting fixed number of $\kDistill$ queries per \emph{item} in $\itemSpace$. 
When selecting a fixed number of items per query, we experiment with $\queryTrainSize \in $ \{1K, 10K, 50K \} and $\kDistill \in \{100, 1000\}$.
When selecting a fixed number of queries per item, we first create a pool of 50K queries and then select $\kDistill$ queries per item for $\kDistill \in \{2, 5, 10\}$.

\textbf{Transductive Matrix Factorization}
For \matrixFact{\transductive}, both $\mathcal{Q}-*$ and $\mathcal{I}-*$ strategies yield similar
Top-$k$-Recall at a given indexing cost on \scidocs as
both strategies result in each item being scored with 
at least a few queries.
However, on \hotpotqa, selecting a fixed number of items
per query may not result in each item being scored against
some queries, and thus $\mathcal{Q}-*$ variants 
yield marginal (if any) improvement over \baseDualEncoder.
$\mathcal{I}-*$ variants perform better than \baseDualEncoder
and corresponding $\mathcal{Q}-*$ variants as each
item is scored against a fixed number of queries.
Note that this performance improvement comes
at the cost of an increase in time required to
compute sparse matrix $\sparseMat$, as shown in Table~\ref{apndx_tab:rq_7_indexing_time_breakdown}.

\textbf{Inductive Matrix Factorization}
For \matrixFact{\inductive}, we observe that $\mathcal{Q}-*$ variants
consistently provide better recall-vs-indexing time trade-offs
as compared to corresponding $\mathcal{I}-*$ variants on both \scidocs and \hotpotqa. 

% YuGiOh
\begin{figure}[!t]
    \centering
    \begin{subfigure}[b]{\textwidth}
    \centering
    \includegraphics[width=\textwidth, trim={0 10.1cm 0 0}, clip]{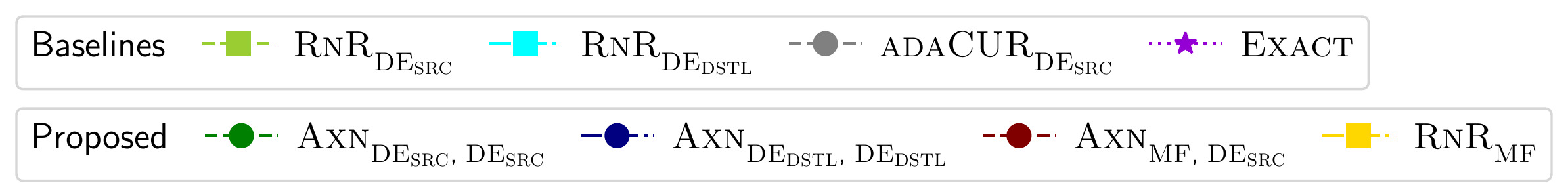}
    \end{subfigure}
    
    \begin{subfigure}[b]{\textwidth}
    \begin{subfigure}[b]{\textwidth}
        \centering
        \includegraphics[width=0.64\textwidth]{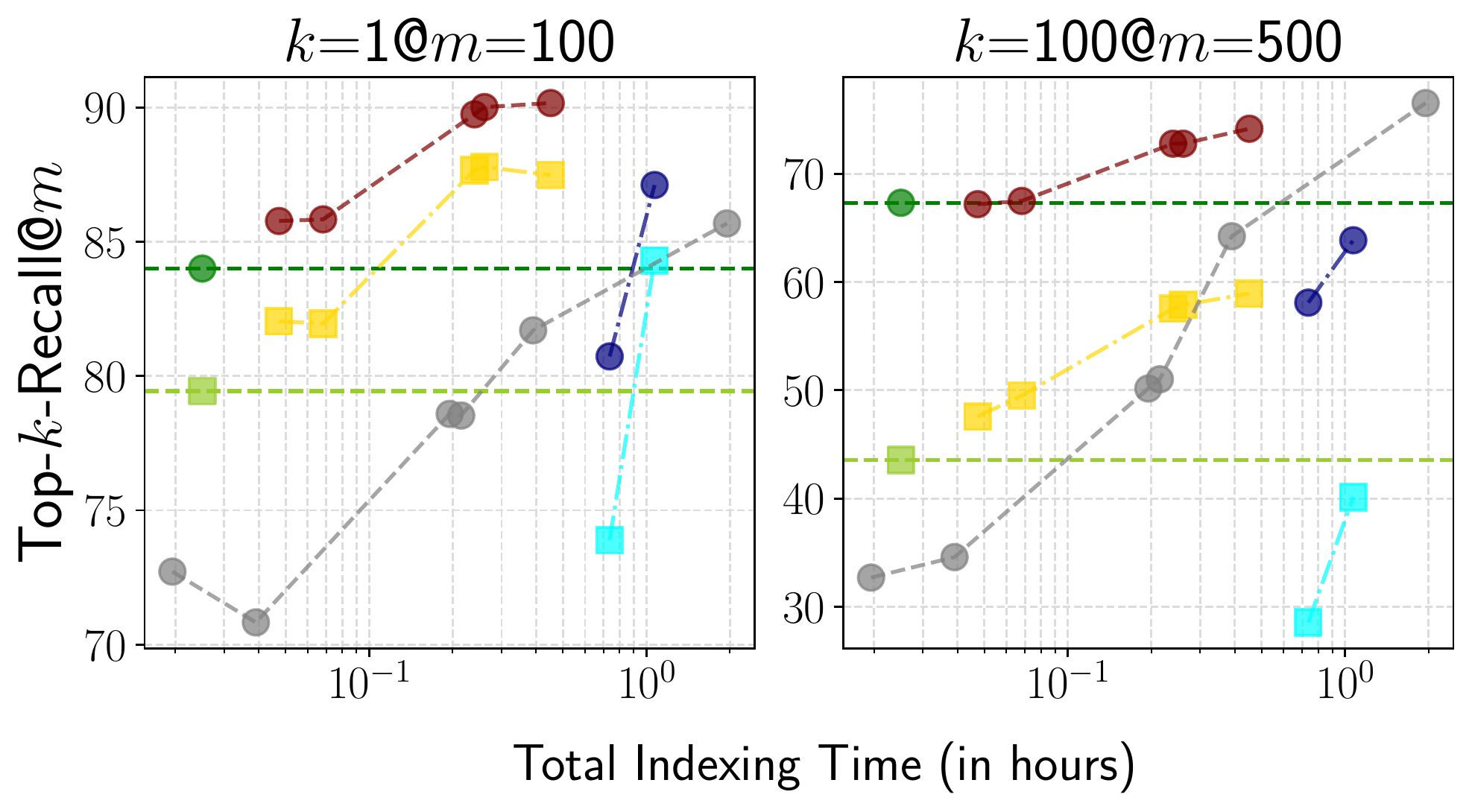}    
        \includegraphics[width=0.32\textwidth]{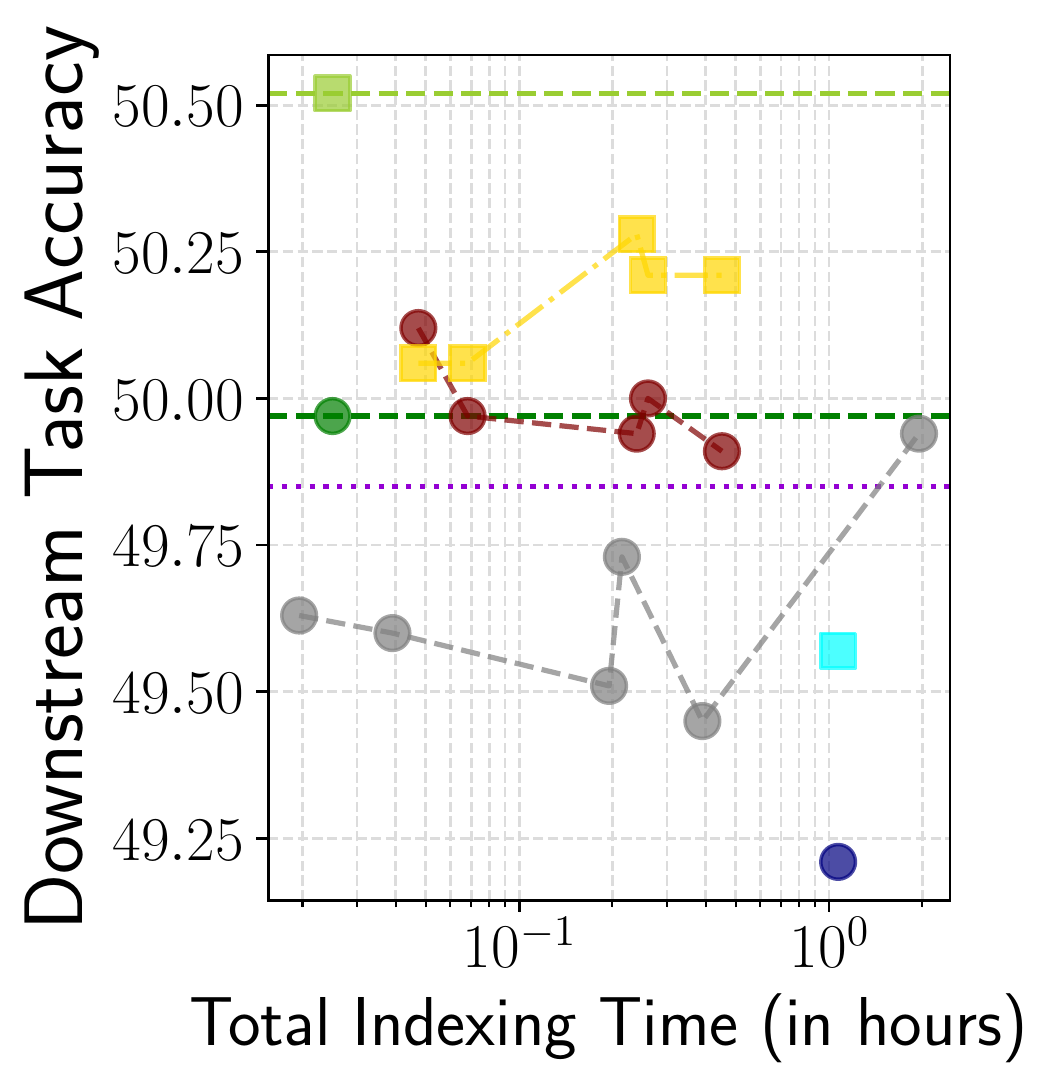}
    \end{subfigure}
    \caption{$\queryTrainSize=100, \queryTestSize=3274$}
    \end{subfigure}

    \begin{subfigure}[b]{\textwidth}
    \begin{subfigure}[b]{\textwidth}
        \centering
        \includegraphics[width=0.64\textwidth]{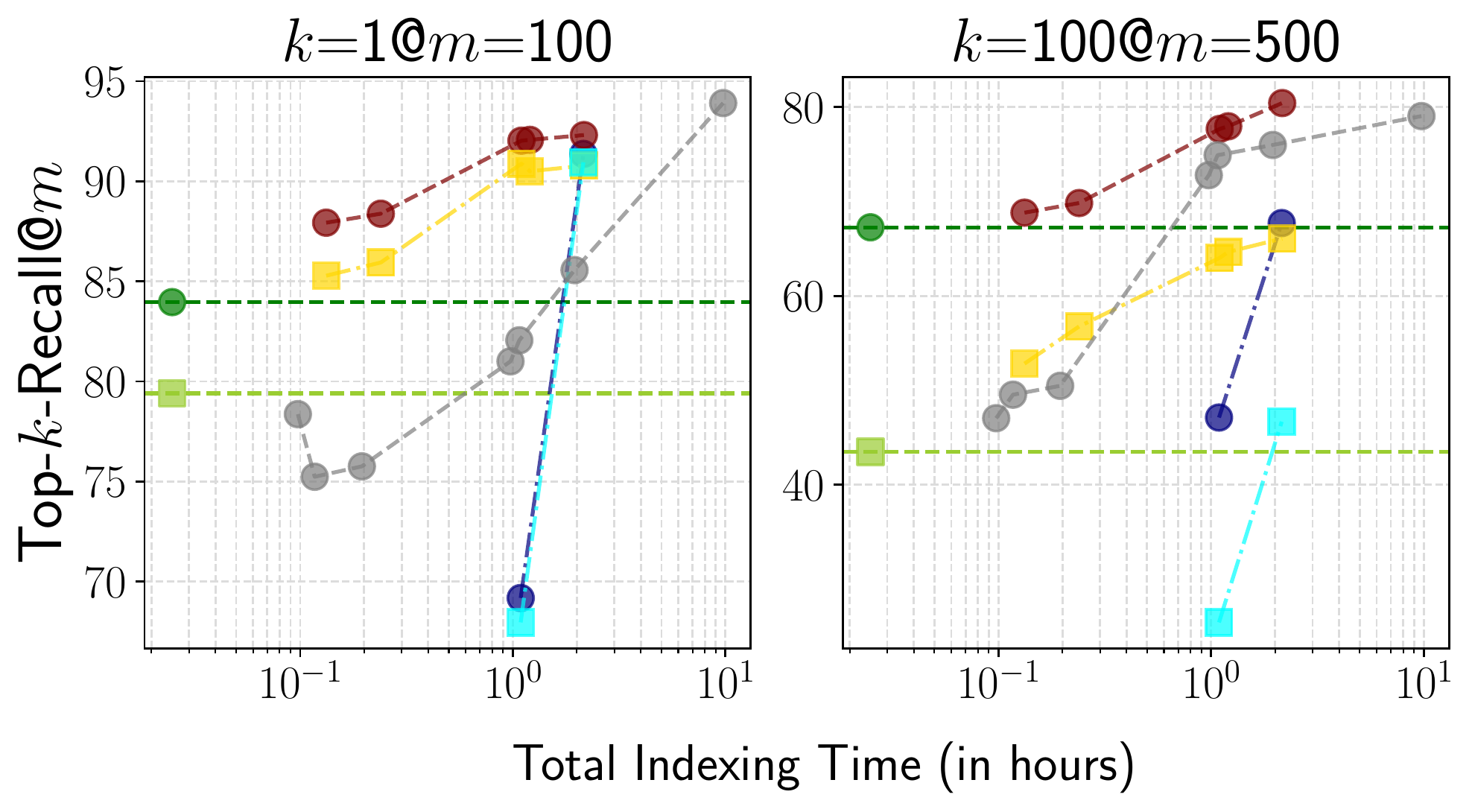}    
        \includegraphics[width=0.32\textwidth]{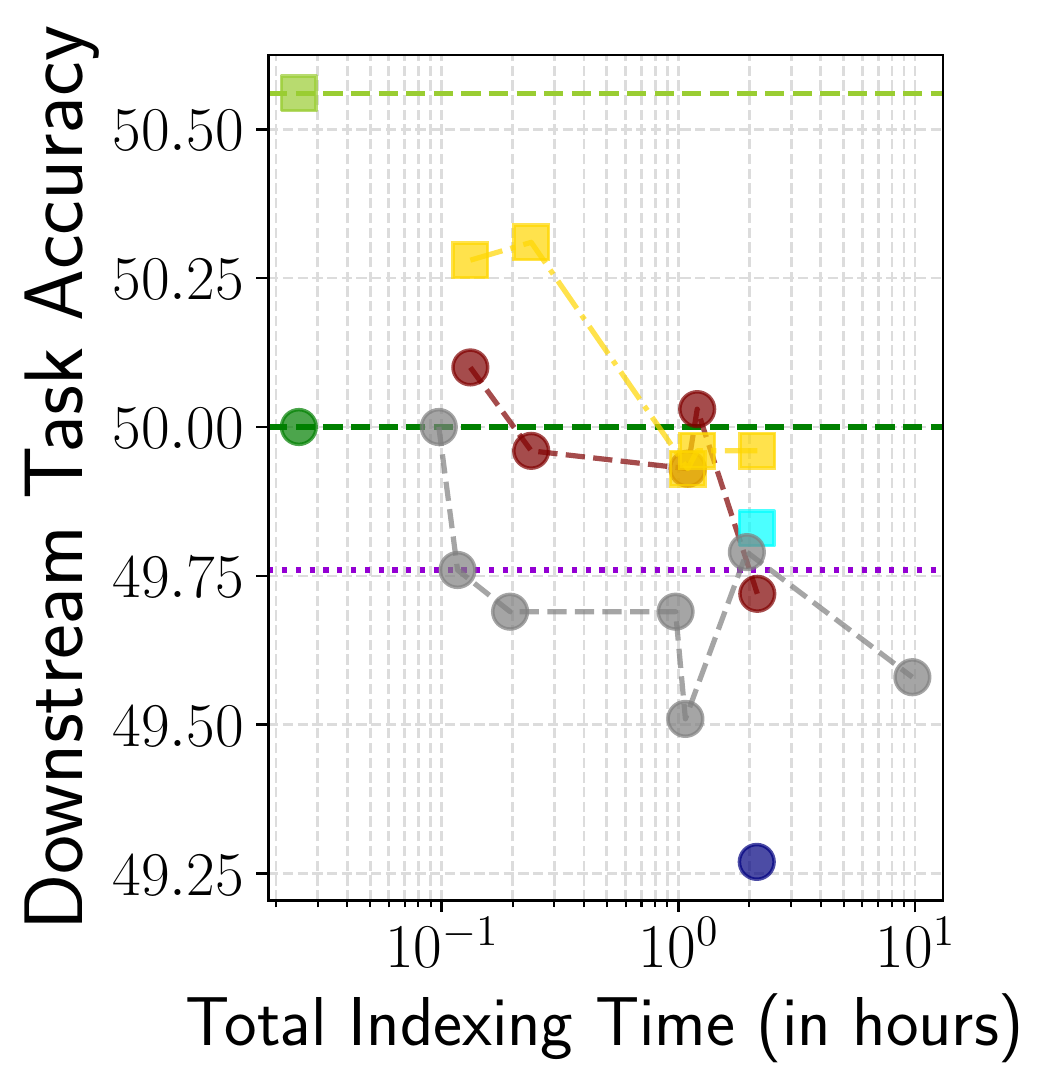}
    \end{subfigure}
    \caption{$\queryTrainSize=500, \queryTestSize=2874$}
    \end{subfigure}

    \begin{subfigure}[b]{\textwidth}
    \begin{subfigure}[b]{\textwidth}
        \centering
        \includegraphics[width=0.64\textwidth]{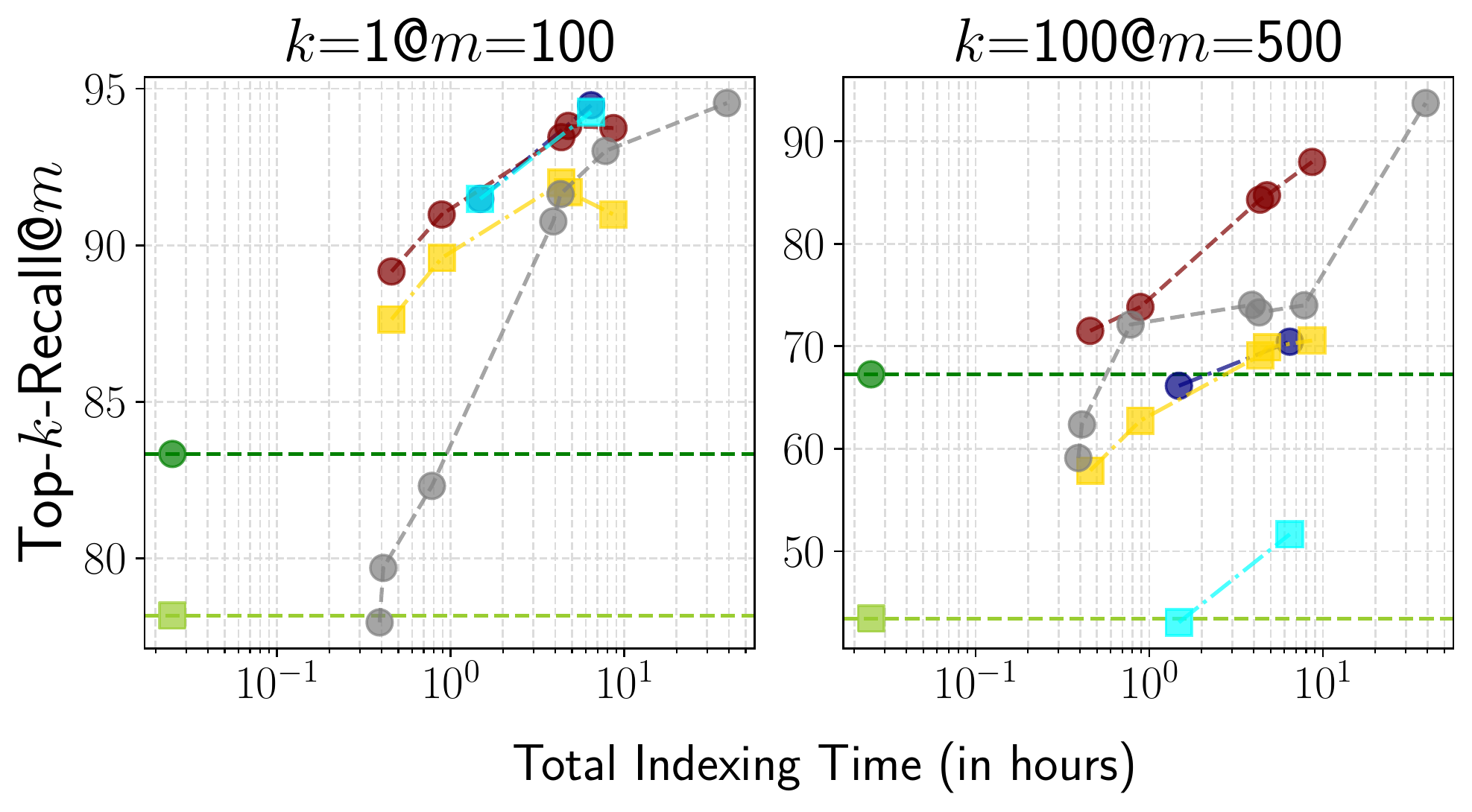}    
        \includegraphics[width=0.32\textwidth]{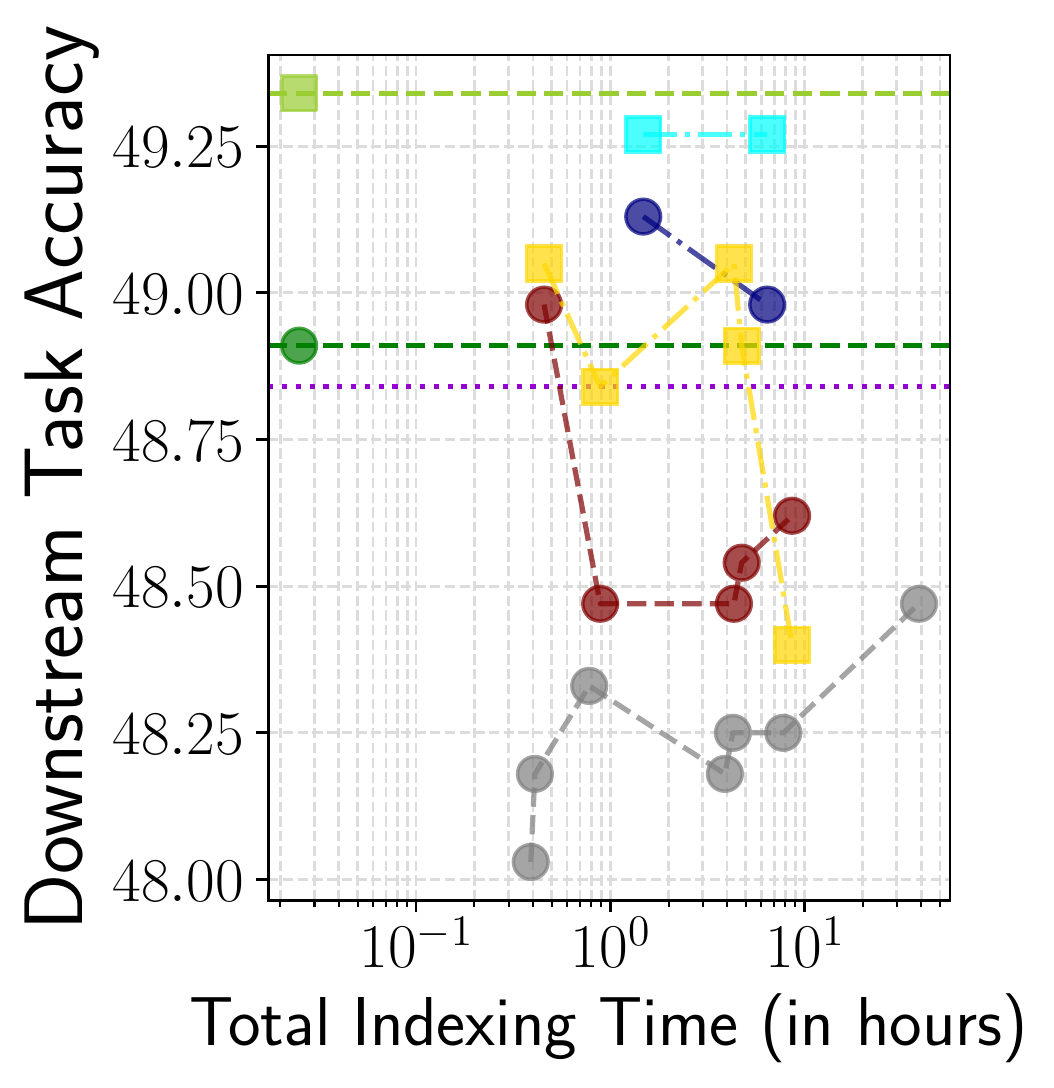}
    \end{subfigure}
    \caption{$\queryTrainSize=2000, \queryTestSize=1374$}
    \end{subfigure}
    \caption{
    Top-$k$-Recall and downstream task accuracy versus indexing time for various approaches on domain=\yugioh. 
    We report Top-1-Recall and Top-100-Recall at fixed inference cost budget ($m$) of 100 and 500 CE calls respectively, and downstream task accuracy for fixed inference cost of 100 CE calls. 
    Each subfigure shows results for different train/test splits.}
    \label{apndx_fig:rq_2a_recall_vs_indexing_cost_wall_clock_time_yugioh}
\end{figure}

% Star-Trek
\begin{figure}[!t]
    \centering
    \begin{subfigure}[b]{\textwidth}
    \centering
    \includegraphics[width=\textwidth, trim={0 10.1cm 0 0}, clip]{apndx/legend_gt_recall_wo_tfidf.pdf}
    \end{subfigure}
    
    \begin{subfigure}[b]{\textwidth}
    \begin{subfigure}[b]{\textwidth}
        \centering
        \includegraphics[width=0.64\textwidth]{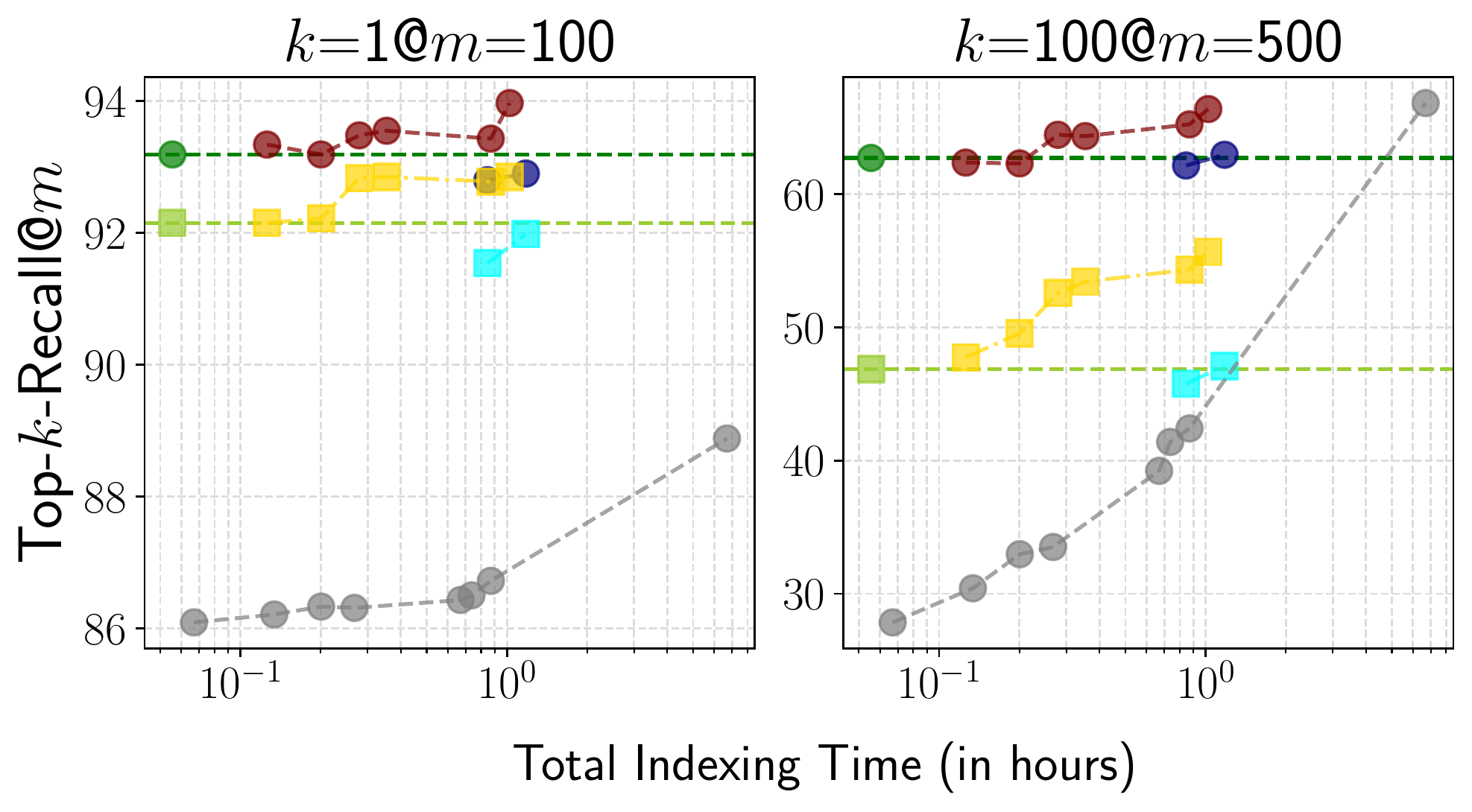}    
        \includegraphics[width=0.32\textwidth]{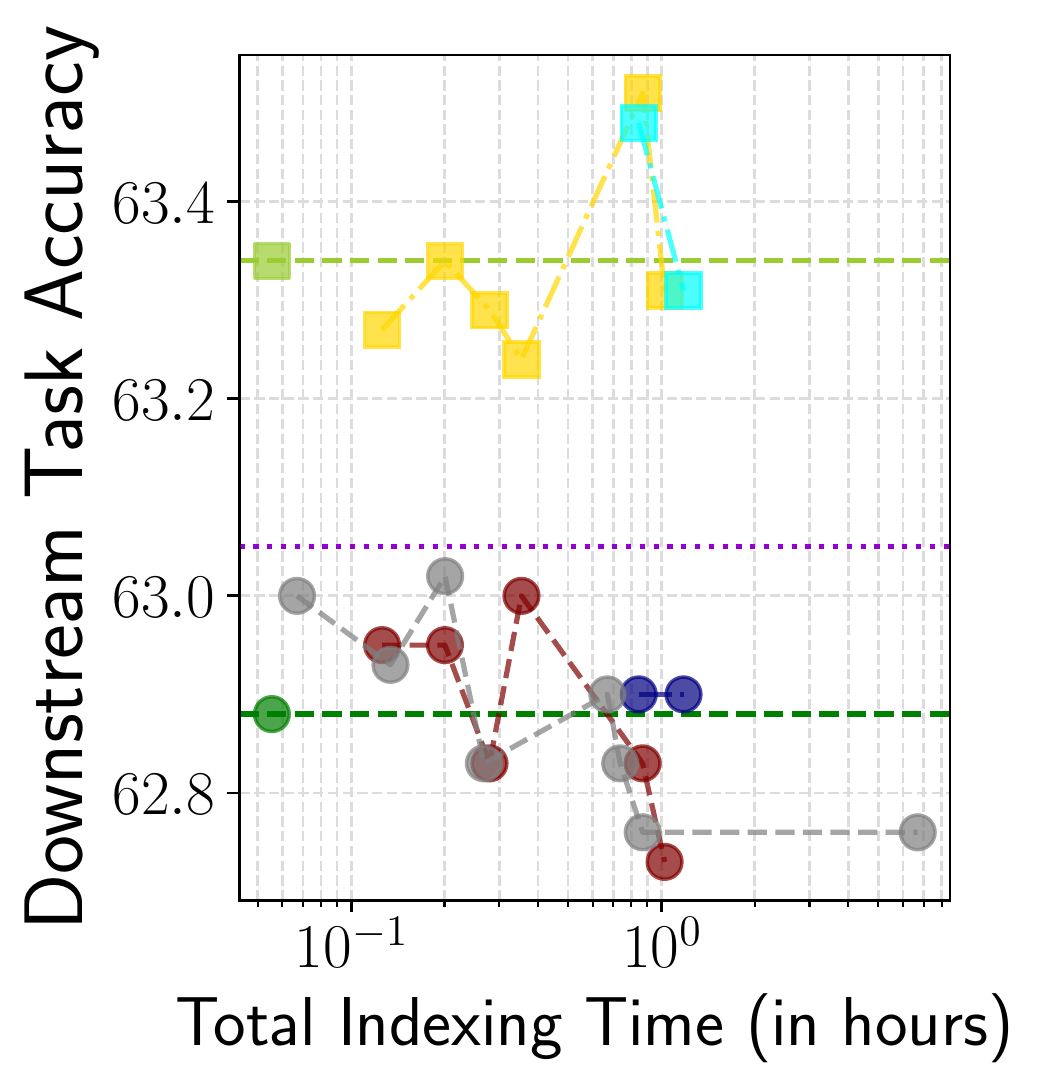}
    \end{subfigure}
    \caption{$\queryTrainSize=100, \queryTestSize=4127$}
    \end{subfigure}

    \begin{subfigure}[b]{\textwidth}
    \begin{subfigure}[b]{\textwidth}
        \centering
        \includegraphics[width=0.64\textwidth]{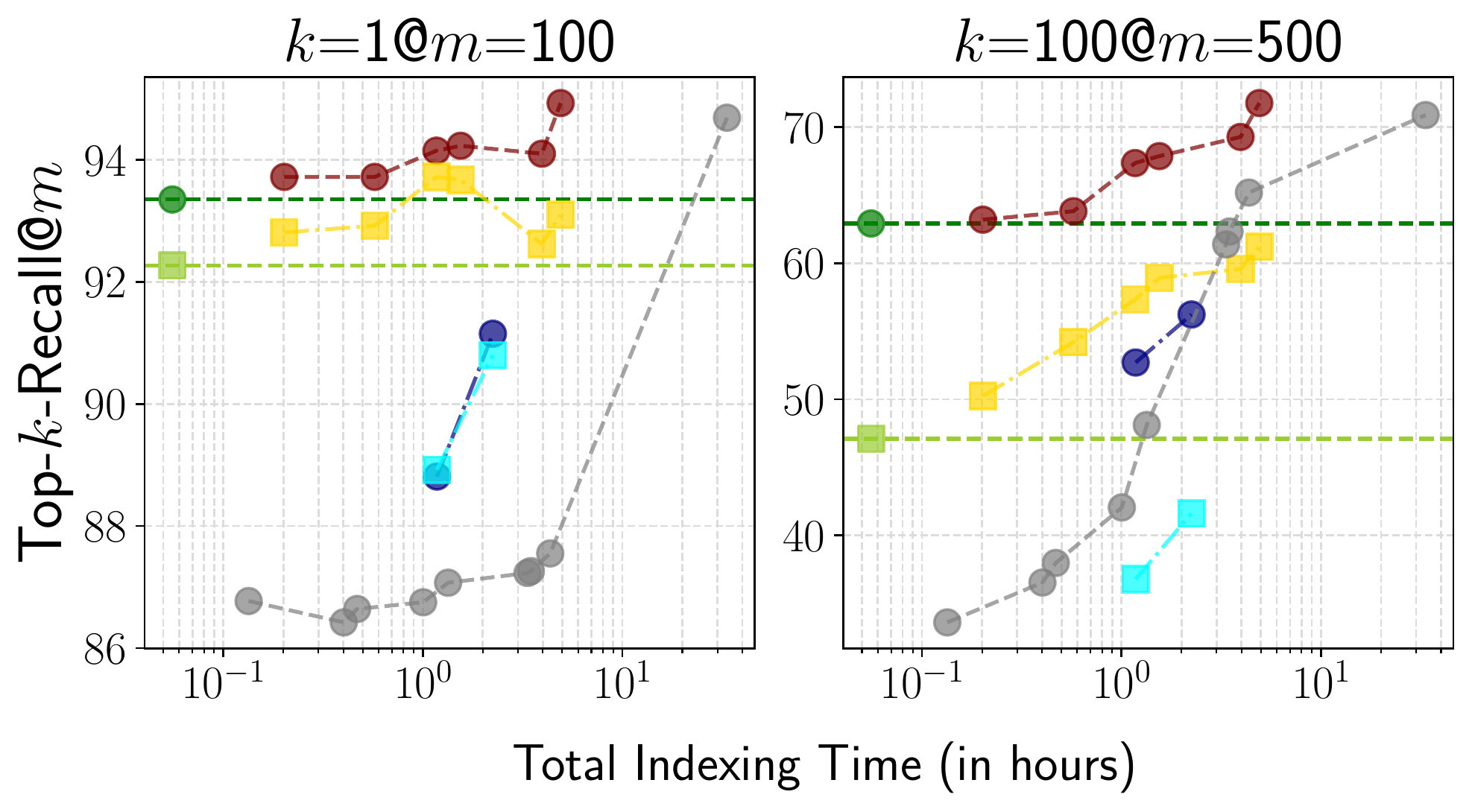}    
        \includegraphics[width=0.32\textwidth]{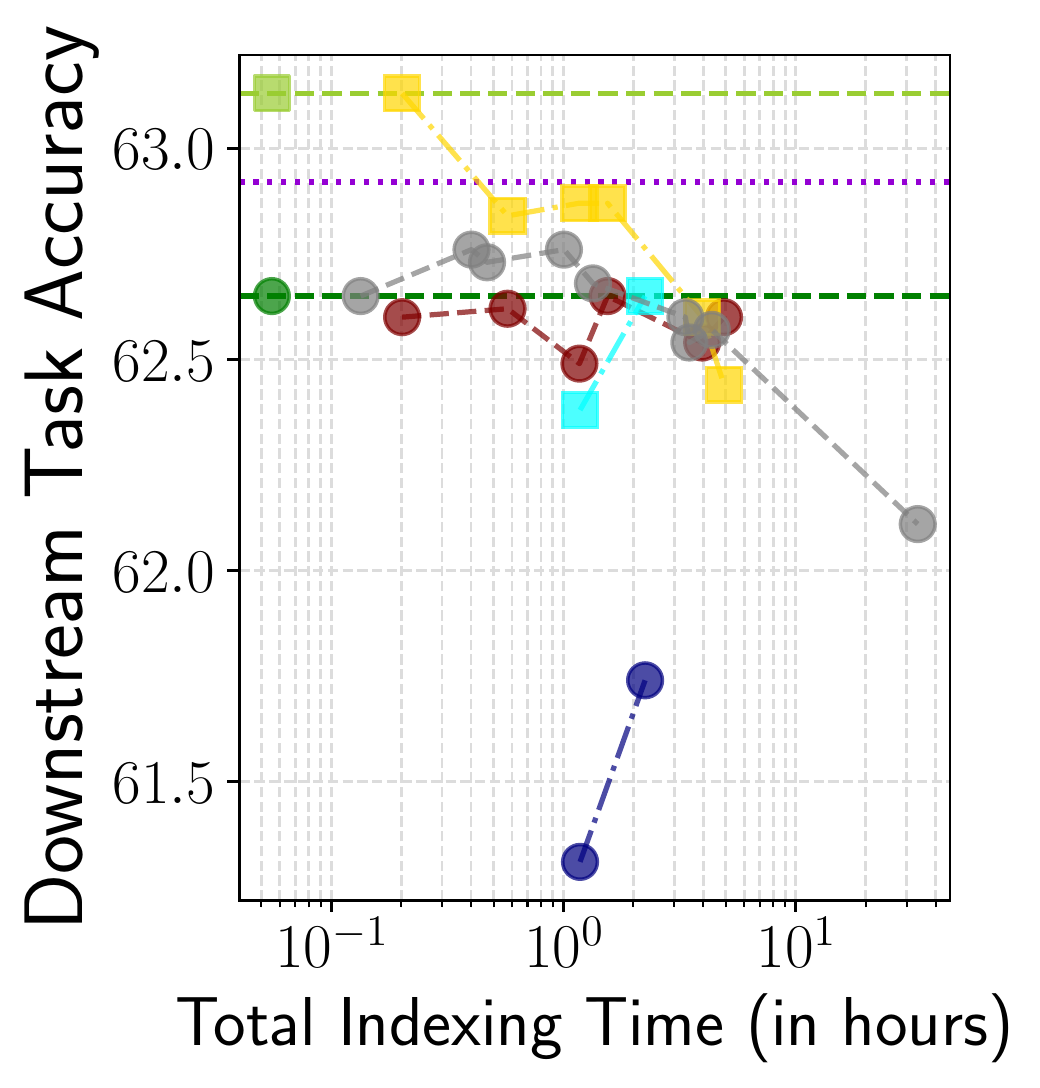}
    \end{subfigure}
    \caption{$\queryTrainSize=500, \queryTestSize=3727$}
    \end{subfigure}

    \begin{subfigure}[b]{\textwidth}
    \begin{subfigure}[b]{\textwidth}
        \centering
        \includegraphics[width=0.64\textwidth]{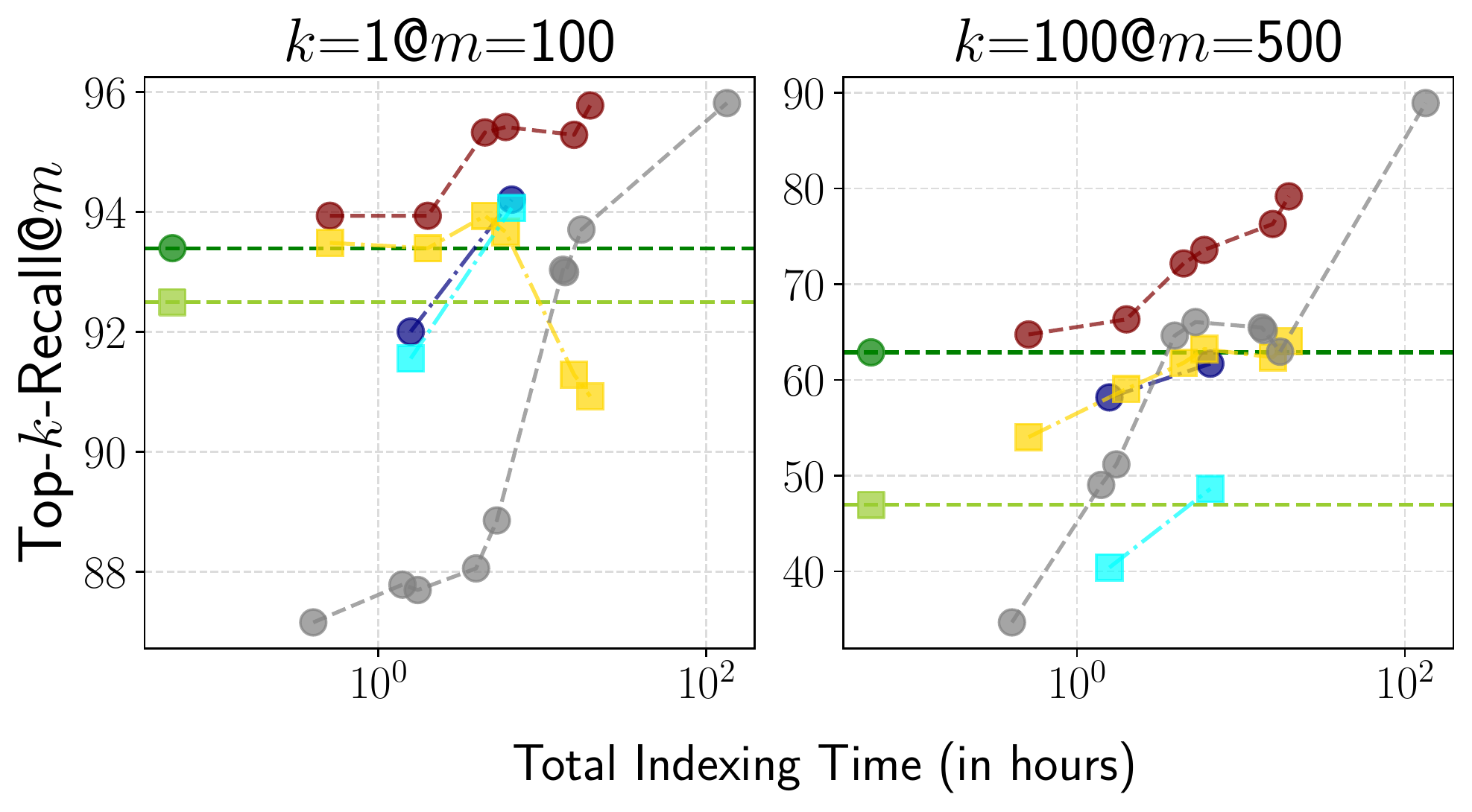}    
        \includegraphics[width=0.32\textwidth]{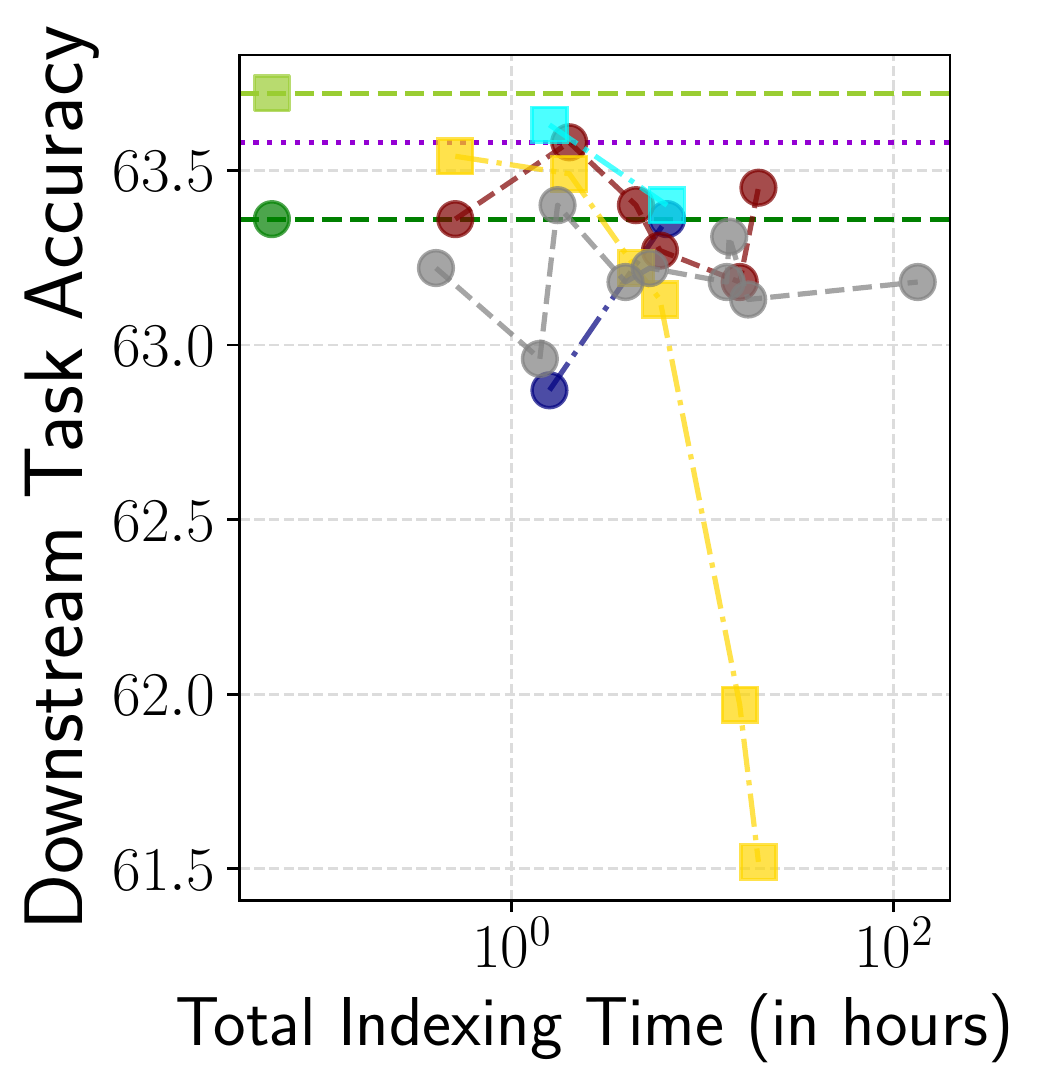}
    \end{subfigure}
    \caption{$\queryTrainSize=2000, \queryTestSize=2227$}
    \end{subfigure}
    \caption{
    Top-$k$-Recall and downstream task accuracy versus indexing time for various approaches on domain=\starTrek. 
    We report Top-1-Recall and Top-100-Recall at fixed inference cost budget ($m$) of 100 and 500 CE calls respectively, and downstream task accuracy for fixed inference cost of 100 CE calls. 
    Each subfigure shows results for different train/test splits.}
    \label{apndx_fig:rq_2a_recall_vs_indexing_cost_wall_clock_time_star_trek}
\end{figure}

\begin{figure}[!t]
    \centering
    \begin{subfigure}[b]{\textwidth}
    \centering
    \includegraphics[width=\textwidth, trim={0 10.1cm 0 0}, clip]{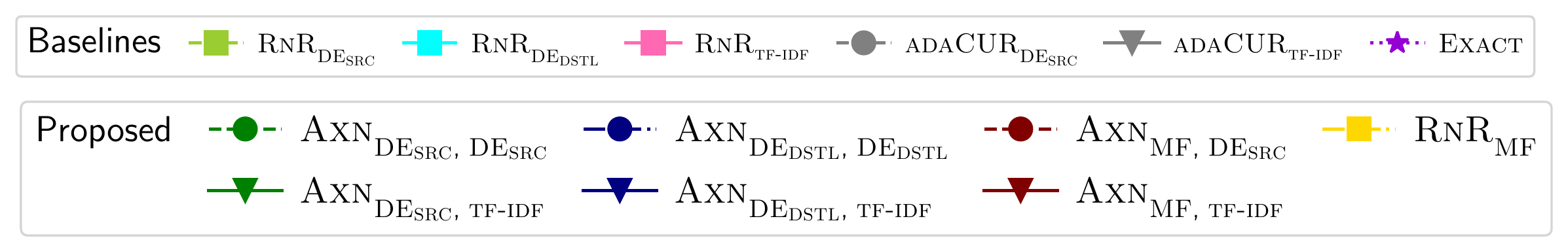}
    \end{subfigure}

    \includegraphics[width=0.9\textwidth]
    {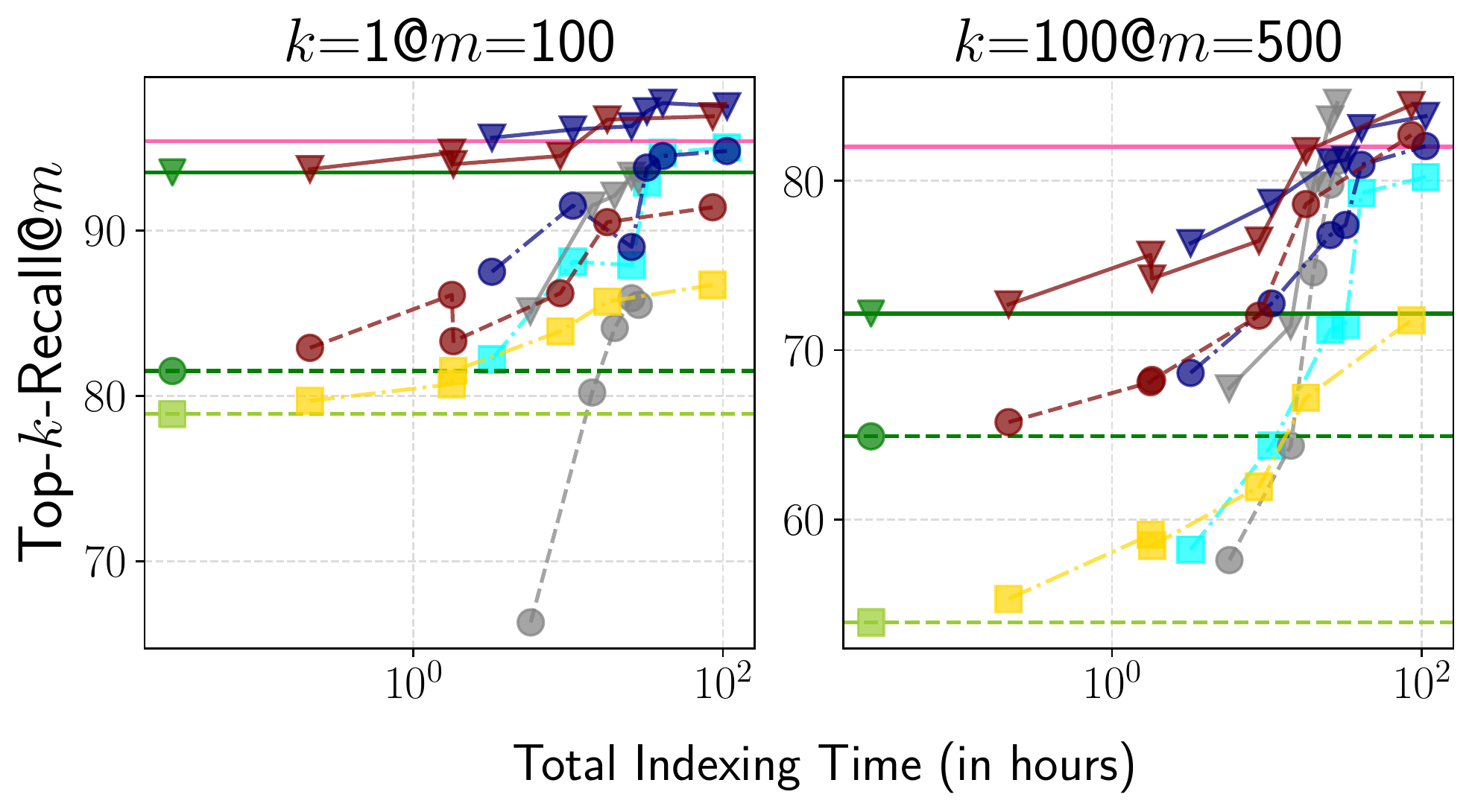}    
    \includegraphics[width=0.45\textwidth]{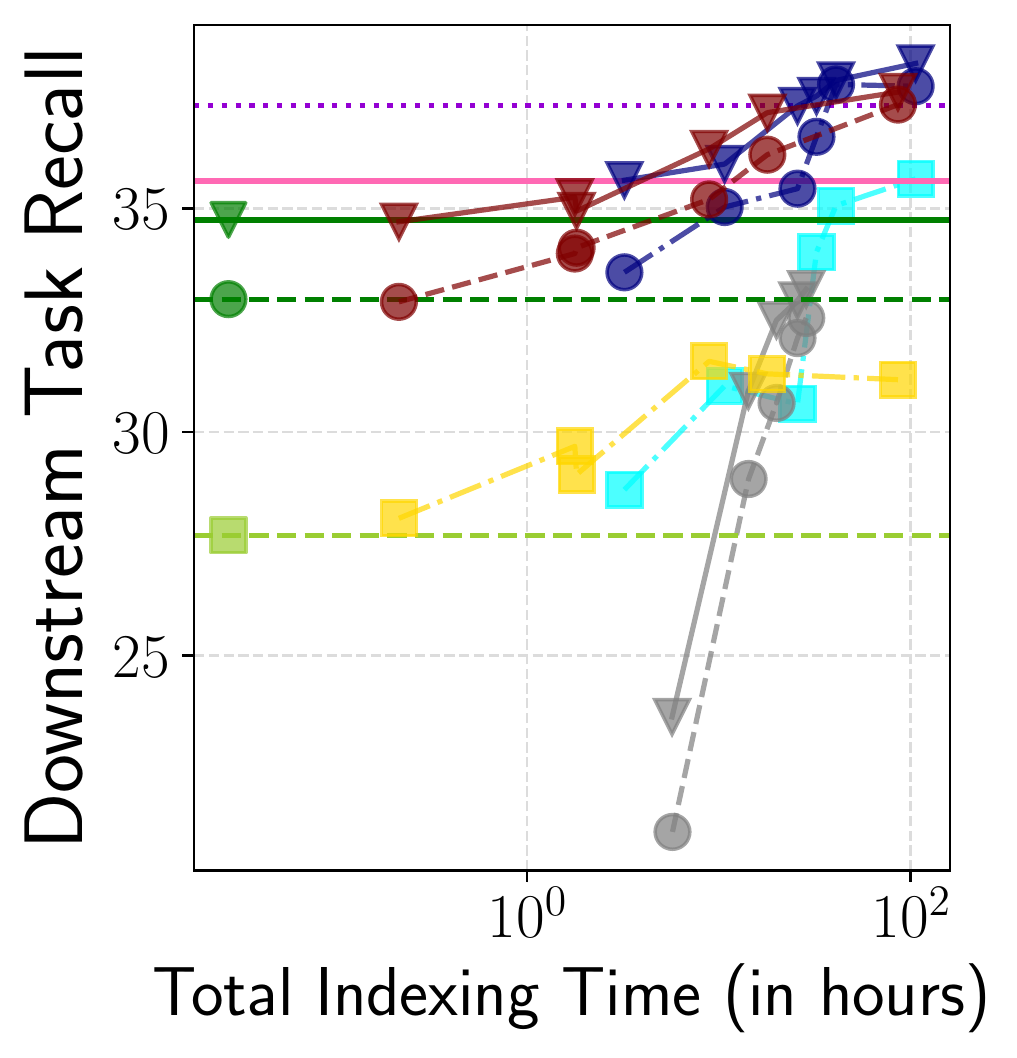}
    \includegraphics[width=0.45\textwidth]{figs/rq_2_test_q=scidocs_all_0-1000_testq_method=all_nq_train=-1_gt_ndcg_at_10_vs_index_cost=time_taken_total_topk=1_at_100.pdf}
    \caption{
    Top-$k$-Recall and downstream task performance metrics versus indexing time for various approaches on domain=\scidocs. 
    We report Top-1-Recall and Top-100-Recall at fixed inference cost budget ($m$) of 100 and 500 cross-encoder (CE) calls respectively, and downstream task metrics for fixed inference cost of 100 cross-encoder calls. We report results for transductive matrix factorization~($\matrixFact{\transductive}$) in these plots. The base dual-encoder (\baseDualEncoder) in these plots is a 6-layer distilbert model finetuned on MS-MARCO dataset. The \baseDualEncoder model is available at \href{https://huggingface.co/sentence-transformers/msmarco-distilbert-base-v2}{https://huggingface.co/sentence-transformers/msmarco-distilbert-base-v2}.}
    \label{apndx_fig:rq_2a_recall_vs_indexing_cost_wall_clock_time_scidocs}
\end{figure}

\begin{figure}[!t]
    \centering
    \begin{subfigure}[b]{\textwidth}
    \centering
    \includegraphics[width=\textwidth, trim={0 10.1cm 0 0}, clip]{figs/legend_gt_recall_2_levels.pdf}
    \end{subfigure}

    \includegraphics[width=0.9\textwidth]{figs/rq_2_test_q=hotpotqa_test_0-1000_testq_method=all_nq_train=-1_kNN_recall_vs_index_cost=time_taken_total_topk=1_at_100_100_at_500.pdf}    
    \includegraphics[width=0.45\textwidth]{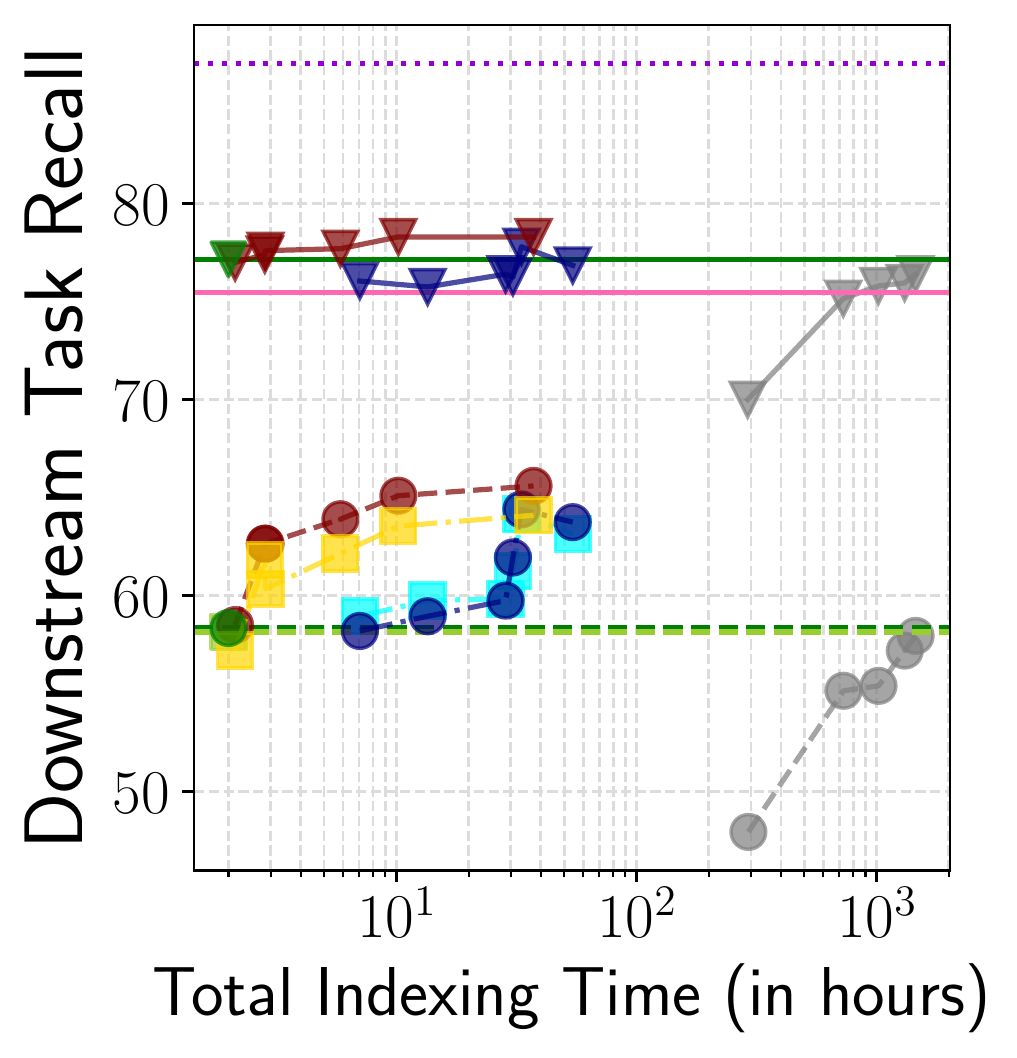}
    \includegraphics[width=0.45\textwidth]{figs/rq_2_test_q=hotpotqa_test_0-1000_testq_method=all_nq_train=-1_gt_ndcg_at_10_vs_index_cost=time_taken_total_topk=1_at_100.pdf}
    \caption{
    Top-$k$-Recall and downstream task performance metrics versus indexing time for various approaches on domain=\hotpotqa. 
    We report Top-1-Recall and Top-100-Recall at fixed inference cost budget ($m$) of 100 and 500 cross-encoder (CE) calls respectively, and downstream task metrics for fixed inference cost of 100 cross-encoder calls. We report results for inductive matrix factorization~($\matrixFact{\inductive}$) in these plots. The base dual-encoder (\baseDualEncoder) in these plots is a 6-layer distilbert model finetuned on MS-MARCO dataset. The \baseDualEncoder model is available at \href{https://huggingface.co/sentence-transformers/msmarco-distilbert-base-v2}{https://huggingface.co/sentence-transformers/msmarco-distilbert-base-v2}.
    }
    \label{apndx_fig:rq_2a_recall_vs_indexing_cost_wall_clock_time_hotpotqa}
\end{figure}

\begin{figure}[!t]
    \centering
    \begin{subfigure}[b]{\textwidth}
    \centering
    \includegraphics[width=\textwidth, trim={0 10.1cm 0 0}, clip]{figs/legend_gt_recall_2_levels.pdf}
    \end{subfigure}
    \includegraphics[width=0.9\textwidth]
    {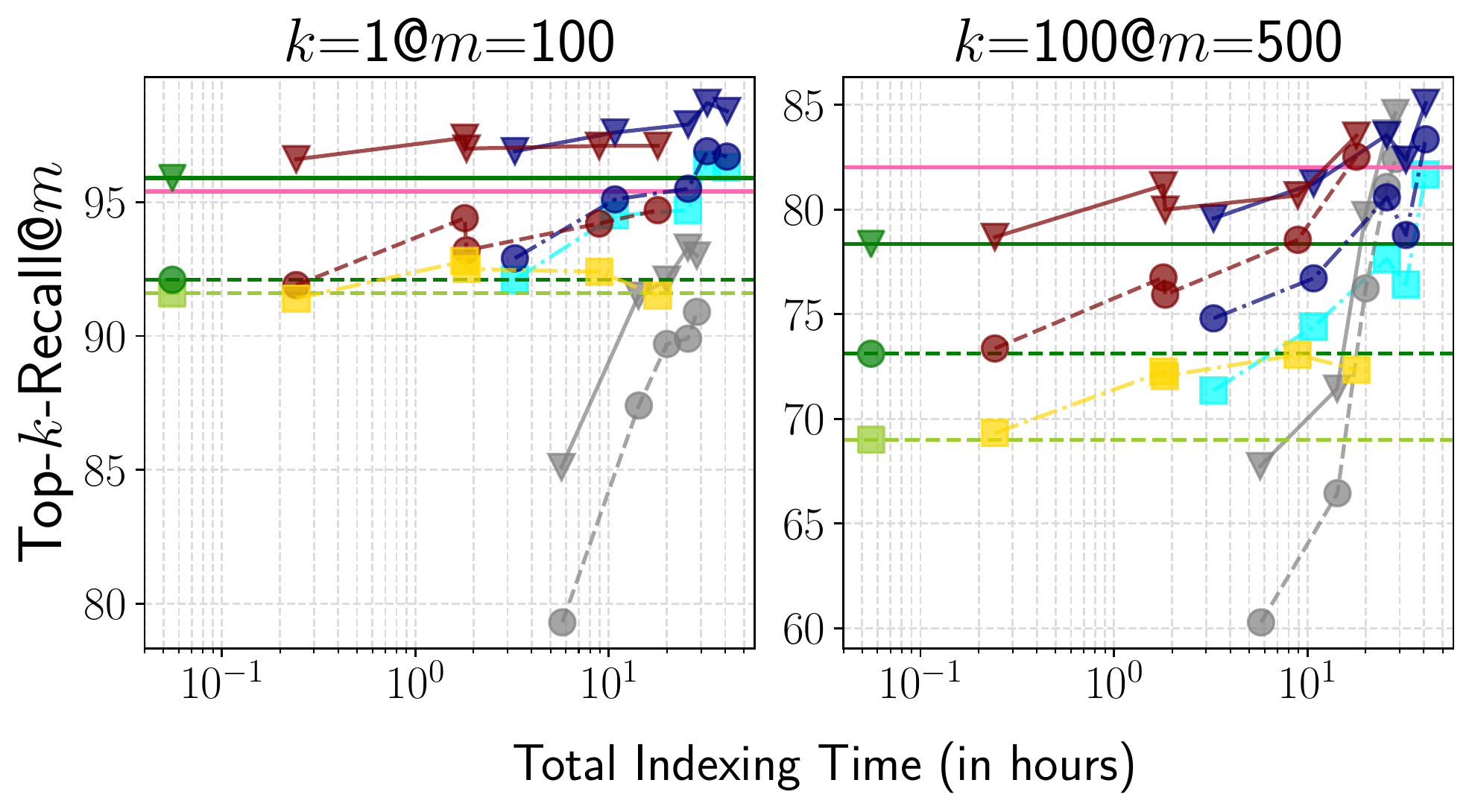}    
    \includegraphics[width=0.45\textwidth]{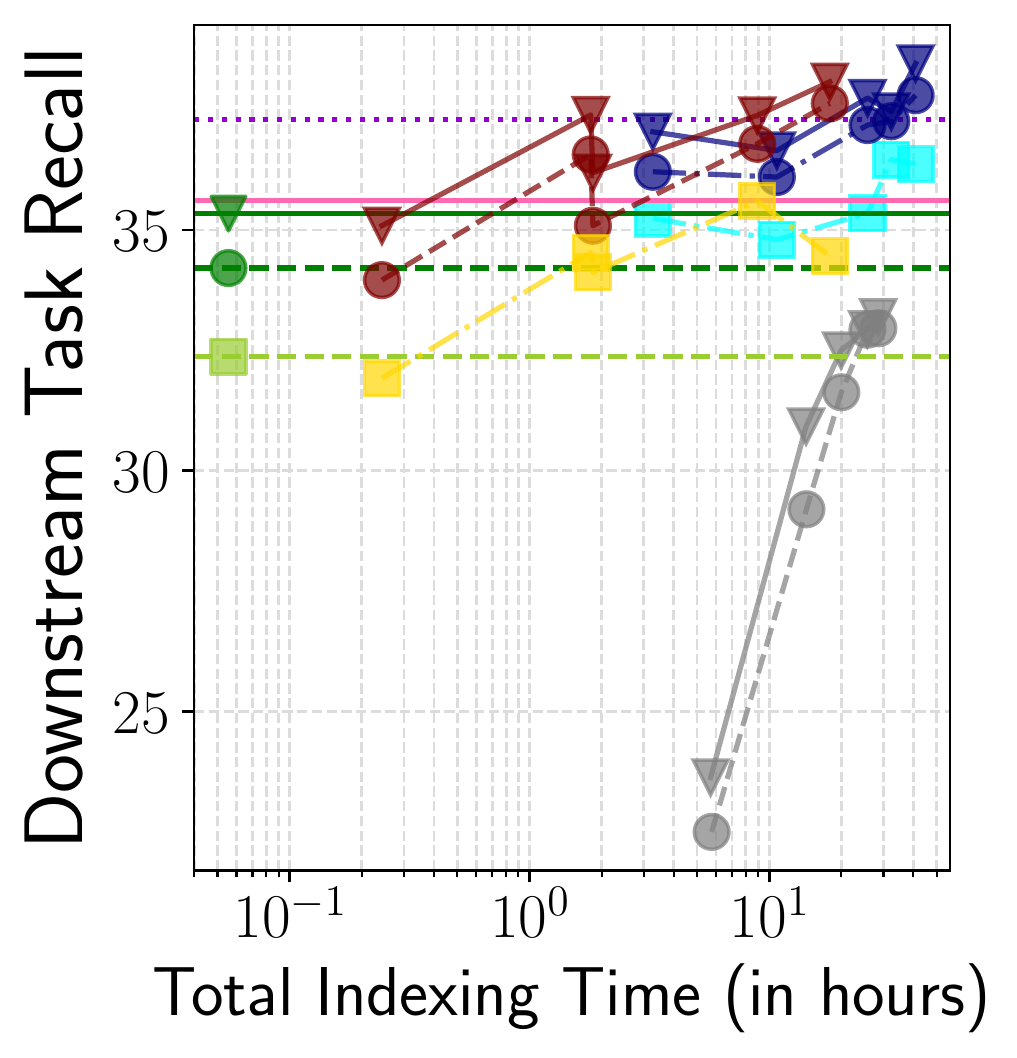}
    \includegraphics[width=0.45\textwidth]{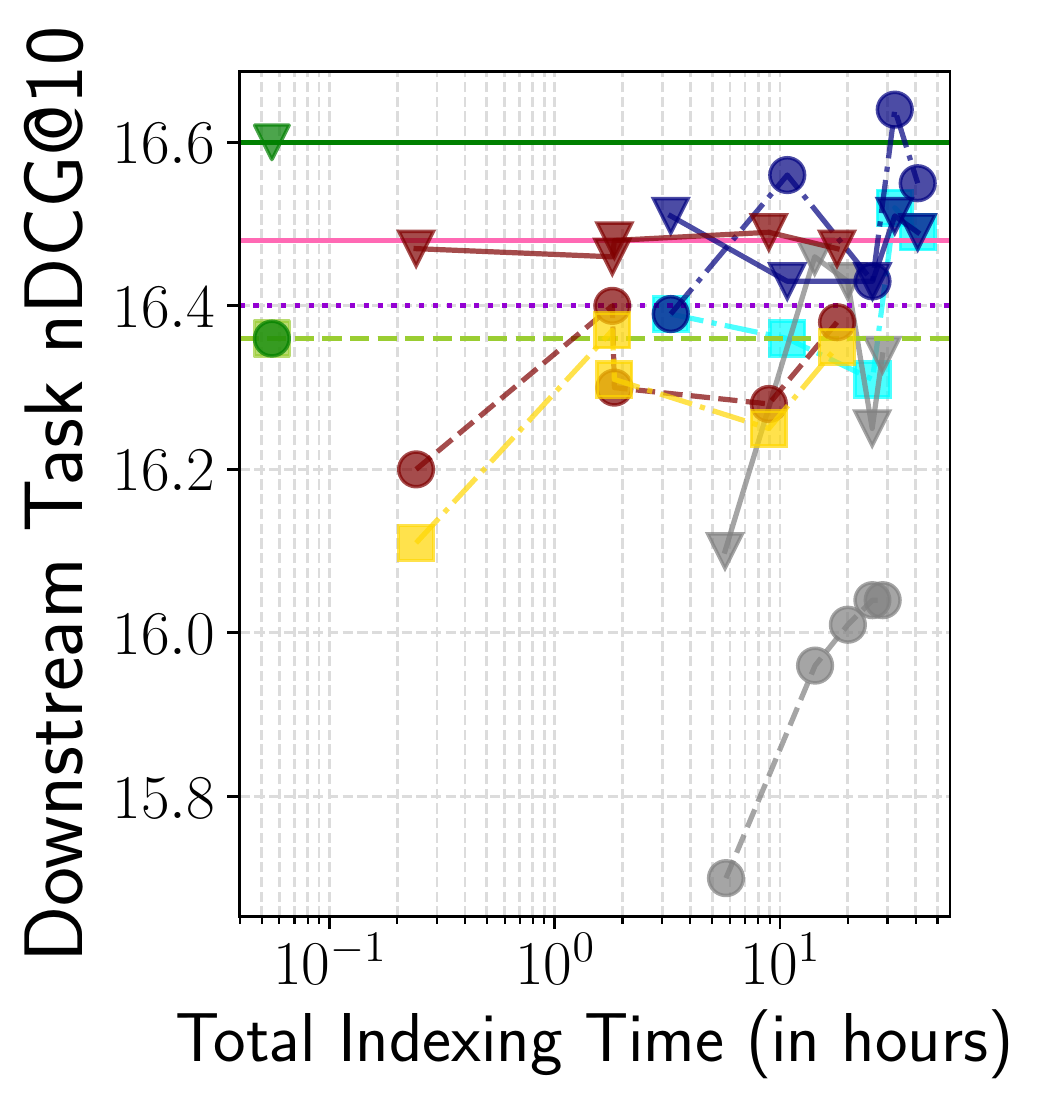}
    \caption{
    Top-$k$-Recall and downstream task performance metrics versus indexing time for various approaches on domain=\scidocs. 
    We report Top-1-Recall and Top-100-Recall at fixed inference cost budget ($m$) of 100 and 500 cross-encoder (CE) calls respectively, and downstream task metrics for fixed inference cost of 100 cross-encoder calls. We report results for transductive matrix factorization~($\matrixFact{\transductive}$) in these plots. The base dual-encoder (\baseDualEncoder) in these plots is a 12-layer bert-base model finetuned on MS-MARCO dataset. The model is available at \href{https://huggingface.co/sentence-transformers/msmarco-bert-base-dot-v5}{https://huggingface.co/sentence-transformers/msmarco-bert-base-dot-v5}. }
    \label{apndx_fig:rq_2a_recall_vs_indexing_cost_wall_clock_time_scidocs_v5_bienc}
\end{figure}

\begin{figure}[!t]
    \centering
    \begin{subfigure}[b]{\textwidth}
    \centering
    \includegraphics[width=\textwidth, trim={0 10.1cm 0 0}, clip]{figs/legend_gt_recall_2_levels.pdf}
    \end{subfigure}

    \includegraphics[width=0.9\textwidth]{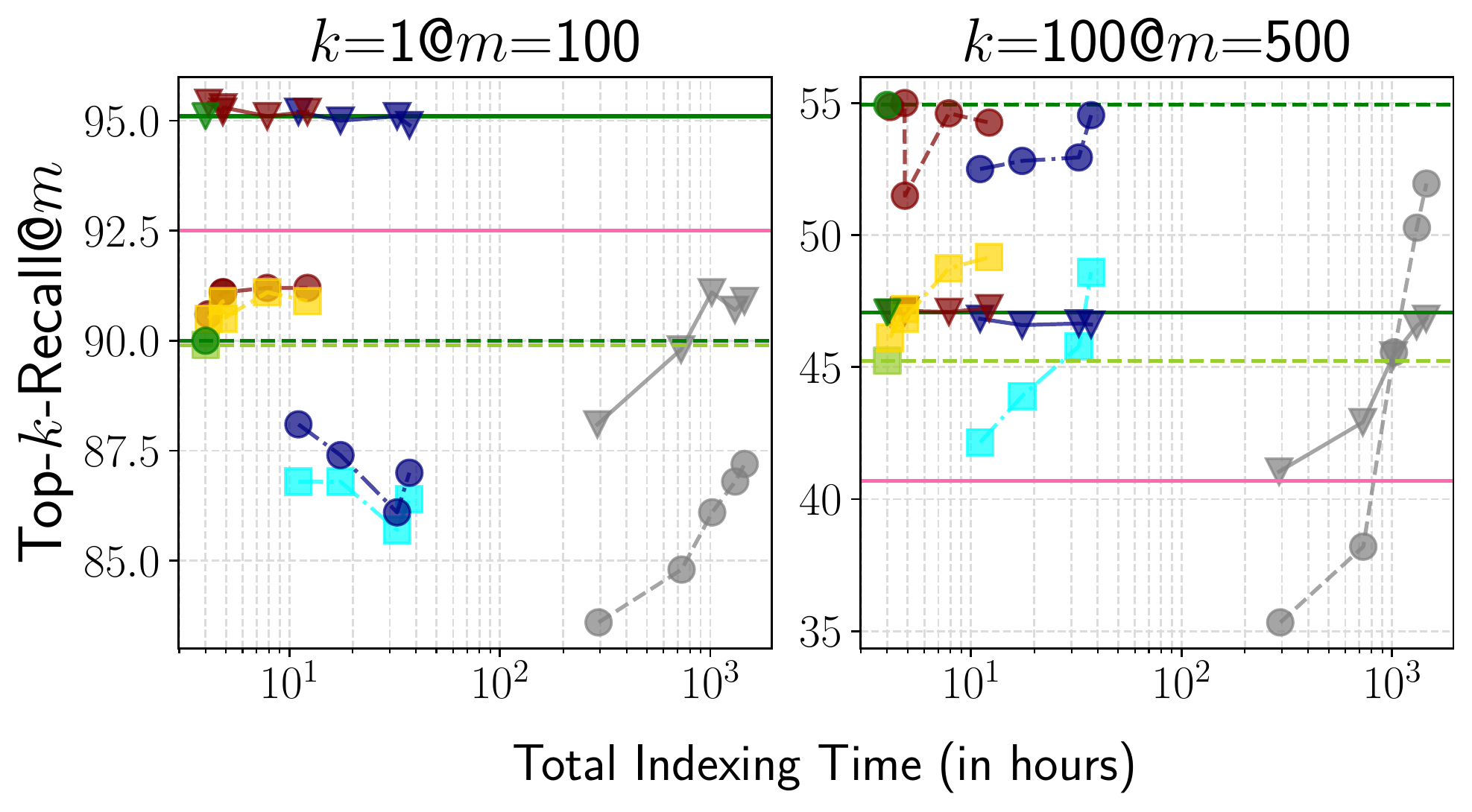}    
    \includegraphics[width=0.45\textwidth]{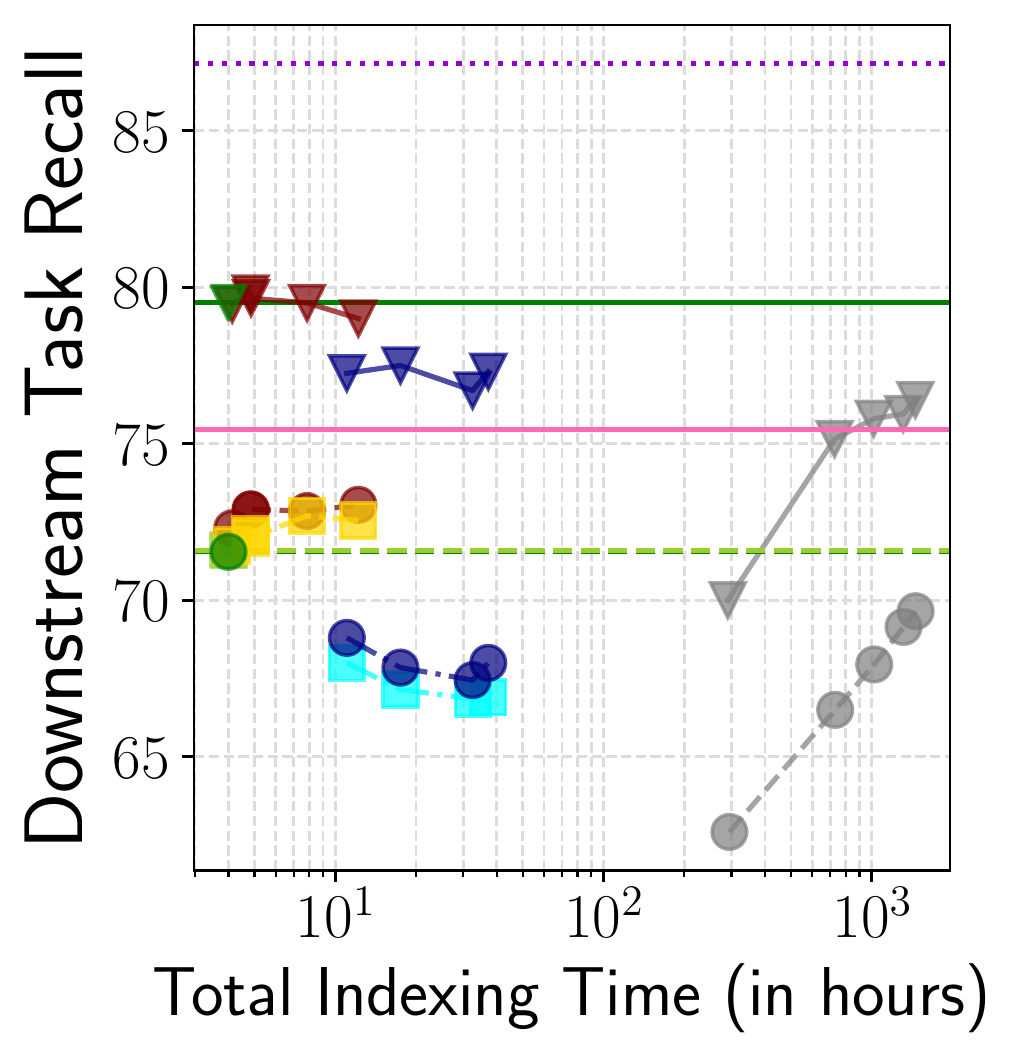}
    \includegraphics[width=0.45\textwidth]{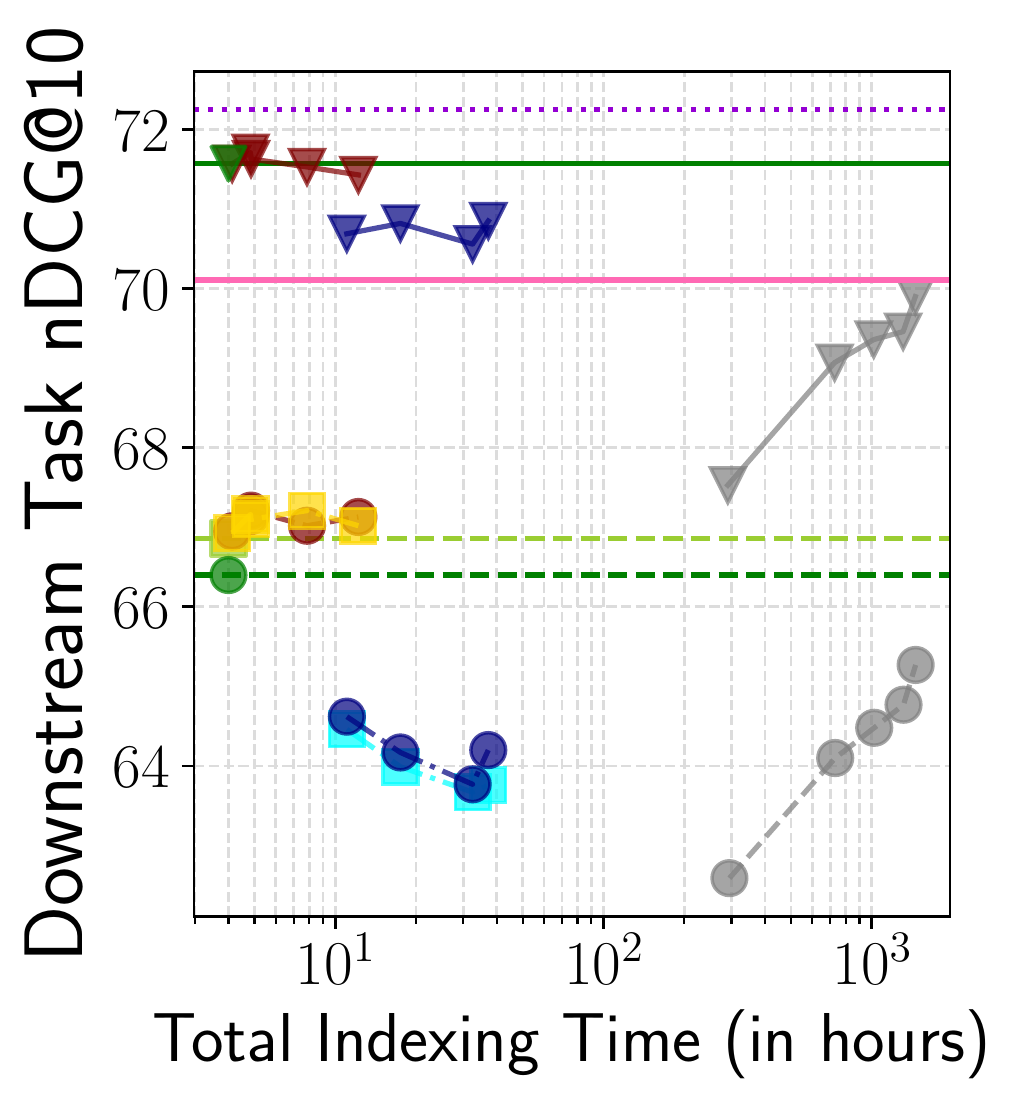}
    \caption{
    Top-$k$-Recall and downstream task performance metrics versus indexing time for various approaches on domain=\hotpotqa. 
    We report Top-1-Recall and Top-100-Recall at fixed inference cost budget ($m$) of 100 and 500 cross-encoder (CE) calls respectively, and downstream task metrics for fixed inference cost of 100 cross-encoder calls. We report results for inductive matrix factorization~($\matrixFact{\inductive}$) in these plots. The base dual-encoder (\baseDualEncoder) in these plots is a 12-layer bert-base model finetuned on MS-MARCO dataset. This \baseDualEncoder model is available at \href{https://huggingface.co/sentence-transformers/msmarco-bert-base-dot-v5}{https://huggingface.co/sentence-transformers/msmarco-bert-base-dot-v5}. }
    \label{apndx_fig:rq_2a_recall_vs_indexing_cost_wall_clock_time_hotpotqa_v5_bienc}
\end{figure}

\end{document}